\documentclass{siamart1116}
\usepackage{indentfirst}
\usepackage{graphicx}
\usepackage{epstopdf}
\usepackage{epsfig}
\usepackage{overpic}
\usepackage{overcite}
\usepackage{array}
\usepackage{color}
\usepackage{multirow}
\usepackage{float}
\usepackage{subfigure} 
\usepackage{amsmath,amssymb}
\usepackage[mathscr]{euscript}
\usepackage{latexsym,bm}
\usepackage{booktabs}
\usepackage{bbm}
\usepackage{diagbox}
 \usepackage[all]{xy}
 \usepackage{color}
\usepackage{diagbox}
\usepackage{algorithm}  
\usepackage{algpseudocode}  
\usepackage{hyperref}
\usepackage{graphicx}
\usepackage{mathtools}
\newcommand\D{\textup{d}}

\def\mi{\mathrm{i}}
\def\me{\mathrm{e}}

\def\bx{\bm{x}}

\def\bk{\bm{k}}
\def\bp{\bm{p}}

\def\by{\bm{y}}

\def\bn{\bm{n}}

\def\bzeta{\bm{\zeta}}
\def\bxi{\bm{\xi}}

\def\pdo{{\rm \Psi} \textup{DO}}
\makeatletter
\@addtoreset{equation}{section}
\makeatother
\renewcommand{\theequation}{\arabic{section}.\arabic{equation}}

\newtheorem{example}{\textup{\textbf{Example}}}
\newtheorem{remark}{Remark}

\begin{document}
\bibliographystyle{unsrt}

\title{A characteristic-spectral-mixed scheme for six-dimensional Wigner-Coulomb dynamics}
\author{Yunfeng Xiong\footnotemark[1], Yong Zhang\footnotemark[2], Sihong Shao\footnotemark[1] \footnotemark[3]
}
\renewcommand{\thefootnote}{\fnsymbol{footnote}}
\footnotetext[1]{CAPT, LMAM and School of Mathematical Sciences, Peking University, Beijing, China. Email addresses: {\tt xiongyf@math.pku.edu.cn} (Y. Xiong). {\tt sihong@math.pku.edu.cn} (S. Shao)}
\footnotetext[2]{The School of Mathematics, Tianjin University, Tianjin, China. Email address: {\tt sunny5zhang@163.com} (Y. Zhang)}
\footnotetext[3]{To whom correspondence should be addressed.}
\maketitle

\begin{abstract}
Numerical resolution for 6-D Wigner dynamics under the Coulomb potential faces with the combined challenges of high dimensionality, nonlocality, oscillation and singularity. In particular, the extremely huge memory storage of 6-D grids hinders the usage of all existing deterministic numerical scheme, which is well-known as the curse of dimensionality. To surmount these difficulties, we propose a massively parallel solver, termed the CHAracteristic-Spectral-Mixed (CHASM) scheme, by fully exploiting two distinct features of the Wigner equation: Locality of spatial advection and nonlocality of quantum interaction. Our scheme utilizes the local cubic B-spline basis to interpolate the local spatial advection. The key is to use a perfectly matched boundary condition to give a closure of spline coefficients, so that distributed pieces can recover the global one as accurately as possible owing to the rapid decay of wavelet basis in the dual space, and communication costs are significantly reduced. To resolve the nonlocal pseudodifferential operator with weakly singular symbol, CHASM further adopts the truncated kernel method to attain a highly efficient approximation. Several typical  experiments including the quantum harmonic oscillator and Hydrogen 1s state demonstrate the accuracy and efficiency of CHASM. The non-equilibrium electron-proton couplings are also clearly displayed and reveal the uncertainty principle and quantum tunneling in phase space. Finally, the scalability of CHASM up to 16000 cores is  presented.

 \vspace*{4mm}





\end{abstract}

\newsavebox{\tablebox}
\setcounter{tocdepth}{3}

\section{Introduction}

The recently burgeoned developments in nano-science and semiconductors, such as the nano-wired FET at 3nm node \cite{DeyJenaMohapatraDashDasMaiti2020}, as well as those in high energy density physics \cite{GrazianiBauerMurillo2014}, quantum tomography \cite{RundleDaviesDwyerToddEveritt2020} and quantum optics \cite{TianWangEberly2017,HanGeFangYuGuoMaDengGongLiu2019}, urgently demand efficient and highly accurate simulations of high-dimensional quantum models. Specifically, the Wigner equation  \cite{Wigner1932} under the Coulomb interaction is of great importance in describing the non-equilibrium electron dynamics in quantum regime, including the electron-proton couplings in hot density matter \cite{GrazianiBauerMurillo2014}, the quantum entanglement in nano-wires \cite{BenamBallicchiaWeinbubSelberherrNedjalkov2021}, the quantum tunneling effects in nanodevices \cite{bk:NedjalkovQuerliozDollfusKosina2011}, strong-field atomic ionization processes \cite{TianWangEberly2017,HanGeFangYuGuoMaDengGongLiu2019} and visualization of quantum states \cite{KurtsieferPfauMlynek1997,DaviesRundleDwyerToddEveritt2019}, owing to its huge advantage in calculating quantum statistics and experimental observability \cite{bk:CurtrightFairlieZachos2013}. However, an investigation of realistic quantum systems in 3-D spatial space requires to solve the Wigner equation in 6-D phase space, so that the curse of dimensionality (CoD) poses a tremendous obstacle to its numerical resolution.

Indeed, it has already taken over thirty years to develop efficient Wigner solvers, including both deterministic and stochastic algorithms. In contrast to the relatively newer branch of particle-based stochastic methods \cite{KosinaNedjalkovSelberherr2003,MuscatoWagner2016,ShaoXiong2019}, which usually exhibit slower convergence rate, grid-based deterministic solvers allow  highly accurate numerical resolutions in the light of their concise principle and solid mathematical foundation, ranging from the finite difference scheme \cite{Frensley1989} and the spectral collocation method combined with the operator splitting \cite{Ringhofer1990,ArnoldRinghofer1995} to the recent advanced techniques such as the spectral element method \cite{ShaoLuCai2011,XiongChenShao2016,ChenShaoCai2019}, the spectral decomposition \cite{VandePutSoreeMagnus2017} and the Hermite spectral method \cite{FurtmaierSucciMendoza2015,ZhanCaiHu2021}, as well as those for advection such as the discontinuous Galerkin method \cite{GambaGualdaniSharp}, WENO scheme \cite{DordaSchurrer2015} and exponential integrators \cite{FurtmaierSucciMendoza2015}. Unfortunately, there still remains a huge gap in terms of the applicability of even the state-of-the-art deterministic scheme to full 6-D problems, and the foremost problem is definitely the storage of 6-D grid mesh. On one hand, the required memory to store a fine 6-D tensor is still prohibitive for a single computer, e.g., the requirement to store a uniform grid mesh of size $81^3 \times 64^3$ in single precision is about $81^3\times64^3\times 4/1000^3 \approx 557$GB. On the other hand, the highly oscillatory structure of the Wigner function poses a severe restriction on the sampling frequency \cite{Frensley1989}, which is further complicated by singular potentials like the Coulomb interaction. As a consequence, it strongly calls for an efficient algorithm that should be highly accurate enough to capture the fine structure of the solutions and suitable for modern high-performance computing platform.

This paper makes the first attempt to simulate the 6-D Wigner equation via a massively parallel deterministic solver. The proposed CHArcteristic-Spectral-Mixed (CHASM) scheme takes advantages of both the parallel semi-Lagrangian scheme \cite{MalevskyThomas1997,KormannReuterRampp2019} and the spectral method, under the same guiding principle in our preceding advective-spectral-mixed (ASM) scheme \cite{XiongChenShao2016}. Specifically, it exploits two distinct features of the Wigner equation:  Locality in spatial advection and nonlocality in quantum interaction. The local cubic B-spline, as a kind of wavelet basis, is applied for interpolating the local advection, while the Fourier basis is adopted to tackle the nonlocal pseudodifferential operator ($\pdo$) due to its intrinsic global and oscillatory nature.

There are two major difficulties to be resolved. The first is how to distribute a global cubic spline into several patches because solving the spline coefficients indeed requires the information from all patches. Owing to a key observation of the rapid decay property of wavelet basis in the dual space \cite{bk:Chui1992,MalevskyThomas1997}, we introduce a perfectly matched boundary condition (PMBC) for patched splines to give a closure of the spline coefficients, which allows the local splines to recover the global one as accurately as possible. Domain decomposition is only performed in the spatial direction so that communications can be restricted in adjacent  processors. 

The second is how to tackle $\pdo$ with a singular Riesz kernel (see Eq.~\eqref{def.pdo}) as the singularity causes  troubles in the convergence of the commonly used Fourier spectral method \cite{Ringhofer1990,Goudon2002}. Motivated from recent progress in fast algorithm for singular convolution \cite{VicoGreengardFerrando2016,GreengardJiangZhang2018,LiuZhangZhang2022}, we utilize the truncated kernel method (TKM) to derive a highly efficient approximation to $\pdo$. With these endeavors, we succeed in simulating 6-D Wigner-Coulomb dynamics of an electron wavepacket attracted by one or two protons. The solutions may help reveal the presence of electron-proton coupling \cite{GrazianiBauerMurillo2014,BenamBallicchiaWeinbubSelberherrNedjalkov2021}, uncertainty principle and quantum tunneling \cite{PakHammesSchiffer2004} in  phase space.

The rest of this paper is organized as follows. In Section \ref{sec.back}, we briefly review the background of the Wigner equation and the characteristic method. In Section \ref{sec.characteristic}, we mainly illustrate the construction of local splines to interpolate the spatial advection.
 Section \ref{spectral} discusses TKM for $\pdo$ with a weakly singular symbol. Several typical numerical experiments are performed in Section \ref{sec.num} to verify the accuracy of CHASM, where a first attempt to simulate quantum Coulomb dynamics in 6-D phase space is obtained. Finally, the conclusion is drawn in Section \ref{sec.discussion}.

\section{Background}
\label{sec.back}
As a preliminary, we make a brief review of the single-body Wigner equation and outline the framework of the characteristic method.

\subsection{The Wigner equation} Quantum mechanics in phase space is rendered by the Wigner function, the Weyl-Wigner transform of a density matrix $\rho(\bx_1, \bx_2, t)$,
\begin{equation}\label{def.Wigner_function}
f(\bx, \bk, t) = \int_{\mathbb{R}^3} \rho(\bx - \frac{\by}{2}, \bx + \frac{\by}{2}, t) \me^{-\mi \bk \cdot \by} \D \by,
\end{equation}
where $\bx$ is the spatial variable and $\bk$ the Fourier conjugated wave vectors. The Wigner function plays a similar role as the probability density function, but allows negative values due to Heisenberg's uncertainty principle. 
The governing equation, known as the Wigner equation, is a partial integro-differential equation,
\begin{equation}\label{eq.Wigner}
\frac{\partial }{\partial t}f(\bm{x}, \bm{k}, t)+ \frac{\hbar \bm{k}}{m} \cdot \nabla_{\bm{x}} f(\bm{x},\bm{k}, t)  = \Theta_V[f](\bx, \bk, t),
\end{equation}
where $m$ is the mass, $\hbar$ is the reduced Planck constant and $\pdo$ reads as
\begin{equation}\label{def.pdo_convolution}
\Theta_V[f](\bx, \bk, t) = \frac{1}{\mi \hbar (2\pi)^3} \iint_{\mathbb{R}^{6}} \me^{-\mi (\bk - \bk^{\prime}) \cdot \by }D_V(\bx, \by, t) f(\bx, \bk^{\prime}, t) \D \by \D \bk^{\prime}
\end{equation}
with $D_V(\bx, \by, t) = V(\bx + \frac{\by}{2}) - V(\bx - \frac{\by}{2})$. 

The Coulomb interaction  in $\bx \in \mathbb{R}^3$ is of great importance in realistic applications.
When the atomic unit $ m = \hbar = e = 1$ is adopted and the attractive Coulomb potential is considered,  $V(\bx) = -{1}/{|\bx -  \bx_A |}$, $\pdo$ is equivalent to
\begin{equation}\label{def.pdo}
\Theta_{V}[f](\bx, \bk, t) = \frac{2}{c_{3, 1}\mi}  \int_{\mathbb{R}^3} \me^{2 \mi (\bx - \bx_A) \cdot \bk^{\prime}} \frac{1}{|\bk^{\prime}|^2}  ( f(\bx, \bk - \bk^{\prime}, t) - f(\bx, \bk+\bk^{\prime}, t) )\D \bk^{\prime}
\end{equation}
with $c_{n, \alpha} = \pi^{n/2} 2^\alpha {\Gamma(\frac{\alpha}{2})}/{\Gamma(\frac{n-\alpha}{2})}$. It is a twisted convolution involving both singular kernel and phase factor.  When the interacting body is torn away from the atom, i.e., $|\bx - \bx_A|$ increases, $\pdo$ decays as the phase factor becomes more oscillating.

Since $\pdo$ is real-valued due to the symmetry $\bk \to - \bk$ and
\begin{equation}\label{mass_conserve}
\int_{\mathbb{R}^3} \Theta_V[f](\bx, \bk, t) \D \bk = 0  \iff  \frac{\D}{\D t} \iint_{\mathbb{R}^3 \times \mathbb{R}^3} f(\bx, \bk, t) \D \bx \D \bk = 0,
\end{equation}
the total mass is conserved. The Wigner equation with $\pdo$ \eqref{def.pdo} have many stationary solutions given by the Weyl-Wigner transform of $\rho(\bx, \by) =\phi(\bx) \phi^\ast(\by)$, with $\phi(\bx)$ being eigenfunction of the corresponding Schr{\"o}dinger equation.

\subsection{The Lawson scheme and the characteristic methods}

A typical numerical scheme for solving Eq.~\eqref{eq.Wigner} is the characteristic method. Its derivation starts from the variation-of-constant formula of \eqref{eq.Wigner},
\begin{equation}\label{variation_of_constant}
f(\bx, \bk, t) = \me^{-\frac{\hbar t}{m} \bk \cdot \nabla_{\bx}} f(\bx, \bk, 0) + \int_0^t  \me^{-\frac{\hbar \tau}{m} \bk \cdot \nabla_{\bx}} \Theta_V[f](\bx, \bk, t - \tau) \D \tau,
\end{equation}
where the semigroup $\me^{-\frac{\hbar \tau}{m} \bk \cdot \nabla_{\bx}}$ corresponds to the advection along the characteristic line, say, $\me^{-\frac{\hbar \tau}{m} \bk \cdot \nabla_{\bx}} f(\bx, \bk, t) =  f(\mathcal{A}_\tau(\bx, \bk), t - \tau)$ with $\mathcal{A}_\tau(\bx, \bk) = (\bx - \frac{\hbar \bk}{m} \tau, \bk)$.

The characteristic method approximates the integral on the right hand side of Eq.~\eqref{variation_of_constant} by polynomial interpolation in the light of the Lawson scheme,
\begin{equation}
f^n(\bx, \bk) = f^{n-1}(\mathcal{A}_\tau(\bx, \bk)) + \tau \sum_{j=0}^{q} \beta_j  \Theta_V[f^{n-j}](\mathcal{A}_{j\tau}(\bx, \bk)).
\end{equation}
We adopt the one-stage Lawson predictor-corrector scheme (LPC1):
\begin{equation*}
\begin{split}
\textup{Predictor}:\widetilde{f}^{n+1}(\bx, \bk) &= f^{n}(\mathcal{A}_\tau(\bx, \bk)) + \tau \Theta_{V}[f^{n}](\mathcal{A}_\tau(\bx, \bk)), \\
\textup{Corrector}: f^{n+1}(\bx, \bk) &=  f^{n}(\mathcal{A}_\tau(\bx, \bk)) + \frac{\tau}{2} \Theta_V[\widetilde{f}^{n+1}](\bx, \bk) + \frac{\tau}{2} \Theta_{V}[f^{n}](\mathcal{A}_\tau(\bx, \bk)).
\end{split}
\end{equation*} 
The Strang splitting is also an efficient strategy for temporal integration and its success in solving 6-D Boltzmann equation was reported in \cite{DimarcoLoubereNarskiRey2018}. However, the non-splitting Lawson scheme is believed to be more advantageous in numerical stability \cite{CrouseillesEinkemmerMassot2020}.

The remaining problem is how to evaluate the exact flow $ f^{n}(\mathcal{A}_{\tau}(\bx, \bk))$ and  $ \Theta_V[f^{n}](\mathcal{A}_{\tau}(\bx, \bk))$  on the shifted grid. In general, they can be interpolated via a specified basis expansion of $f^n$ within the framework of the semi-Lagrangian method, such as the spline wavelets \cite{CrouseillesLatuSonnendrucker2009,Kormann2015}, the Fourier basis and the Chebyshev polynomials \cite{ChenShaoCai2019}.  Regarding that the spatial advection is essentially local, we adopt the cubic B-spline as it is a kind of wavelet basis with low numerical dissipation and the cost scales as $\mathcal{O}(N_x^d)$ ($d$ is dimensionality) \cite{CrouseillesLatuSonnendrucker2009}.

Here we focus on the unidimensional uniform setting, while the multidimensional spline can be constructed by its tensor product (see Section \ref{sec:6d_parallel} below). Suppose the computational domain is $[x_0, x_{N}]$ containing $N+1$ grid points with uniform spacing $h = {(x_{N} - x_0)}/{N}$. The projection of $\varphi(x)$ onto the cubic spline basis is given by
\begin{equation}\label{interpolation}
\varphi(x) \approx s(x) = \sum_{\nu = -1}^{N+1}  \eta_{\nu} B_{\nu}(x) \quad \textup{subject to} \quad \varphi(x_i) = s(x_i), \quad i = 0, \dots, N,
\end{equation}
where $B_\nu$ is the cubic B-spline with compact support over four grid points, 
\begin{equation}
B_{\nu}(x) = 
\left\{
\begin{split}
&\frac{(x - x_{\nu-2})^3}{6h^3}, \quad  x \in [x_{\nu-2}, x_{\nu-1}],\\
&-\frac{(x - x_{\nu-1})^3}{2h^3} + \frac{(x - x_{\nu-1})^2}{2h^2} + \frac{(x - x_{\nu-1})}{2h} + \frac{1}{6}, \quad x \in [x_{\nu-1}, x_{\nu}],\\
&-\frac{(x_{\nu+1} - x)^3}{2h^3} +\frac{(x_{\nu+1} - x)^2}{2h^2} + \frac{(x_{\nu+1} - x)}{2h} + \frac{1}{6}, \quad x \in [x_{\nu}, x_{\nu+1}],\\
&\frac{(x_{\nu+2} - x)^3}{6h^3}, \quad x \in [x_{\nu+1}, x_{\nu+2}],\\
&0, \quad \textup{otherwise},
\end{split}
\right.
\end{equation}
implying that $B_{\nu - 1}, B_{\nu}, B_{\nu+1}, B_{\nu+2}$ overlap a grid interval $(x_{\nu}, x_{\nu+1})$ \cite{MalevskyThomas1997}.

Denote by $\bm{\eta} = (\eta_{-1}, \dots, \eta_{N+1})$. By taking derivatives of $B_{\nu}(x)$, it reads that
\begin{equation}
s^{\prime}(x_i) = - \frac{1}{2h} \eta_{i-1} +  \frac{1}{2h} \eta_{i+1}, \quad s^{\prime\prime}(x_i) =  \frac{1}{h^2} \eta_{i-1} - \frac{2}{h^2} \eta_i +  \frac{1}{h^2} \eta_{i+1}.
\end{equation}
Since $B_{i \pm 1}(x_i) = \frac{1}{6}$ and $B_i(x_i) = \frac{2}{3}$, it yields $N+1$ equations for $N+3$ variables,
\begin{equation}\label{three_term_relation}
\varphi(x_i) = \frac{1}{6} \eta_{i-1} + \frac{2}{3} \eta_{i} + \frac{1}{6} \eta_{i+1}, \quad 0 \le i \le N.
\end{equation}
Two additional equations are needed to solve a unique $\bm{\eta}$ and can be completed by specified boundary conditions at both ends.  For instance, consider the Hermite boundary condition (also termed the clamped spline) \cite{CrouseillesLatuSonnendrucker2009}, $s^{\prime}(x_0) = \phi_L, s^{\prime}(x_{N}) = \phi_R$, where $\phi_L$ and $\phi_R$ are parameters to be determined, it is equivalent to add two constraints,
\begin{equation}\label{Hermite_boundary}
\phi_L =  -\frac{1}{2h} \eta_{-1} + \frac{1}{2h} \eta_1,\quad \phi_R = -\frac{1}{2h} \eta_{N-1} + \frac{1}{2h} \eta_{N+1}.
\end{equation} In particular, when $\phi_L = \phi_R = 0$, it reduces to the Neumann boundary condition on both ends. Alternative choice is the natural boundary condition for cubic spline, which requires $s^{\prime \prime}(x_0) = 0, s^{\prime \prime}(x_N) = 0$, or equivalently,
\begin{equation}\label{natural_boundary}
 \frac{1}{h^2} \eta_{-1} - \frac{2}{h^2} \eta_0 +  \frac{1}{h^2} \eta_{1} = 0, \quad  \frac{1}{h^2} \eta_{N-1} - \frac{2}{h^2} \eta_N +  \frac{1}{h^2} \eta_{N+1} = 0.
\end{equation}
Combining Eqs.~\eqref{three_term_relation} and \eqref{Hermite_boundary} (or \eqref{natural_boundary}) yields an algebraic equations
\begin{equation}\label{dual_equation}
{A} \bm{\eta}^{T} = (\phi_L, \varphi(x_0), \dots, \varphi(x_{N}), \phi_R)^T, 
\end{equation}
with a tridiagonal matrix $A$, which can be solved by the sweeping method \cite{CrouseillesLatuSonnendrucker2009}.

\begin{remark}
In our preceding ASM scheme, we suggested to use three-stage characteristic method and investigated its convergence and mass conservation property \cite{XiongChenShao2016}. However, after a thorough comparison among various integrators as well as the Strang splitting scheme, we have found that LPC1 outperforms others in both numerical accuracy and stability, as it avoids both multi-stage interpolations and splitting errors. In particular,  LPC1 requires spatial interpolation once and calculations of $\pdo$ twice  per step, so that its complexity is definitely lower than multi-stage ones. For details, the readers can refer to Section 4 of our supplementary material \cite{XiongZhangShao2022}.
\end{remark}

\section{Local spatial advection and local spline interpolation}
\label{sec.characteristic}

When we shift to a full 6-D simulation, the foremost problem encountered is to represent the Wigner function on a $N_x^3 \times N_k^3$ grid mesh, which is usually prohibitive for single machine and has to be distributed into multiple ones. This may cause some troubles in solving Eq.~\eqref{dual_equation} as  it requires the information of all interpolated points, so that its efficiency on a distributed-memory environment is dramatically hindered by high communication costs. Fortunately, the cubic B-spline can be essentially constructed in a localized manner,  laying the foundation for the parallel semi-Lagrangian scheme \cite{MalevskyThomas1997,CrouseillesLatuSonnendrucker2009,KormannReuterRampp2019}. 

The local cubic spline basis seems to be very suitable to tackle the local advection mainly for two reasons. First, it is possible for local splines to recover the global one as accurately as possible by imposing some effective boundary conditions on local pieces, which may potentially avoid global communications. Second, the constant advection on 3-D equidistributed grid mesh can be interpolated by a convolution with a $4\times 4\times 4$ window function with relatively small computational cost of about $4^3 N_x^3 N_k^3$. In particular, when $\hbar k_{\max} \tau/\hbar \le h$, it can avoid non-adjacent communications.

%
%
%
%

\subsection{Perfectly matched boundary condition for local spline}

 Without loss of generality, we divide $N+1$ grid points on a line into $p$ uniform parts, with $M = N/p$,
\begin{align}
\underbracket{x_0 < x_1 < \cdots < x_{M-1}}_{\textup{the first processor}} < \underbracket{x_{M}}_{\textup{shared}} < \cdots < \underbracket{ x_{(p-1)M}}_{\textup{shared}} < \underbracket{x_{(p-1)M+1} < \cdots < x_{pM}}_{\textup{$p$-th processor}},
\end{align}
where the $l$-th processor manipulates $M+1$ grid points $\mathcal{X}_l = (x_{(l-1)M}, \dots, x_{l M})$, $l = 1, \dots, p$ and
$x_{M}, x_{2M}, \dots, x_{(p-1)M}$ are shared by the adjacent patches. Denote by $\bm{\eta}^{(l)} = (\eta_{-1}^{(l)}, \dots, \eta_{M+1}^{(l)})$ the local spline coefficients for $l$-th piece. The target is to approximate the global spline coefficients $(\eta_{-1 +(l-1)M}, \dots, \eta_{M+1+ (l-1)M})$ by $\bm{\eta}^{(l)}$.
 

There are two approaches to solving $\bm{\eta}^{(l)}$ without global communications. One is based on a key observation that the off-diagonal elements of the inverse spline matrix $A^{-1}$ decay exponentially away from the main diagonal \cite{MalevskyThomas1997}, so that the coefficients shared by adjacent patches can be calculated by merging the left and right truncated sequences with only local communications. The other is to impose effective Hermite boundary conditions on local pieces and to approximate the unknown first derivatives on the shared grid points by finite difference stencils \cite{CrouseillesLatuSonnendrucker2009}. The former is more preferable in consideration of accuracy and a benchmark can be found in Section 2.3 of our supplementary note \cite{XiongZhangShao2022}, while the latter seems more friendly to implementations. Our PMBC combines the advantages of both approaches and provides a unified framework for different boundary conditions imposed on the global spline.

\subsubsection{Truncation of off-diagonal elements}

\begin{figure}[h]
\centering
\subfigure[Distribution of $\log_{10}(|b_{ij}|)$ for $N = 33$.\label{plot_inverse_matrix_element}]
{\includegraphics[width=0.48\textwidth,height=0.27\textwidth]{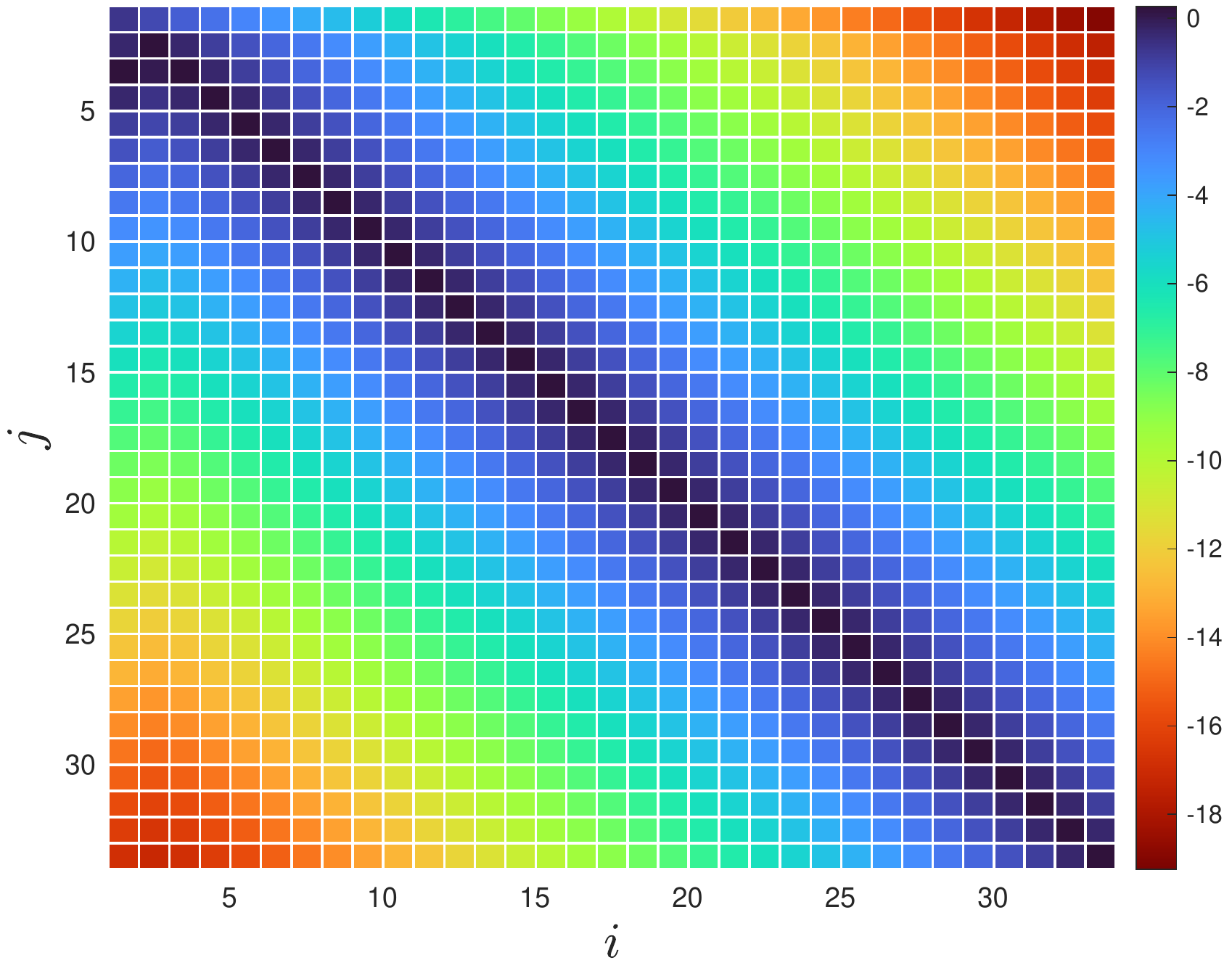}}
\subfigure[Rapid decay of off-diagonal elements. \label{plot_inverse_matrix_element_slice}]
{\includegraphics[width=0.48\textwidth,height=0.27\textwidth]{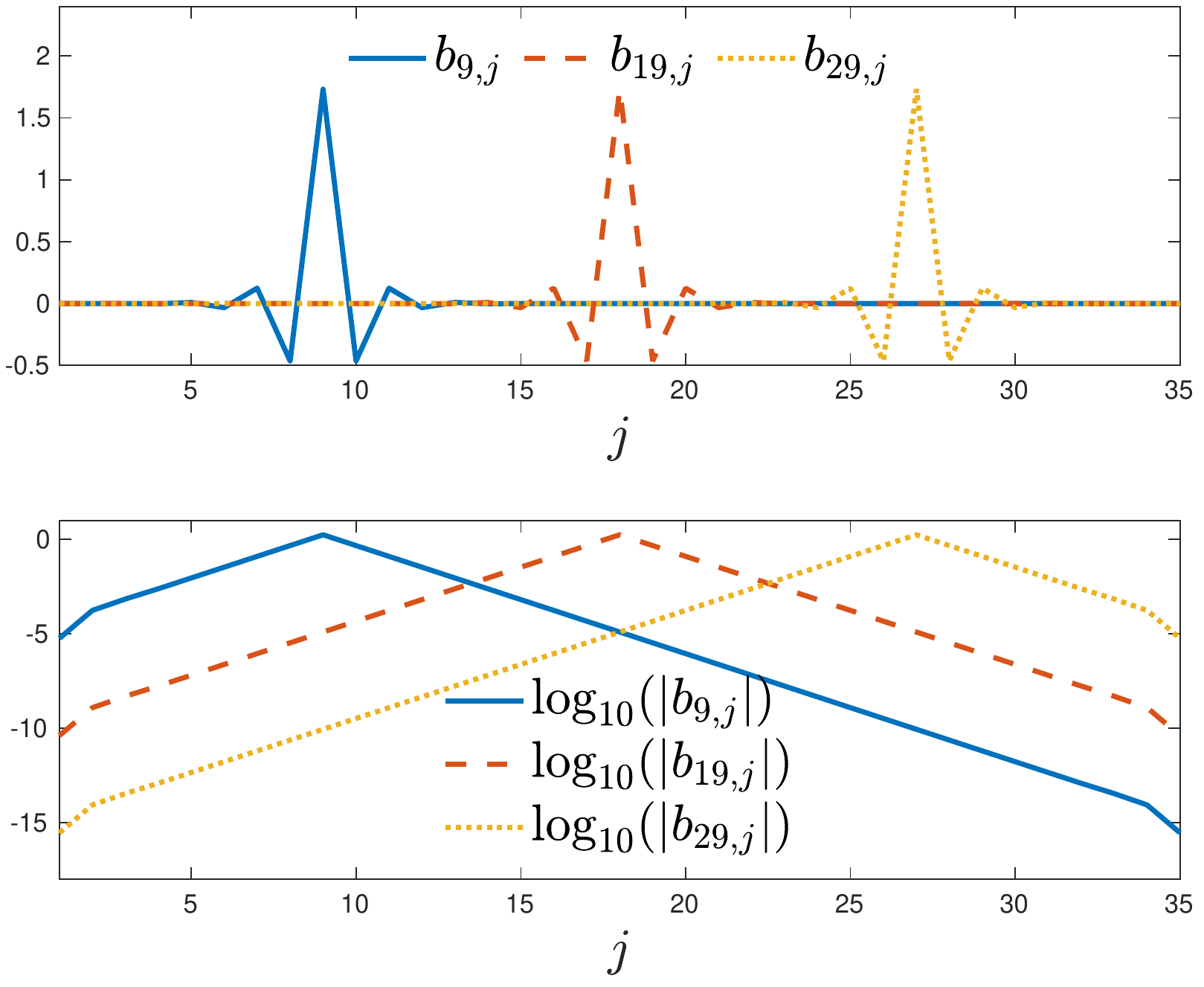}}
\caption{\small The distribution of elements in inverse spline transform matrix $A^{-1}$: The off-diagonal elements exhibit a rapid and monetone decay away from the main diagonal. }
\end{figure}

Denote $A^{-1} = (b_{ij})$, $-1 \le i, j \le pM+1$. The solutions of global set of equations \eqref{dual_equation} are represented as 
\begin{equation}\label{exact_solution_A}
\eta_i = b_{ii} \varphi(x_i) + \sum_{j=-1}^{i-1} b_{ij} \varphi(x_j) + \sum_{j = i+1}^{pM+1} b_{i j} \varphi(x_{j}), \quad i = -1,\dots, p M +1,
\end{equation}
with the convention $\varphi(x_{-1}) = \phi_L$, $\varphi(x_{pM+1}) = \phi_R$.
Despite the inverse spline matrix $A^{-1}$ is a full matrix, its off-diagonal elements exhibit a rapid and monetone decay away from the diagonal element \cite{MalevskyThomas1997} (see Figure \ref{plot_inverse_matrix_element}), which is a well-known fact in the wavelet theory \cite{bk:Chui1992}.  One can see in Figure \ref{plot_inverse_matrix_element_slice} that the elements $b_{ij}$ decays exponentially as $|i-j|$ increases.

This fact allows us to truncate Eq.~\eqref{exact_solution_A} and throw away the terms $|i-j| \ge n_{nb}$,
\begin{equation}\label{truncation}
\eta_i \approx b_{ii} \varphi(x_i) + \sum_{j= i-n_{nb}+1}^{i-1} b_{ij} \varphi(x_j) + \sum_{j = i+1}^{i+n_{nb}-1} b_{ij} \varphi(x_j), \quad i = -1, \dots, pM+1.
\end{equation} 
In particular, when $n_{nb} \le M$, the coefficients $\bm{\eta}^{(l)} = (\eta_{-1}^{(l)}, \dots, \eta_{M+1}^{(l)})$ can be well approximated when $\mathcal{X}_{l-1}$ and $\mathcal{X}_{l+1}$ are known, without information of $\mathcal{X}_1, \dots, \mathcal{X}_{l-2}$ and $\mathcal{X}_{l+2}, \dots, \mathcal{X}_p$ \cite{MalevskyThomas1997}. Thus  the spline transform is localized as data exchanges are only needed in adjacent processors and global communications are completely avoided.  

\subsubsection{Construction of PBMC}

Essentially, the role of spline boundary conditions is to give a closure of coefficients $\bm{\eta}$. Therefore, for $l$-th patch, it is equivalent to impose effective Hermite boundary conditions on both ends of the local spline, 
\begin{equation}
\begin{split}
-\frac{1}{2h} \eta_{-1}^{(l)} + \frac{1}{2h} \eta_{1}^{(l)} &= \phi_{L}^{(l)}(\varphi(x_0), \dots, \varphi(x_{pM+1})), \quad l = 2, \dots, p, \\
-\frac{1}{2h} \eta_{M-1}^{(l)} + \frac{1}{2h} \eta_{M+1}^{(l)} &= \phi_{R}^{(l)}(\varphi(x_0), \dots, \varphi(x_{pM+1})), \quad l = 1, \dots, p-1,
\end{split}
\end{equation}
where $-\frac{1}{2h} \eta_{-1}^{(l+1)} + \frac{1}{2h} \eta_{1}^{(l+1)}  = -\frac{1}{2h} \eta_{M-1}^{(l)} + \frac{1}{2h} \eta_{M+1}^{(l)}$, implying that $\phi_{R}^{(l)} = \phi_{L}^{(l+1)}$,  $1\le l \le p-1$. Using the truncated stencils \eqref{truncation}, it yields the formulation of PMBC
\begin{equation*}
\begin{split}
\phi_{R}^{(l)} =  \phi_{L}^{(l+1)} \approx & \underbracket{\frac{1}{2}c_{0,l} \varphi(x_{lM}) +  \sum_{j = 1}^{n_{nb}} c_{j, l}^{-} \varphi(x_{lM-j})}_{\textup{stored in left processor}} + \underbracket{\frac{1}{2}c_{0,l} \varphi(x_{lM}) + \sum_{j = 1}^{n_{nb}} c_{j, l}^{+} \varphi(x_{lM+j})}_{\textup{stored in right processor}},
\end{split}
\end{equation*}
where $c_{0, l} = -\frac{b_{lM-1, lM}}{2h} + \frac{b_{lM+1, lM}}{2h}$ and 
\begin{equation}\label{PMBC_coeffcients}
\begin{split}
&c_{j, l}^- =  -\frac{b_{lM-1, lM-j}}{2h} + \frac{b_{lM+1, lM-j}}{2h}, \quad c_{j, l}^+ =  -\frac{b_{lM-1, lM+j}}{2h} + \frac{b_{lM+1, lM+j}}{2h}.
\end{split}
\end{equation}

Following the same idea, one can represent all kinds of spline boundary condition by PMBC. For instance, when the natural boundary conditions \eqref{natural_boundary} are adopted and denote $\widetilde{A}$ the corresponding coefficient matrix, $(\widetilde{b}_{ij}) = \widetilde{A}^{-1}, -1\le i, j \le N+1$, then the equation $\widetilde A\bm{\eta}^T= (0, \varphi(x_0), \dots, \varphi(x_N),  0)^T$ can be transformed into $A\bm{\eta}^T= (\phi_{L}^{(1)}, \varphi(x_0), \dots, \varphi(x_N),  \phi_{R}^{(p)})^T$ with
\begin{equation}\label{true_boundary_truncate}
\phi_{L}^{(1)} = \frac{\eta_1-\eta_{-1}}{2h}  \approx \underbracket{\sum_{j=0}^{n_{nb}} c_{j, 0}^{+} \varphi(x_j),}_{\textup{stored in first processor}} ~  \phi_{R}^{(p)}= \frac{\eta_{N+1}-\eta_{N-1}}{2h} \approx \underbracket{\sum_{j=0}^{n_{nb}} c_{j, p}^{-}\varphi(x_{N-j}),}_{\textup{stored in last processor}}
\end{equation}
where  $c_{j, 0}^{+} =  \frac{1}{2h}(-\widetilde{b}_{-1, j} + \widetilde{b}_{1, j})$ and $c_{j, p}^{-} = \frac{1}{2h}(-\widetilde{b}_{pM-1, pM-j} + \widetilde{b}_{pM+1, pM-j})$.
\begin{figure}[!h]
\centering
\includegraphics[width=1\textwidth,height=0.30\textwidth]{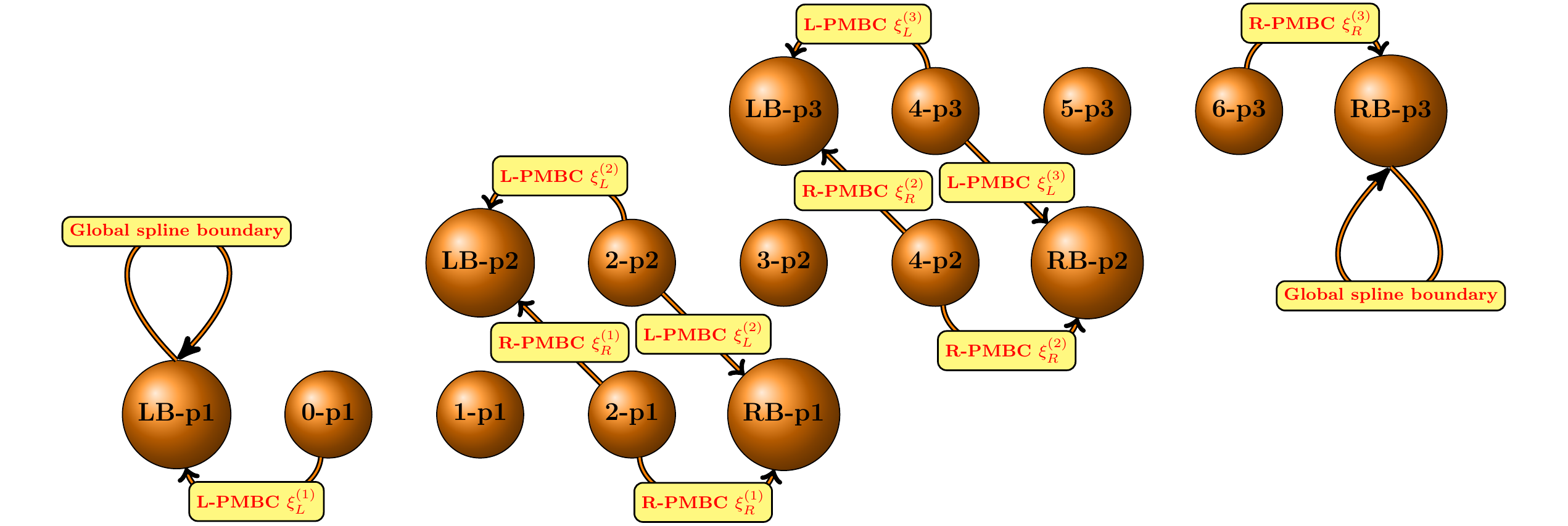}
\caption{An illustration of the cubic spline coefficients in the distributed setting: Seven grid points are distributed evenly in three processors. For each processor, PMBCs are assembled by exchanging and merging the stencils in the adjacent neighborhood. The  boundary condition for global spline can also be realized by imposing effective Hermite boundary conditions on the first and last processors. 
 \label{fig_spline_illustration}}
\end{figure}

Figure \ref{fig_spline_illustration} illustrates the construction of three local splines by seven grid points $\mathcal{X} = (x_0, \dots, x_6)$, with $\mathcal{X}_1 = (x_0, x_1, x_2)$, $\mathcal{X}_2 = (x_2, x_3, x_4)$ and $\mathcal{X}_3 = (x_4, x_5, x_6)$. 
\begin{itemize}

\item[(1)] The left boundary $\phi_L^{(1)}$ for the first processor (LB-p1) and the right boundary $\phi_R^{(p)}$ for the last processor (RB-p3) are calculated by Eq.~\eqref{true_boundary_truncate}. 

\item[(2)] The $l$-th processor calculates the following quantities,
\begin{equation*}
\begin{split}
\textup{L-PMBC}: \quad &\xi_L^{(l)} = \frac{1}{2}c_{0,l} \varphi(x_{(l-1)M}) + \sum_{j = 1}^{n_{nb}} c_{j, l}^{+} \varphi(x_{(l-1)M+j}), \\
\textup{R-PMBC}: \quad &\xi_{R}^{(l)} = \frac{1}{2}c_{0,l} \varphi(x_{l M}) +  \sum_{j = 1}^{n_{nb}} c_{j, l}^{-} \varphi(x_{l M-j}).
\end{split}
\end{equation*}

\item[(3)] The $l$-th processor transfers $\xi_L^{(l)}$ to its left neighbor: $(l$-$1)$-th processor $(l > 1)$, and transfers $\xi_R^{(l)}$ to its right neighbor: $(l$+$1)$-th processor $(l < p)$.

\item[(4)] For $l$-th processor, $\phi_L^{(l)} = \xi_L^{(l)} + \xi_R^{(l-1)} ~(l > 1)$ and $\phi_R^{(l)} =\xi_L^{(l+1)} + \xi_R^{(l)} ~(l < p)$.

\item[(5)] Each patch solves spline coefficients $\bm{\eta}^{(l)}$ via the exact LU decomposition of $(M+3)\times(M+3)$ tridiagonal matrix $A^{(l)}$,
\begin{equation*}
A^{(l)} (\bm{\eta}^{(l)})^T = LU (\bm{\eta}^{(l)})^T = (\phi_{L}^{(l)}, \varphi(x_{(l-1)M}), \dots, \varphi(x_{lM}), \phi_{R}^{(l)})^T.
\end{equation*}

\end{itemize}

\subsubsection{Interpolation and correction for constant advection}

Once the spline coefficients  $\bm{\eta}^{(l)}$ are determined, interpolating $\varphi(x - \alpha h)$ with a constant shift $\alpha h$ can be realized by taking a weighted summation of $B_{\nu}(x - \alpha h)$ over indices $\nu$ with the whole cost being $\mathcal{O}(4N)$. Suppose all grid points are shifted by $\alpha h$,
\begin{equation}
\varphi(x_j - \alpha h) = \sum_{\nu = -1}^{N +1} \eta_{\nu} B_{\nu}(x_j - \alpha h), \quad 0 \le j \le N,
\end{equation}
where
$B_{\nu}(x_j)$ only takes five possible values $b_1, b_2, b_3, b_4$ and $0$, and
\begin{equation}
\begin{split}
&b_1 = \frac{(1 - \alpha)^3}{6}, \quad b_2 = - \frac{(1- \alpha)^3}{2} + \frac{(1-\alpha)^2}{2} + \frac{1- \alpha}{2} + \frac{1}{6}, \\
&b_3 = - \frac{\alpha^3}{2} + \frac{\alpha^2}{2} + \frac{\alpha}{2} + \frac{1}{6}, \quad b_4 = \frac{\alpha^3}{6}.
\end{split}
\end{equation}
As the shifted grid point may move outside the domain $[x_0, x_{N}]$, it shall add ghost splines $B_{-2}(x)$ and $B_{N+2}(x)$ with coefficients $\eta_{-2} = \eta_{N+2}$.

When $0< \alpha < 1$, $x_{j} - \alpha h \in [x_{j-1}, x_j]$, a simple calculation yields that
\begin{equation}\label{interpolation_left}
\begin{split}
\varphi(x_j - \alpha h) = &\eta_{j-2} B_{j-2}(x_j - \alpha h)  + \eta_{j-1} B_{j-1}(x_j - \alpha h)  \\
&+ \eta_{j} B_{j}(x_j - \alpha h) + \eta_{j+1} B_{j+1}(x_j - \alpha h).
\end{split}
\end{equation}
Similarly, one can tackle the case $-1 < \alpha < 0$, $x_j - \alpha h \in [x_{j}, x_{j+1}]$, yielding that
\begin{equation}\label{interpolation_vector}
\varphi(x_j - \alpha h)  = 
\left\{
\begin{split}
&(\eta_{j-2}, \eta_{j-1}, \eta_{j}, \eta_{j+1}) \cdot (b_4, b_3, b_2, b_1), ~ &0< \alpha< 1,\\
& (\eta_{j-1}, \eta_{j}, \eta_{j+1}, \eta_{j+2}) \cdot  (b_1, b_2, b_3, b_4), ~&-1 < \alpha < 0.
\end{split}
\right.
\end{equation}

\begin{figure}[!h]
\centering
\includegraphics[width=1\textwidth,height=0.22\textwidth]{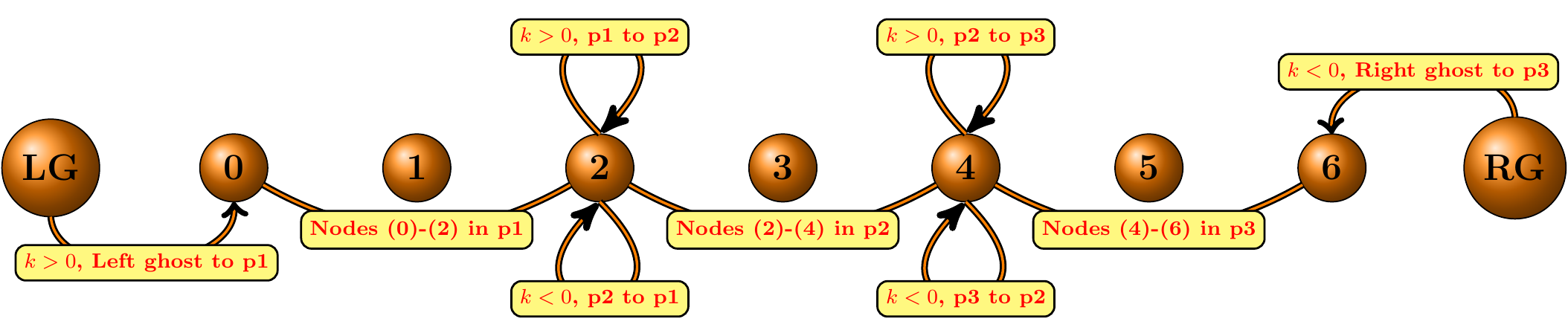}
\caption{Illustration of the local cubic spline interpolation of the constant advection. The shifted grid points are first interpolated within each processor independently. Then the boundary nodes that shifts to other local pieces are corrected from the adjacent  neighborhood. The ghost regions are added on the first and last processors for imposing specified boundary condition on the global spline. \label{fig_interpolation_illustration}}
\end{figure}

The interpolation procedure under the parallel setting is almost the same except a correction procedure. Since the ghost splines with $\eta^{(l)}_{-2} = \eta^{(l)}_{M+2} = 0$ have to be added on both sides of all local splines, the shifted grid points outside the subdomain might not be properly interpolated. Therefore, the correct interpolated values need to be transferred from its adjacent processor. Figure \ref{fig_interpolation_illustration} illustrates the interpolation of the constant advection under the distributed environment. Again, seven grid points are distributed into three clusters, with $p = 3$ and $N = 6$.

\begin{itemize}

\item[(1)] When $\alpha > 0$, $(x_0 - \alpha h) < x_0$, the interpolation of $\varphi(x_0 - \alpha h)$ uses the left ghost spline. Similarly, when $ \alpha < 0$, $(x_N - \alpha h) > x_N$,  the interpolation of  $\varphi(x_N - \alpha h)$ uses the right ghost spline. 

\item[(2)] For the shared grid points $x_{2l}$, e.g., $l = 1, 2$,  when $\alpha > 0$, $(x_{2l} - \alpha h) < x_{2l}$, the left processor interpolates $\varphi(x_{2l} - \alpha h)$ correctly and sends the value  to its right neighbor. Similarly, when $\alpha < 0$, $(x_{2l} - \alpha h) > x_{2l}$,  the right processor interpolates $\varphi(x_{2l} - \alpha h)$ correctly and sends the value  to its left neighbor.

\end{itemize}

\subsection{Parallel implementation in 6-D phase space}
\label{sec:6d_parallel}

For a 6-D problem, the Wigner function is expanded as the tensor product of cubic splines in three directions,
\begin{equation}
f(\bx, \bk, t) \approx \sum_{\nu_1 = -1}^{N_x+1}   \sum_{\nu_2 = -1}^{N_x+1}  \sum_{\nu_3 = -1}^{N_x+1} \eta_{\nu_1, \nu_2, \nu_3} (\bk, t)  \prod_{j=1}^3 B_{\nu_j}(x_j).
\end{equation}
Hereafter we take a $(N_x+1)^3 \times N_k^3$ uniform grid mesh for 6-D phase space. Because $\bk$-domain involves nonlocal interaction, the domain decomposition is only performed in $\bx$-space to split the whole domain into $p^3$ mutually disjoint rectangular patches, where $p$ divides into $N_x$. Each processor manipulates $(\frac{N_x}{p} + 1)^3 \times N_k^3$ grid points.

The 3-D cubic splines can be constructed in each direction successively, but each `grid point' to be interpolated is a long vector of length $N_k^3$, and PMBC turns out to be a $(\frac{N_x}{p}+1)^2 N_k^3$ tensor.  Thus for each processor, the cost of constructing the cubic spline is $\mathcal{O}((\frac{N_x}{p}+1)^3 N_k^3)$ and that of exchanging six PMBCs is about $6 (\frac{N_x}{p}+1)^2 N_k^3$.

For the constant advection $\bm{\alpha}h = (\alpha_1 h, \alpha_2 h, \alpha_3 h)$, interpolating $f(\bx_j - \bm{\alpha}h , \bk, t)$ is a convolution of $64$ grid points with a $4\times 4 \times 4$ window function since
\begin{equation}\label{3d_advection}
 f(\bx - \bm{\alpha}h, \bk, t) \approx \sum_{\nu_1 = -1}^{N_x+1}   \sum_{\nu_2 = -1}^{N_x+1}  \sum_{\nu_3 = -1}^{N_x+1} \eta_{\nu_1, \nu_2, \nu_3}(\bk, t) \prod_{j=1}^3 B_{\nu_j}(x_j -  \alpha_j h )
\end{equation}
has only $4^3$ nonzero terms $B_{\nu_j}(x_j -  \alpha_j h )$ obtained by Eqs.~\eqref{interpolation_left} and \eqref{interpolation_vector}.
Thus interpolating one point involves 64 multiplications and 64 summations,  and the computational and communication costs are $64 (\frac{N_x+1}{p})^3 N_k^3$ and $(\frac{N_x+1}{p})^2 N_k^3$, respectively.

\section{Nonlocal quantum interaction and truncated kernel method}	
\label{spectral} 
Once CoD is alleviated via the local  cubic spline construction, the remaining challenge is to seek a highly efficient approximation to $\pdo$ with a weakly singular symbol, as it has to be calculated twice per LPC1 evolution.
To this end, we borrow the idea of TKM \cite{VicoGreengardFerrando2016,GreengardJiangZhang2018,LiuZhangZhang2022} to derive a spectrally accurate approximation for smooth and rapidly decreasing Wigner function, with its implementation greatly accelerated by FFTs.

\subsection{Truncated kernel method}
Here we omit the time variable for brevity. By a change of variables,  we can rewrite  \eqref{def.pdo} as follows
\begin{equation*}
\begin{split}
\label{twistedConv} \Theta_{V}[f](\bx, \bk) &= \frac{2}{c_{3, 1} \mi}  \int_{\mathbb{R}^3} \frac{ \me^{2 \mi (\bx - \bx_A) \cdot \bk^{\prime}}-\me^{-2 \mi (\bx - \bx_A) \cdot \bk^{\prime}} }{ |\bk^{\prime}|^2}  f(\bx, \bk - \bk^{\prime}) \D \bk^{\prime} \coloneqq (I^{+} - I^{-}), \\
I^{\pm}(\bx, \bk) &= \frac{2}{c_{3, 1} \mi}  \int_{\mathbb{R}^3} \frac{ \me^{\pm 2 \mi (\bx - \bx_A) \cdot \bk^{\prime}} }{ |\bk^{\prime}|^2}  f(\bx, \bk - \bk^{\prime}) \D \bk^{\prime}. 
\end{split}
\end{equation*}

 Note that $I^+ - I^- = 2\Re(I^+ )$ for real-valued function $f(\bx, \bk)$, therefore, the above integral can be reduced to 
the computation of $I^{+}$. It notes that
 \begin{equation}
 \begin{split} \label{i1inte}
 I^{+}(\bx,\bk) &= \frac{2}{c_{3, 1} \mi }  \me^{2 \mi (\bx - \bx_A) \cdot \bk} \int_{\mathbb{R}^3}
\frac{1}{{ |\bk^{\prime}|^2}}  \me^{-2 \mi (\bx - \bx_A) \cdot (\bk-\bk^{\prime})}  f(\bx, \bk - \bk^{\prime}) \D \bk^{\prime} \\
&= \frac{2}{c_{3, 1} \mi }  \me^{2 \mi (\bx - \bx_A)\cdot \bk } \left( |\bk|^{-2} \ast  f^{s} \right)(\bx,\bk),
 \end{split}
 \end{equation}
 where $f^s(\bx,\bk):=  f(\bx,\bk)\me^{-2 \mi (\bx - \bx_A) \cdot \bk}$ is a smooth and fast-decaying complex-valued function.  The twisted convolution evaluation boils down to the standard convolution of singular kernel $|\bk|^{-2}$ with smooth fast-decaying function $f^s(\bx,\bk)$. For brevity, we shall omit $\bx$ and focus on the following convolution 
 \begin{eqnarray*} 
 \Phi(\bk) = ( U \ast f^s ) (\bk) :=\int_{\mathbb R^{3}}  U(\bk-\bk^{\prime}) f^s(\bk^{\prime}) {\rm d} {\bk}^{\prime}, 
 \end{eqnarray*} where the kernel $U(\bk) =|\bk|^{-2}$ is singular and the Wigner function $f(\bk)$ is assumed to be smooth and fast-decaying.  It is reasonable to assume the density to be {\sl numerically} supported on a bounded domain, for example,  a rectangular $\Omega :=[-L_k,L_k]^3 \subset  \mathbb R^{3}$,  and to utilize Fourier spectral method. 
To compute $\Phi$ on the same domain $\Omega$, we choose to apply TKM \cite{GreengardJiangZhang2018,VicoGreengardFerrando2016} which is an $O(N\log N )$ fast algorithm, implemented with FFT, and achieves spectral accuracy. 

The basic idea is to screen the unnecessary interaction and apply trapezodial quadrature to the smooth-integrand Fourier transform, i.e., for $\bk \in \Omega$, it has that
\begin{eqnarray*} 
 \Phi(\bk) &= &\int_{\mathbb R^{3}}  U(\bk-\bk^{\prime}) f^s(\bk^{\prime}) {\rm d} {\bk}^{\prime} 
 \approx \int_{\Omega}  U(\bk-\bk^{\prime}) f^s(\bk^{\prime}) {\rm d} {\bk}^{\prime} = \int_{\mathbb R^{3}}  U_{D}(\bk-\bk^{\prime}) f^s(\bk^{\prime}) {\rm d} {\bk}^{\prime},
  \end{eqnarray*}
where the truncated kernel $U_{D}(\bk)$ is defined as  \begin{equation}
U_{D}(\bk):=\left\{
\begin{array}{ll}
U(\bk),  & |\bk| \leq D,\\
0, &  |\bk| > D,
\end{array}
\right.
 \end{equation} with $D= \text{diam }{\Omega} := \max_{\bk,\bk^{\prime}\in \Omega}|\bk-\bk^{\prime}|$.
 The second equality holds because $U_D(\bk-\bk^{\prime})= 0, ~\forall~ \bk\in \Omega, ~\bk^{\prime} \in \Omega^{c}$. 
By the Paley-Wiener Theorem \cite{bk:Rudin1991},  we know that the Fourier transform of $U_{D}$ is smooth, therefore, it is convenient to compute the convolution's Fourier transform as follows
\begin{equation}\label{ktmFourier}
\Phi(\bk) = \frac{1}{(2\pi)^{3}}\int_{\mathbb R^{3}}  \widehat U_{D}(\bxi) \widehat f^s(\bxi) ~\me^{\mi\bk \cdot \bxi }~{\rm d} {\bxi},\quad \bk \in \Omega,
\end{equation} 
with $\widehat f^s(\bxi)= \mathcal{F}_{\bk \to \bxi} f^s(\bk) = \int_{\mathbb R^{3}}  f^s(\bk) ~\me^{-\mi\bk \cdot \bxi }~{\rm d} {\bk}$ with its inverse denoted by $\mathcal{F}_{\bxi \to \bk}^{-1}$ and
\begin{eqnarray}
\nonumber \widehat U_{D}(\bxi) &= &\int_{\mathbb R^{3}} U_{D}(\bk) ~\me^{-\mi\bk \cdot \bxi }~{\rm d} {\bk}  = 4\pi \int_{0}^{D} U(\bk) k^{2} \frac{\sin( k |\bxi|)}{k |\bxi|} {\rm d} k 
\\
&=&  \frac{4\pi}{|\bxi|} \int_{0}^{|\bxi| D}  \frac{\sin t}{t} {\rm d} t =  \frac{4\pi}{|\bxi|} {\rm Si}(|\bxi| D),
\end{eqnarray} 
with ${\rm Si}(x) := \int_{0}^{x} \sin t /t  ~{\rm d } t$ being the sine integral function. The asymptotic  is 
$\widehat U_{D}(\bxi) \approx 4 D\pi - \frac{2}{9} (D^{3}\pi )|\bxi|^{2} + O(|\bxi|^{4})$ as $|\bxi| \to 0$.

As is seen, there is {\sl not} any singularity in $\widehat U_D(\bxi)$. However, the kernel truncation brings in extra oscillations ${\rm Si}(|\bxi| D)$ to the integrand. To resolve such oscillations, we need a fine mesh in the frequency space $\bxi$, which, by the duality argument, corresponds to a large computational domain in the physical space $\bk$.
Recently, Liu {\sl et al} proved that a {\bf threefold}, instead of fourfold, zero-padding of $f^s(\cdot, \bk)$ is sufficient to resolve such extra oscillation in \eqref{ktmFourier},  and we refer the readers to \cite{LiuZhangZhang2022} for more details.

To sum up,  we derived a discretized approximation $ \Theta_V^{T}[f]$ to $\Theta_V[f]$ as follows
\begin{equation}\label{discrete_quantization}
\begin{split}
\Theta_V^{T}[f](\bx, \bk_{\bp}) =& \frac{2}{c_{3,1}\mi} \me^{2\mi \widetilde \bx  \cdot \bk_{\bp}} \mathscr{F}^{-1}_{\bxi_{\bn} \to \bk_{\bp}}\left[ \widehat U_D(\bxi_{\bn}) \mathscr{F}_{\bk_{\bp} \to \bxi_{\bn}}\left(\me^{-2\mi\widetilde \bx\cdot \bk_{\bp}} f(\bx, \bk_{\bp})\right)\right] \\
 &-  \frac{2}{c_{3,1}\mi}  \me^{-2\mi\widetilde \bx\cdot \bk_{\bp}} \mathscr{F}^{-1}_{\bxi_{\bn} \to \bk_{\bp}} \left[\widehat U_D(\bxi_{\bn}) \mathscr{F}_{\bk_{\bp} \to \bxi_{\bn}}\left(\me^{2\mi\widetilde \bx\cdot \bk_{\bp}} f(\bx, \bk_{\bp})\right)\right],
\end{split}
\end{equation}
where $\widetilde \bx = \bx-\bx_{A}$,  $\bk_{\bp} = \bk_{ijl}$ is the discrete grid point evenly spaced in each spatial direction of $\Omega$, and $\mathscr{F}_{\bk_{\bp} \to \bxi_{\bn}}$ and $\mathscr{F}^{-1}_{\bxi_{\bn} \to \bk_{\bp}}$ denote the forward and backward discrete Fourier transform of size $(3N_k)^3$ with threefold zero-padding of $f(\cdot, \bk_{\bp})$, respectively.

\begin{remark}
Before moving to the detailed implementation, let us make a comparison between TKM and the commonly used pseudo-spectral method \cite{Ringhofer1990,Goudon2002}. In fact, $\Theta_V^T[f](\bx, \bk_{\bp})$ can be rewritten as
\begin{equation}\label{lattice_pdo}
\Theta_V^T[f](\bx, \bk_{\bp}) = \mathscr{F}^{-1}_{\bxi_{\bn} \to \bk_{\bp}} \left(\sigma_D(\bx, \bxi_{\bn}) \mathscr{F}_{\bk_{\bp} \to \bxi_{\bn}} f(\bx, \bk_{\bp})\right),
\end{equation}
with a {\bf non-singular} symbol $\sigma_D(\bx, \bxi)$ given by
\begin{equation*}
\sigma_D(\bx, \bxi) =  \frac{2}{c_{3,1} \mi} \left(\mathcal{S}_{2 \widetilde\bx} ~ \widehat{U}_D(\bxi) ~\mathcal{S}_{-2 \widetilde\bx} - \mathcal{S}_{-2\widetilde\bx} ~\widehat{U}_D(\bxi)~ \mathcal{S}_{2\widetilde\bx}\right), \quad\widetilde\bx = \bx - \bx_{A},
\end{equation*}
and $S_{\bm{\alpha}} g(\bxi) = g(\bxi - \bm{\alpha})$ is the shift operator, while $\pdo$ \eqref{def.pdo} in $\mathbb{R}^3 \times \mathbb{R}^3$ reads that 
\begin{equation}\label{pdo_definition_2}
\Theta_V[f](\bx, \bk) = \mathcal{F}^{-1}_{\bm{\bxi} \to \bk} (\sigma(\bx, \bxi)\widehat f(\bx, \bxi) ),
\end{equation}
 with a {\bf singular} symbol $ \sigma(\bx, \bxi) = \frac{2}{c_{3,1} \mi} (\mathcal{S}_{2\widetilde\bx} ~\widehat U(\bxi) ~\mathcal{S}_{-2\widetilde\bx} - \mathcal{S}_{-2\widetilde\bx} ~ \widehat U(\bxi) ~\mathcal{S}_{2\widetilde\bx})$. When $f$ is approximated by a truncated Fourier series in $\bk$-space, the formula \eqref{lattice_pdo} is almost the same as the pseudo-spectral approach except the difference between $\sigma_D(\bx, \bxi)$ and $\sigma(\bx, \bxi)$, as well as zero-padding. In other words, the difficulty induced by singular symbol is resolved by exploiting an elegant fact the Fourier conjugate of truncated kernel $U_{D}$ removes the singularity at origin. By contrast, the widely used pseudo-spectral method suffers from large errors near singularity and numerical instability, which can be alleviated by TKM. Details are referred to Section 3 of our supplementary note \cite{XiongZhangShao2022}.
 \end{remark}

In practice, with a precomputation technique, the above quadrature can be implemented only with {\sl twofold} zero-padding of the source function $f^s(\cdot, \bk_{\bp})$. As pointed out in \cite{VicoGreengardFerrando2016}, after plugging the finite Fourier series approximation of size $(3N_k)^3$ into \eqref{ktmFourier}, reducing zero-padding terms and utilizing the symmetry of $\widehat U_D$, we can reformulate the above quadrature \eqref{discrete_quantization} into the following discrete convolution 
\begin{equation}\label{DisConvPhi}
\Phi(\bk_{ijl}) \approx \Phi_{ijl} = \sum_{i^{\prime}=1}^{N_k}\sum_{j^{\prime}=1}^{N_k}\sum_{l^{\prime}=1}^{N_k} T_{i-i^{\prime},j-j^{\prime},l-l^{\prime}  } f^s_{i^{\prime}j^{\prime}l^{\prime}},
\end{equation} 
where $f^s_{ijl}$ is the numerical approximation of function $f^s(\cdot,\bk_{\bp}), \bp \in \Lambda$ with index set $\Lambda:=\left\{(i,j,l)\in \mathbb Z^{3}\big| 1\leq i,j,l \leq N_{k} \right\}$.
The convolution tensor $T_{i,j,l}$ is symmetric in each direction, e.g., $T_{i,j,l} = T_{-i,j,l}$, and is given explicitly as follows
\begin{equation}\label{convolution_tensor}
T_{\bp} \coloneqq  \frac{1}{(3N_k)^{3}} \sum_{\bn\in {\mathcal I }} \widehat{U}_{D}(\xi_{\bn}) \me^{\frac{2\pi\mi \bp \cdot \bn}{3N_k}}, \quad \bp \in \Lambda,
\end{equation} 
where $\xi_{\bn} = \frac{2\pi}{6L_k} \bn, ~\bn \in {\mathcal I }$ is the Fourier mode and the dual index set ${\mathcal I}$ is defined 
\begin{equation}
\label{InSetI}
{\mathcal I }:= \left\{(n_{1},n_{2},n_{3})\in \mathbb Z^{3} \big | n_{j} = -3 N_{k}/2,\ldots, 3N_{k}/2 \!-\!1\right\}.
\end{equation}
 
It is clear that the tensor \eqref{convolution_tensor} can be calculated with a backward FFT of length $(3N_k)^{3}=27N_k^{3}$, which inevitably requires a quite large memory. Fortunately, compared with the original fourfold zero-padding TKM \cite{VicoGreengardFerrando2016,GreengardJiangZhang2018},  the minimal memory requirement of our algorithm is reduced further by a factor of $(\frac{4}{3})^{3}=\frac{64}{27} \approx 2.37$, and it shall bring about a 
significant improvement in real simulations, especially when the mesh size is large enough. Therefore, our algorithm grants a much easier access even on a personal computer.  More importantly, the tensor is of size $(2N_k)^{3}$ and independent of the position variable $\bx$ and time variable $t$, therefore, it can be precomputed only once for the whole lifetime. That is, the convolution \eqref{DisConvPhi} can be accelerated within $O(8N_k^{3} \log( 8N_k^{3}))$ flops with FFT as long as the tensor \eqref{convolution_tensor} is available.

\subsection{Error estimates} 


Our error estimates focus on the TKM approximation to the nonlocal convolution potential $\Phi = U\ast f^{s}$ with the singular kernel $U(\bx) =  |\bx|^{-2}$ and the effective density function $f^{s}(\bx,\bk) =f(\bx,\bk)\me^{-2 \mi (\bx - \bx_A) \cdot \bk}$.


\begin{theorem}
\label{thm_1}
Suppose that Wigner function $f(\bx,\bk)$ is a smooth and fast-decaying function of $\bk$ and has a $\bx$-independent common compact support,
i.e., ${\rm supp}(f(\bx,\cdot))\subsetneq \Omega = [-L_k, L_k]^3$, then we have for any integer $m\in \mathbb Z^{+}$,
\begin{eqnarray}
\Vert \Theta_V[f] -  \Theta^T_V[f] \Vert_{\infty} &\lesssim& ~ C~ |\bx-\bx_{A}|^{m}N_{k}^{-(m-\frac{3}{2})} \|f(\bx,\cdot) \|_m, \quad m \geq 2,\\
\Vert \Theta_V[f] -  \Theta^T_V[f] \Vert_{2}& \lesssim & ~C ~ |\bx-\bx_{A}|^{m} N_{k}^{-m}  \|f(\bx,\cdot) \|_m ,  \quad m \geq 1,
\end{eqnarray} where constant $C = C(L_{k},m)$ is independent of $\bk$ and $\|f(\bx,\cdot)\|_m$ is the standard Sobolev norm with respect to $\bk$.
\end{theorem}

The proof is based on the recent error estimates of TKM given by Liu {\sl et al} \cite{LiuZhangZhang2022}. For brevity, we choose not to repeat the lengthy and technical proof but to directly quote them, and refer the readers to \cite{LiuZhangZhang2022} for more details. Here $H_p^m(\Omega)$ denotes the subspace of $H^m(\Omega)$ with derivatives of order up to $m-1$ being $\Omega$-periodic.
\begin{lemma}[\cite{LiuZhangZhang2022}]
Suppose $\rho(\bx) \in H_p^m(\Omega)$ associated with the semi-norm 
\begin{equation}
|\rho|_m = \left(\sum_{k_1= -\infty}^{\infty} \sum_{k_2= -\infty}^{\infty} \sum_{k_3= -\infty}^{\infty} |\bk|^{2m} |\widehat \rho_{\bk}|^2\right)^{1/2}
\end{equation}
and $\Phi_{N}$ is the numerical approximation to Eq.~\eqref{ktmFourier} with $N^{3}$ uniform grid points, then it has that
\begin{eqnarray} \label{infyNormEsti}
\|\Phi_{N}-\Phi\|_{{\infty}}  &\leq &C~ N^{-(m-\frac{3}{2})} |\rho|_m, \quad m \geq 2, \\
\label{2NormEsti} \|\Phi_{N}-\Phi\|_{2}&\leq&C~ N^{-m} | \rho|_m,\quad m \geq 1,
\end{eqnarray} 
where $C$ depends only on domain size $L_k$ and $m$.
\end{lemma}

\begin{proof}[Proof of Theorem \ref{thm_1}]
 The nonlocal potential is given by a similar convolution $\Phi = U\ast \rho$ where the density function $\rho$ is also smooth and fast decaying with a compact support  and the kernel $U$ is singular. Since the Wigner function is smooth and fast decaying in $\bk$ and shares a common compact support, substituting $f^{s}(\bx,\bk) $ for $\rho$ in \eqref{infyNormEsti}-\eqref{2NormEsti}, and computing its $m$-th semi-norm, we have 
 \begin{equation}
| f^{s}(\bx,\bk) |_{m} \lesssim  C~ |\bx-\bx_{A}|^{m} \|f^{s}(\bx,\cdot)\|_{m}, \quad \forall ~m \in \mathbb Z^{+}.
 \end{equation}
Plugging back into \eqref{i1inte}, we have 
 \begin{eqnarray*}
 \|I^{+}- I^{+}_{N_k}\|_{\infty} &\lesssim &~C~ |\bx-\bx_{A}|^{m}N_{k}^{-(m-\frac{3}{2})} \|f(\bx,\cdot) \|_m,\quad m \geq 2,\\
 \|I^{+}- I^{+}_{N_k}\|_{2} &\lesssim &~ C~ |\bx-\bx_{A}|^{m}N_{k}^{-m} \|f(\bx,\cdot) \|_m, \quad m \geq 1,
  \end{eqnarray*} where $I^{+}_{N_k}$ denotes the numerical approximation of $I^{+}$ using TKM.
Obviously from \eqref{i1inte}, the desired twisted convolution \eqref{def.pdo} is effectively reduced to the real part of $I^{+}$, which immediately completes the proof. 
\end{proof}

Next we present the numerical errors and computational time (in seconds) in Table \ref{TKM_convergence_data} to confirm the spectral convergence and efficiency of TKM with a localized Gaussian function $f(\bk)$, from which we can see clearly that our algorithm converges spectrally fast and the errors approach the machine precision as $N_{k}$ increases. 
\begin{example}\label{tkmGaussian}
\textup{
For a symmetric Gaussian function $f(\bk)= e^{- |\bk|^2}, \bk \in \mathbb R^3$, the convolution potential $\Phi$ is symmetric  and reads explicitly as follows
\begin{equation}\label{test}
\Phi(\bk) = \left(\frac{1}{|\bk|^2} \ast f\right)(\bk)= 2  \pi^{\frac{3}{2}}   \frac{  \textrm{DawsonF}( k)}{k}, \quad k = |\bk|,
\end{equation}
with $\textrm{DawsonF}(k) :=\int_0^\infty \sin (k r) ~ e^{-\frac{k^2}{4}} {\rm d} k$ \cite{ZaghloulAli2011}. Then, for a scaled and shifted  Gaussian function $f_\alpha(\bk) = f(\alpha (\bk-\bk_0)), ~\bk_0 \in \mathbb R^3,\alpha >0 $, we have $\Phi_\alpha (\bk) = {\alpha}^{-1} \Phi(\alpha (\bk-\bk_0))$.
}
\end{example}


\begin{table}[!h]
  \centering
  \caption{\small {Numerical errors and computational time of TKM in Example~\ref{tkmGaussian}.}}
  \label{TKM_convergence_data}
   \begin{tabular}{c|c|c|c|c}
    \hline\hline
   Convergence & $N_k$ & $l^\infty$-error & $l^2$-error & Time(s)\\
   \hline
    \multirow{6}{*}{\includegraphics[scale=0.18]{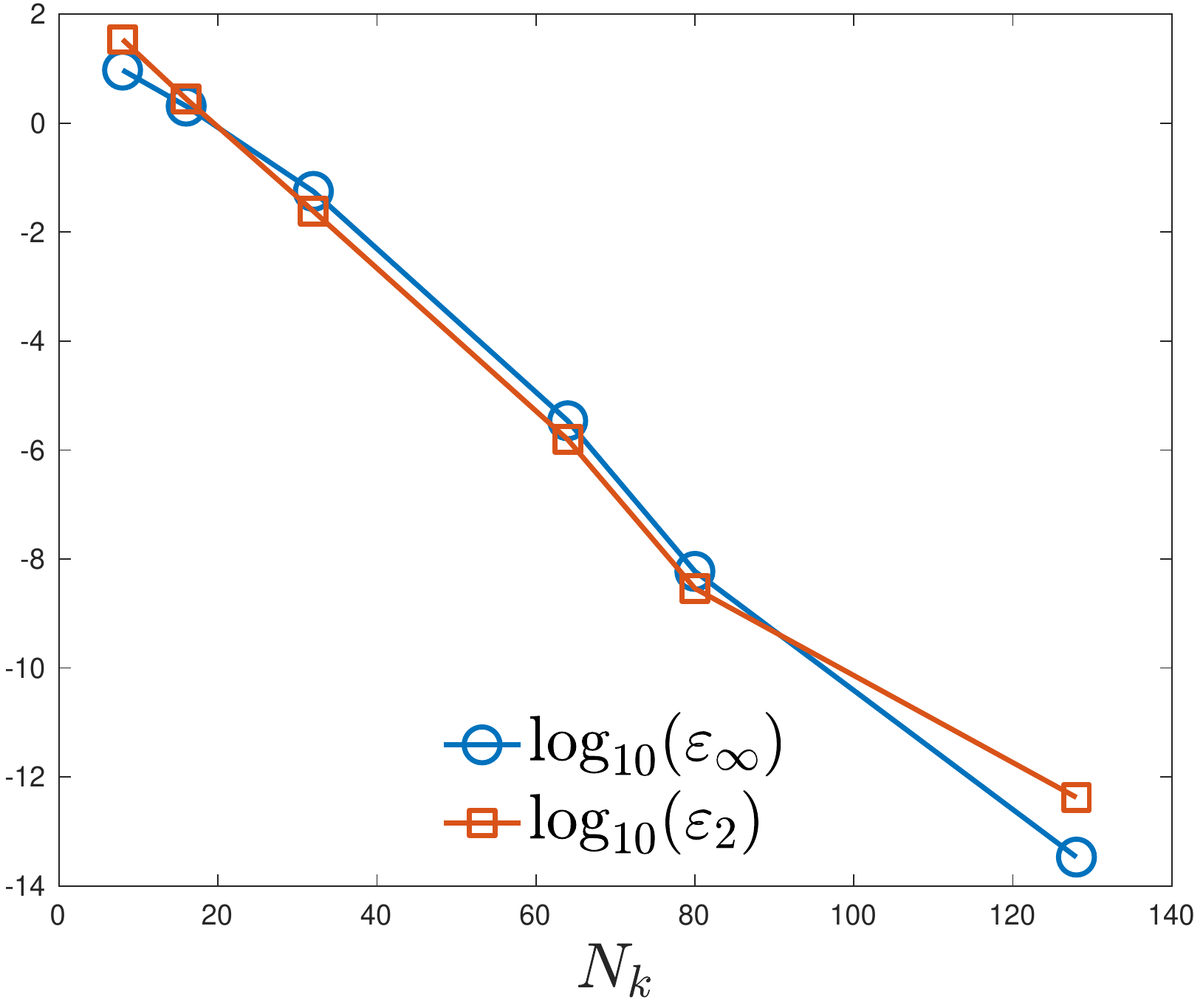}}  
&8	&	9.380				&	 34.209				&	8.300$\times10^{-5}$	\\
&16	&	2.044				&	2.784				&	8.500$\times10^{-4}$	\\
&32	&	5.575$\times10^{-2}$	&	2.423$\times10^{-2}$	&	8.424$\times10^{-3}$	\\
&64	&	3.434$\times10^{-6}$	&	1.556$\times10^{-6}$	&	8.624$\times10^{-2}$\\
&80	&	5.918$\times10^{-9}$	&	2.879$\times10^{-9}$	&	1.960$\times10^{-1}$	\\
&128	&	3.197$\times10^{-14}$	&	4.205$\times10^{-13}$	&	8.142$\times10^{-1}$\\
   \hline\hline
 \end{tabular}  
  \end{table}

\section{Numerical experiments}
\label{sec.num}

From this section, it begins to perform a series of benchmark tests and make a thorough performance evaluation of CHASM. 
The scalability of our scheme up to 16000 cores is also presented, with details of parallel implementations and computational environments given in Section \ref{sec:parallel}.

As the first step, we need to investigate the convergence, stability and mass conservation property of CHASM. 
To this end, we test the quantum harmonic oscillator in 2-D phase space, where  the Wigner dynamics reduces to the classical Liouville systems and the exact solutions are obtained by solving the Hamiltonian trajectories. We will show that the setting of PMBC brings in very small errors for a nonlocal problem and have only a slight influence on the mass conservation when the stencil length $n_{nb}\ge 15$.

After that, we turn to evaluate the performance of TKM. The stationary Hydrogen Wigner function of 1s state, which can be well approximated by FFTs, will be adopted as the initial and reference solution for the Wigner-Coulomb dynamics. Once the numerical accuracy is tested, it is able to study some typical quantum systems, such as the electron dynamics interacting with one or two protons, and reveal the presence of electron-proton coupling, quantum tunneling and uncertainty principle. 

The maximal error $\varepsilon_{\infty}(t) =\max_{(\bm{x},\bm{k})\in\mathcal{X}\times \mathcal{K}}\big |f^{\textup{ref}}\left(\bm{x},\bm{k},t\right)-f^{\textup{num}}\left(\bm{x},\bm{k},t\right) \big |$, the $L^2$-error $\varepsilon_{2}(t)= [\iint_{\mathcal{X}\times \mathcal{K}} \left(f^{\textup{ref}}\left(\bm{x},\bm{k},t\right)-f^{\textup{num}}\left(\bm{x},\bm{k},t\right)\right)^{2}\textup{d}\bm{x}\textup{d} \bm{k}]^{\frac{1}{2}}$, 
and the deviation of total mass $\varepsilon_{\textup{mass}}(t)= |\iint_{\mathcal{X}\times \mathcal{K}} (f^{\textup{num}}\left(\bm{x},\bm{k},t\right)-  f^{\textup{ref}}\left(\bm{x},\bm{k},t=0\right))\textup{d}\bm{x}\textup{d} \bm{k}|$ are adopted as the performance metrics, 
with $f^{\textup{ref}}$ and $f^{\textup{num}}$ the reference and numerical solution, respectively, and $\mathcal{X}\times \mathcal{K}$ denotes the computational domain. In practice, the integral can be replaced by the average over all grid points.

For a 6-D problem, we adopt the reduced Wigner function onto $(x_j$-$k_j$) plane, say, $W_j(x, k, t)  = \iint_{\mathbb{R}^2 \times \mathbb{R}^2} f(\bx, \bk, t) \D \bx_{\{1,2,3\}\setminus\{j\}}  \D \bk_{\{1,2,3\}\setminus\{j\}}$,  and the spatial marginal distribution $P(x_1, x_2, t) = \iint_{\mathbb{R} \times \mathbb{R}^3} f(\bx, \bk, t) \D x_3 \D \bk$ for visualizations.

\subsection{2-D Quantum harmonic oscillator}
The first example is the quantum harmonic oscillator  $V(x) =  {m \omega x^2}/{2}$ and its $\pdo$ reduces to the first-order derivative,
\begin{equation}\label{Wigner_harmonic}
\frac{\partial }{\partial t} f(x, k, t) + \frac{\hbar k}{m}  \nabla_{x} f(x, k, t) - \frac{1}{\hbar}\nabla_{x} V(x)  \nabla_{k} f(x, k, t) = 0.
\end{equation}
The exact solution can be solved by $f(x, k, t) = f(x(t), k(t), 0)$,
where $(x(t), k(t))$ obey a (reverse-time) Hamiltonian system ${\partial x}/{\partial t} =  -{\hbar k}/{m}, {\partial k}/{\partial t} = {m\omega x}/{\hbar}$,
and reads 
\begin{equation}
\begin{split}
&x(t) = \cos \left(\sqrt{\omega} t\right) x(0) - \frac{\hbar}{m \sqrt{\omega}} \sin \left(\sqrt{ \omega}t\right) k(0), \\
&k(t) =\frac{m\sqrt{\omega}}{\hbar}\sin \left(\sqrt{\omega} t\right) x(0) +  \cos \left(\sqrt{ \omega}t\right) k(0).
\end{split}
\end{equation}

\begin{example} 
\textup{
Consider a quantum harmonic oscillator  $V(x) =  m \omega x^2/2$ and an initial Gaussian wavepacket $f_0(x, k) = \pi^{-1} \me^{-\frac{1}{2}(x-1)^2 - 2k^2}$.  We choose $\omega = (\pi/5)^2$ so that the wavepacket returns back to the initial state at the final time $T = 10$.
}
\end{example}

The computational domain is $\mathcal{X} \times \mathcal{K} = [-12, 12] \times [-6.4, 6.4]$, which is evenly decomposed into 4 patches for MPI implementation. The natural boundary condition is adopted at two ends so that there is a slight loss of mass (about $10^{-13}$) up to $T=10$, while the Neumann boundary condition may lead to artificial wave reflection and exhibits a rapid growth of errors when the wavepacket moves close to the boundary (see Section 2.4 of our supplementary material \cite{XiongZhangShao2022}).

\begin{figure}[!h]
\centering
\subfigure[$\varepsilon_{\infty}(t)$ under $n_{nb}=10$ and different $\Delta x$.\label{harmonic_comp_dx_nb10}]{
{\includegraphics[width=0.48\textwidth,height=0.27\textwidth]{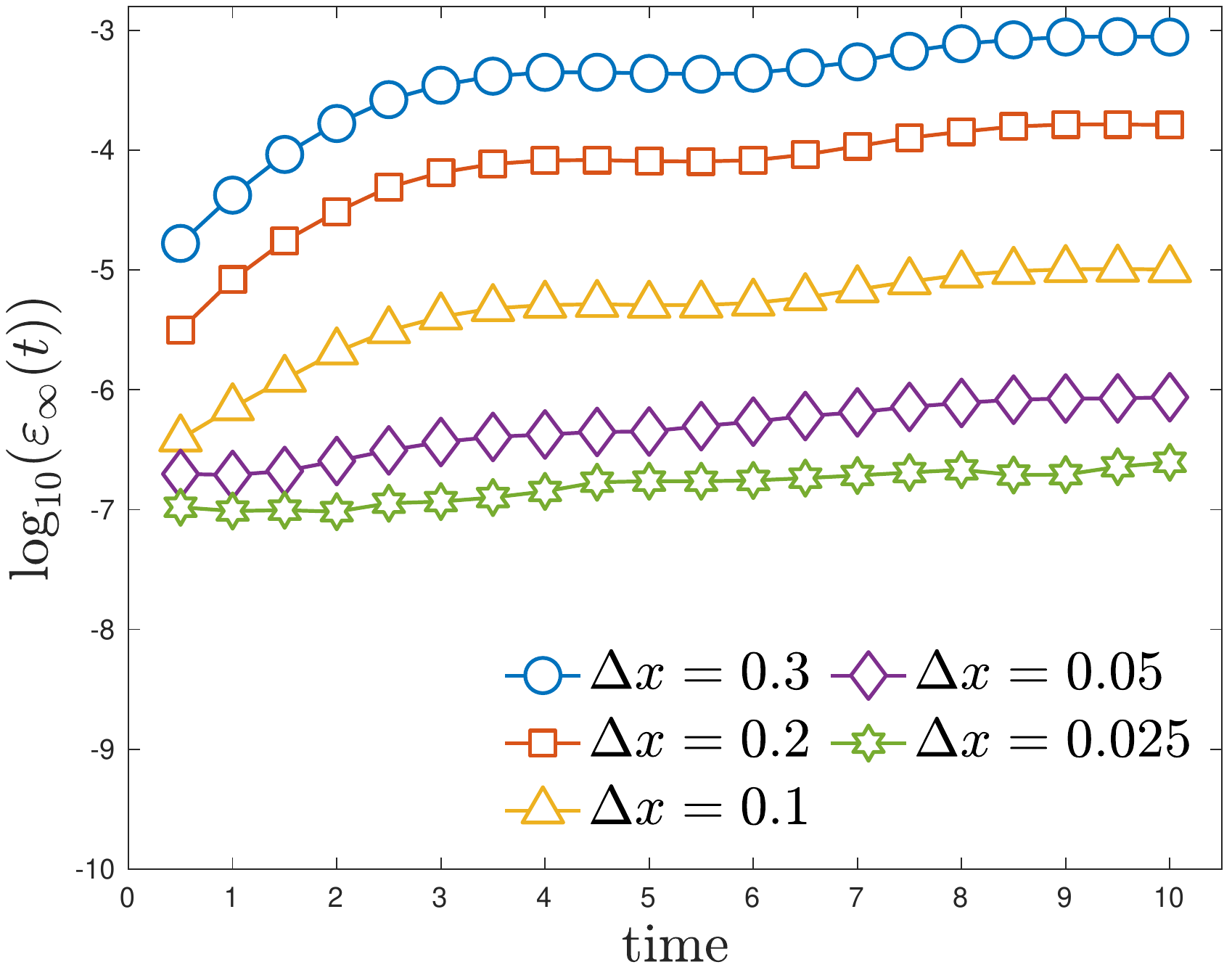}}}
\subfigure[$\varepsilon_{\infty}(t)$ under $n_{nb}=20$ and different $\Delta x$.\label{harmonic_comp_dx_nb20}]
{\includegraphics[width=0.48\textwidth,height=0.27\textwidth]{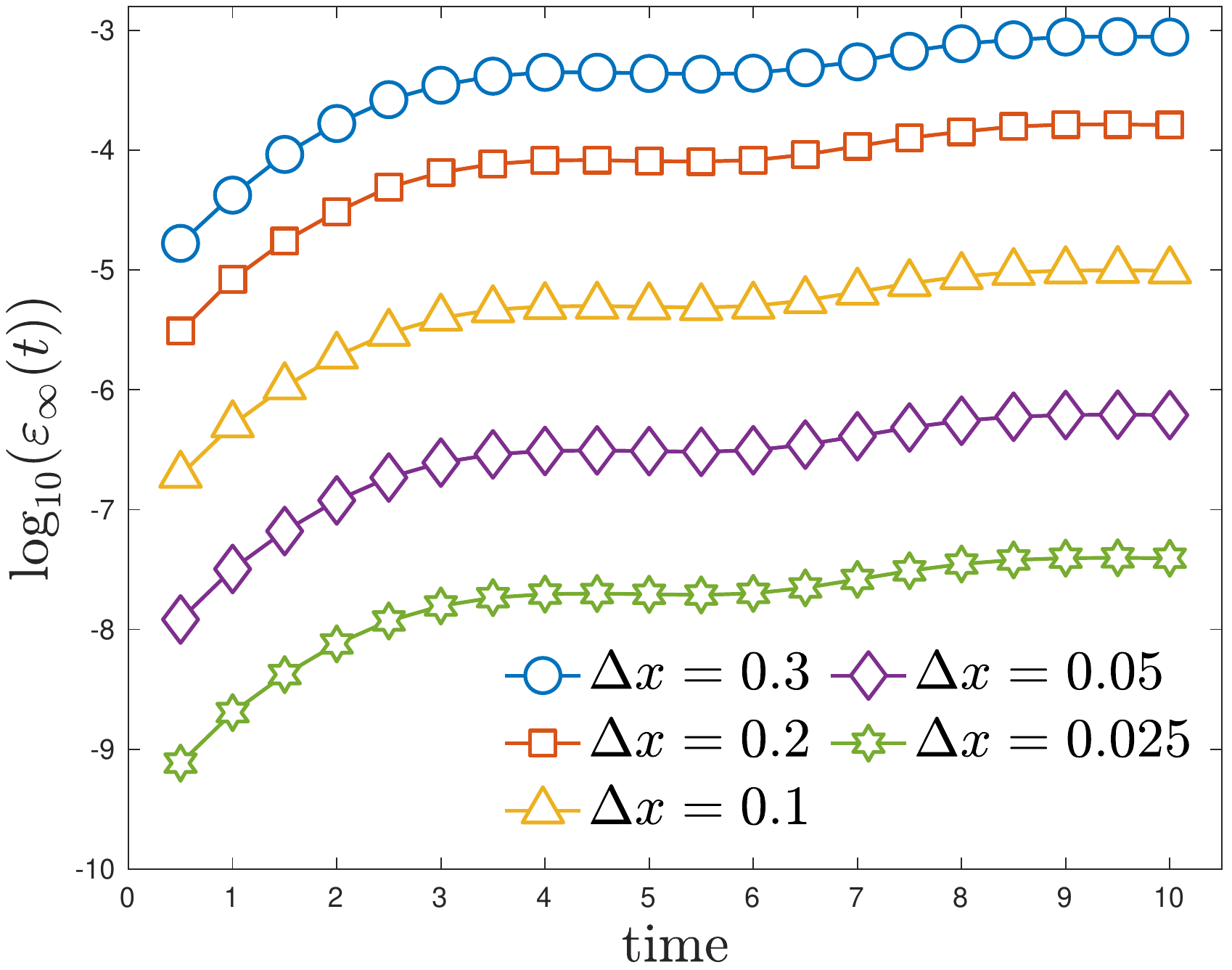}}
\\
\centering
\subfigure[$\varepsilon_{\infty}(t)$ under $\Delta x= 0.1$ and different $n_{nb}$.\label{1s_maxerr}]{
{\includegraphics[width=0.48\textwidth,height=0.27\textwidth]{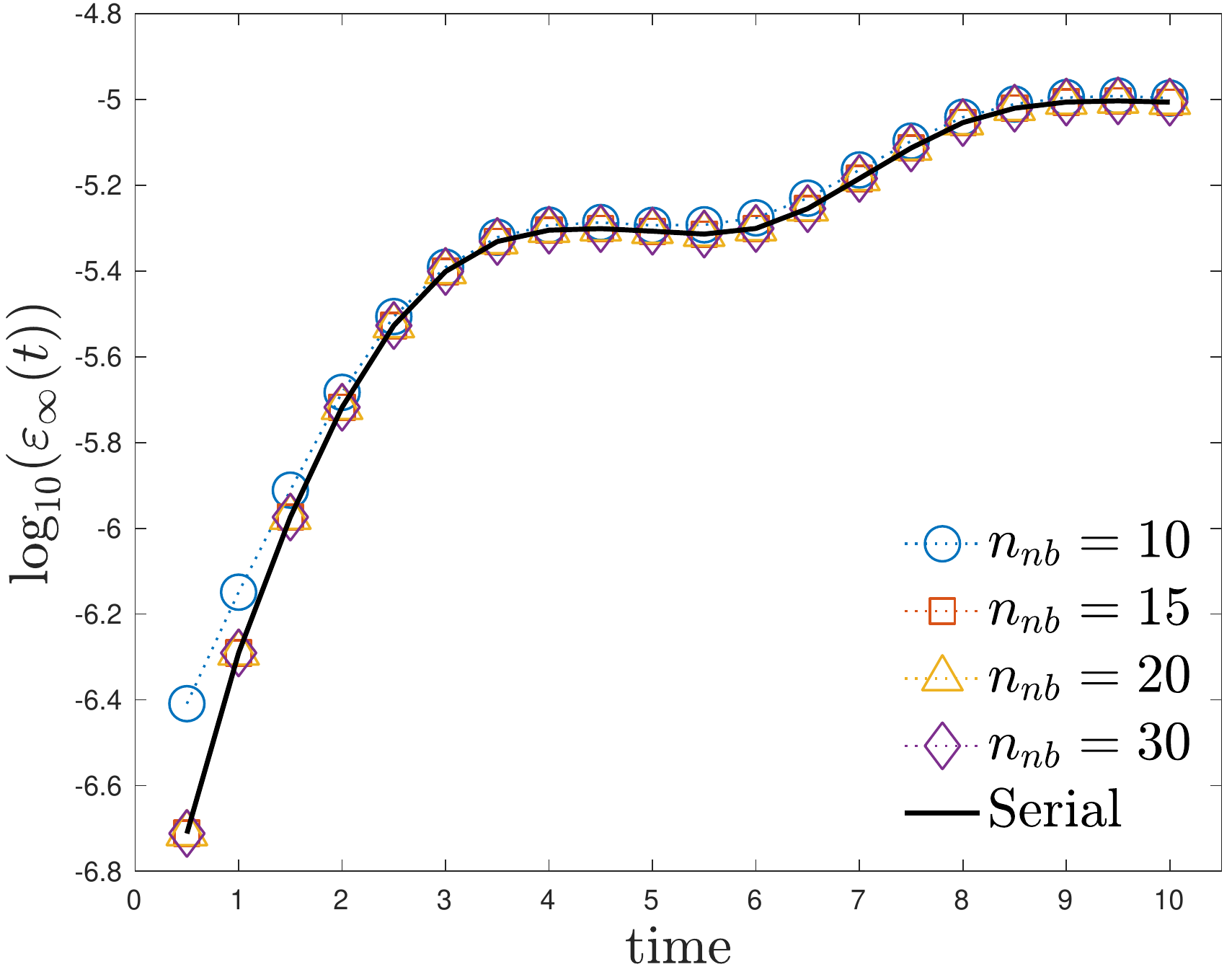}}}
\subfigure[$\varepsilon_{2}(t)$ under $\Delta x= 0.1$ and different $n_{nb}$.\label{1s_L2err}]
{\includegraphics[width=0.48\textwidth,height=0.27\textwidth]{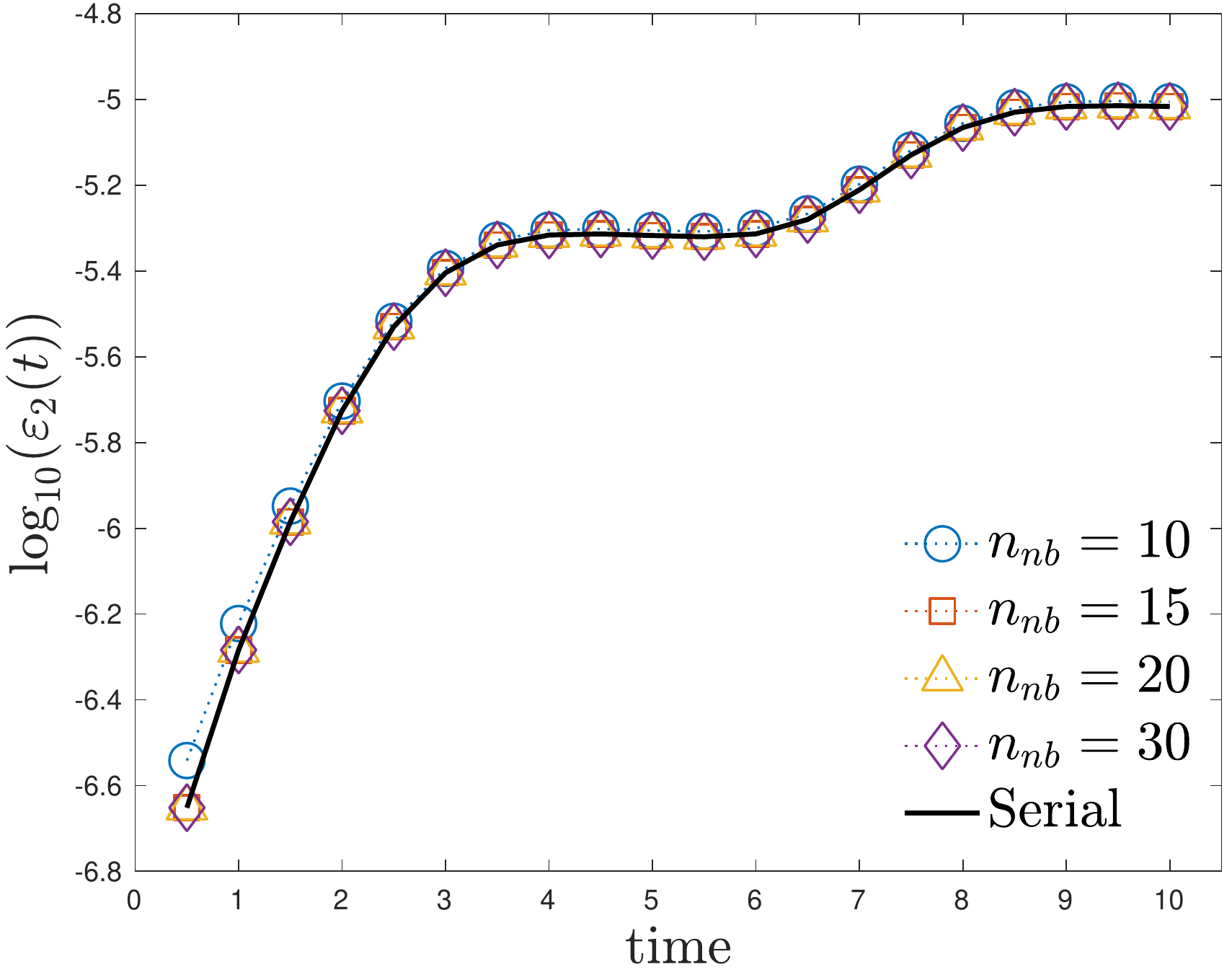}}
\\
\centering
\subfigure[Convergence with respect to $\Delta x$.\label{convergence_LPC1}]
{\includegraphics[width=0.48\textwidth,height=0.27\textwidth]{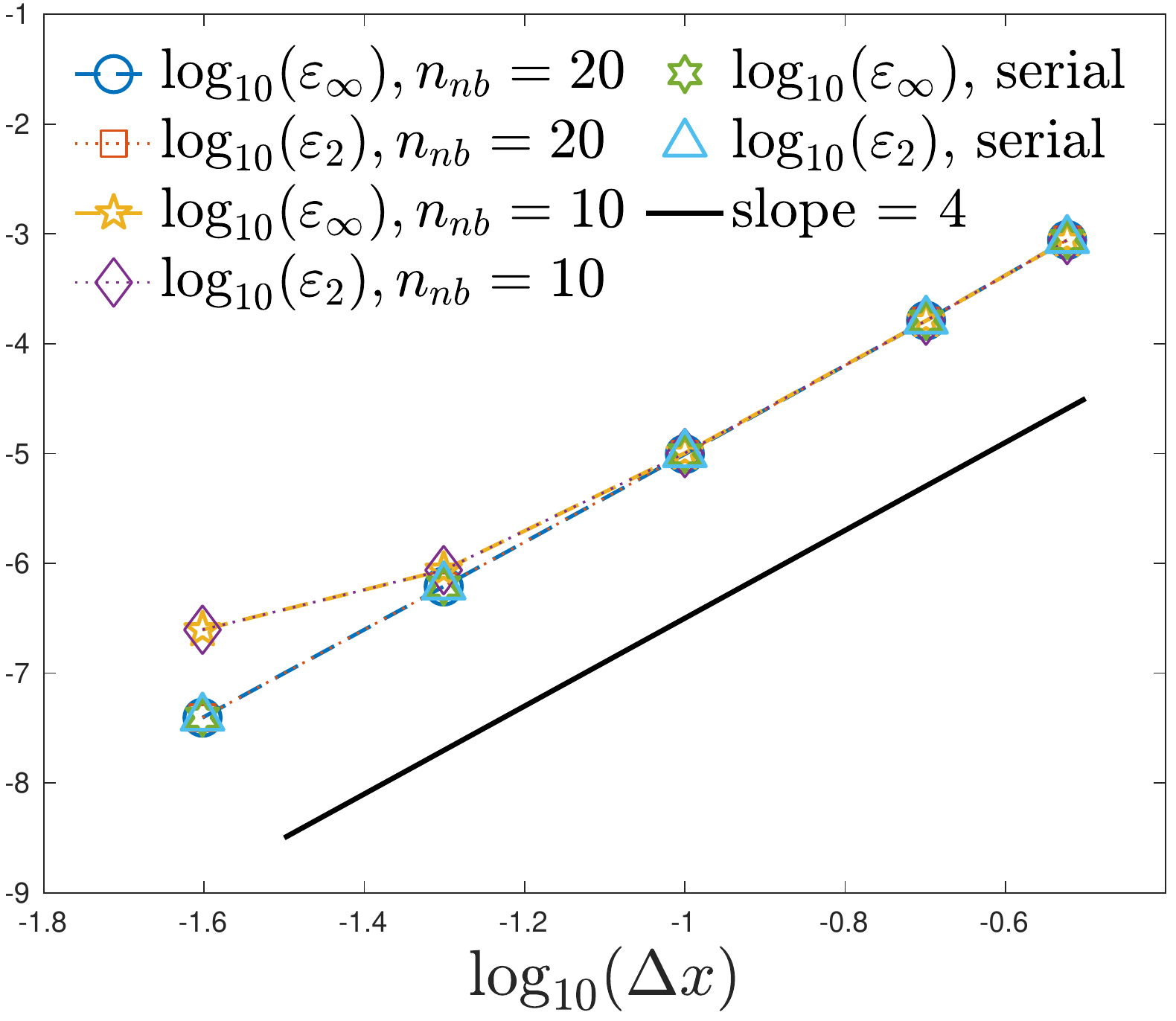}}
\subfigure[Evolution of  $\varepsilon_{\textup{mass}}(t)$.\label{mass_LPC1}]
{\includegraphics[width=0.48\textwidth,height=0.27\textwidth]{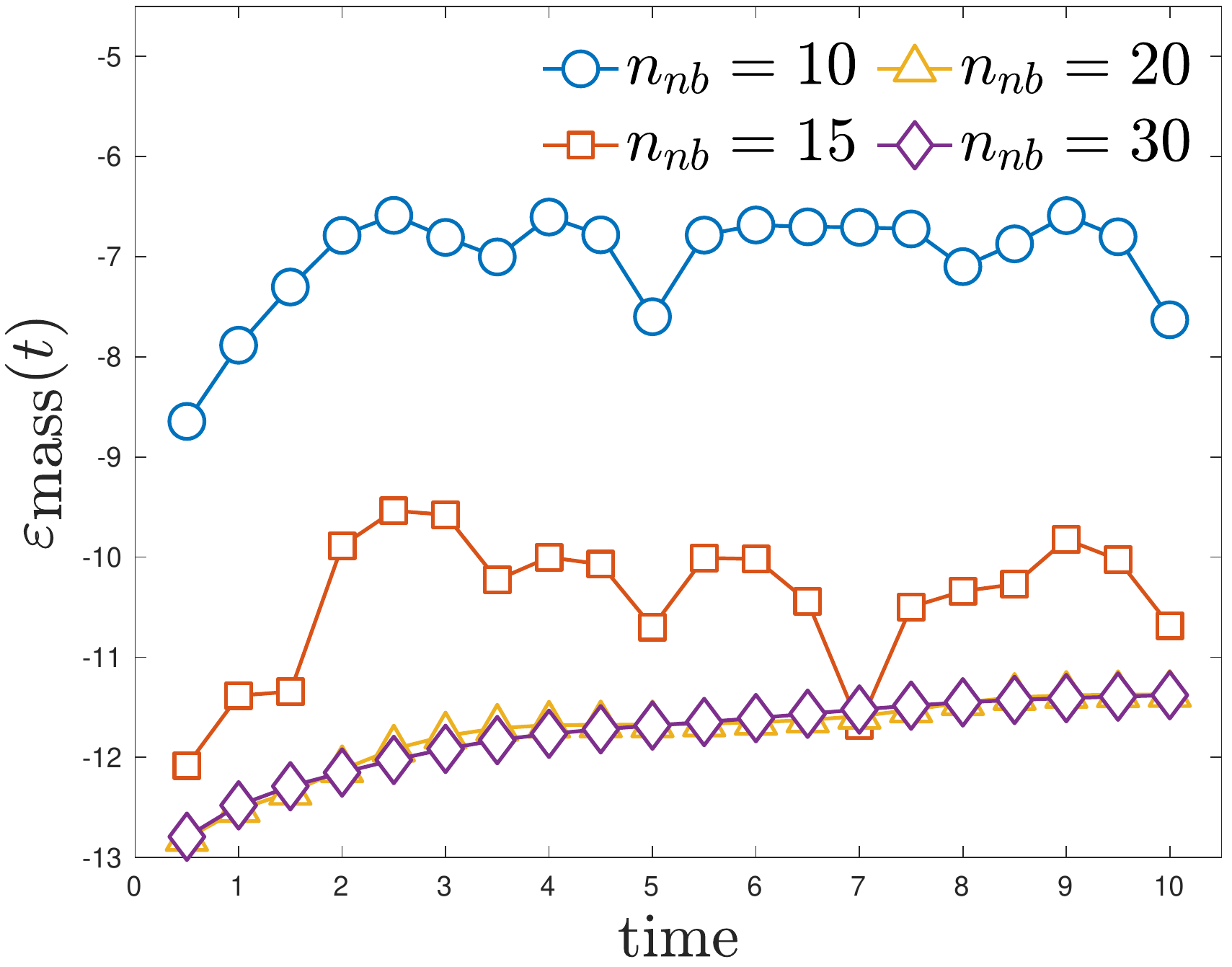}}
\caption{\small 2-D quantum harmonic oscillator: The convergence and mass conservation of LPC1. LPC1 can achieve fourth-order convergence in $\Delta x$. PMBC brings in smaller errors and causes a slight loss of mass, but fortunately they are almost  eliminated when $n_{nb} \ge 20$.\label{harmonic_convergence_LPC1}}
\end{figure}

Since we mainly focus on the convergence with respect to $\Delta x$ and $n_{nb}$, several groups of  simulations under $\Delta x = 0.025, 0.05, 0.1, 0.2,0.3$ and $n_{nb}=10, 15,20,30$ are performed, where other parameters are set as: the time step $\tau  = 0.00002$ and $\Delta k = 0.025$ to achieve spectrally accurate approximation to $\pdo$. The convergence with respect to $\Delta x$ and the mass conservation under different $n_{nb}$ are given in Figure \ref{harmonic_convergence_LPC1}. From the results, we can make the following observations.

{\bf Convergence with respect to $\Delta x$:} The convergence rate is plotted in Figure \ref{convergence_LPC1}. LPC1 can achieve spatial fourth order convergence when $n_{nb} \ge 15$, according with the theoretical value of the cubic spline interpolation. While a reduction in convergence is observed when $n_{nb} = 10$ because of the truncated stencils in Eq.~\eqref{truncation}.

{\bf Influence of PMBCs:} From Figures \ref{harmonic_comp_dx_nb10} and \ref{harmonic_comp_dx_nb20}, one can see that $n_{nb} = 10$ only bring in additional errors about $10^{-5}$. Such errors seem to be negligible when $n_{nb} \ge 15$, which coincides with the observations made in \cite{MalevskyThomas1997}. However, the truncation of stencils indeed has a great influence on the mass conservation as seen in Figure  \ref{mass_LPC1}, where $\varepsilon_{\textup{mass}}$ is about $10^{-6}$ when $n_{nb}=10$ or $10^{-9}$ when $n_{nb}=15$. Fortunately, its influence on total mass can be nearly eliminated when $n_{nb} \ge 20$.

{\bf Numerical stability:} The first-order derivative in Eq.~\eqref{Wigner_harmonic} brings in strong numerical stiffness and puts a severe restriction on the time step $\tau$ in CHASM. Nevertheless, we have observed in \cite{XiongZhangShao2022} that LPC1 is more stable than the splitting scheme, which has also been pointed out in \cite{CrouseillesEinkemmerMassot2020}, as well as the multi-stage non-splitting scheme.  Actually, LPC1 turns out to be stable up to $T=20$ under a much larger time step $\tau =0.0005$, while the Strang operator splitting becomes unstable under such setting (see Section 4.1 of our supplementary material \cite{XiongZhangShao2022}).


%
%
%
%
%

\subsection{Hydrogen Wigner function: 1s state}
\begin{figure}[!h]
\centering
\subfigure[$P(x_1, x_2)$ for 1s orbital.\label{1s_xdist}]{
{\includegraphics[width=0.48\textwidth,height=0.27\textwidth]{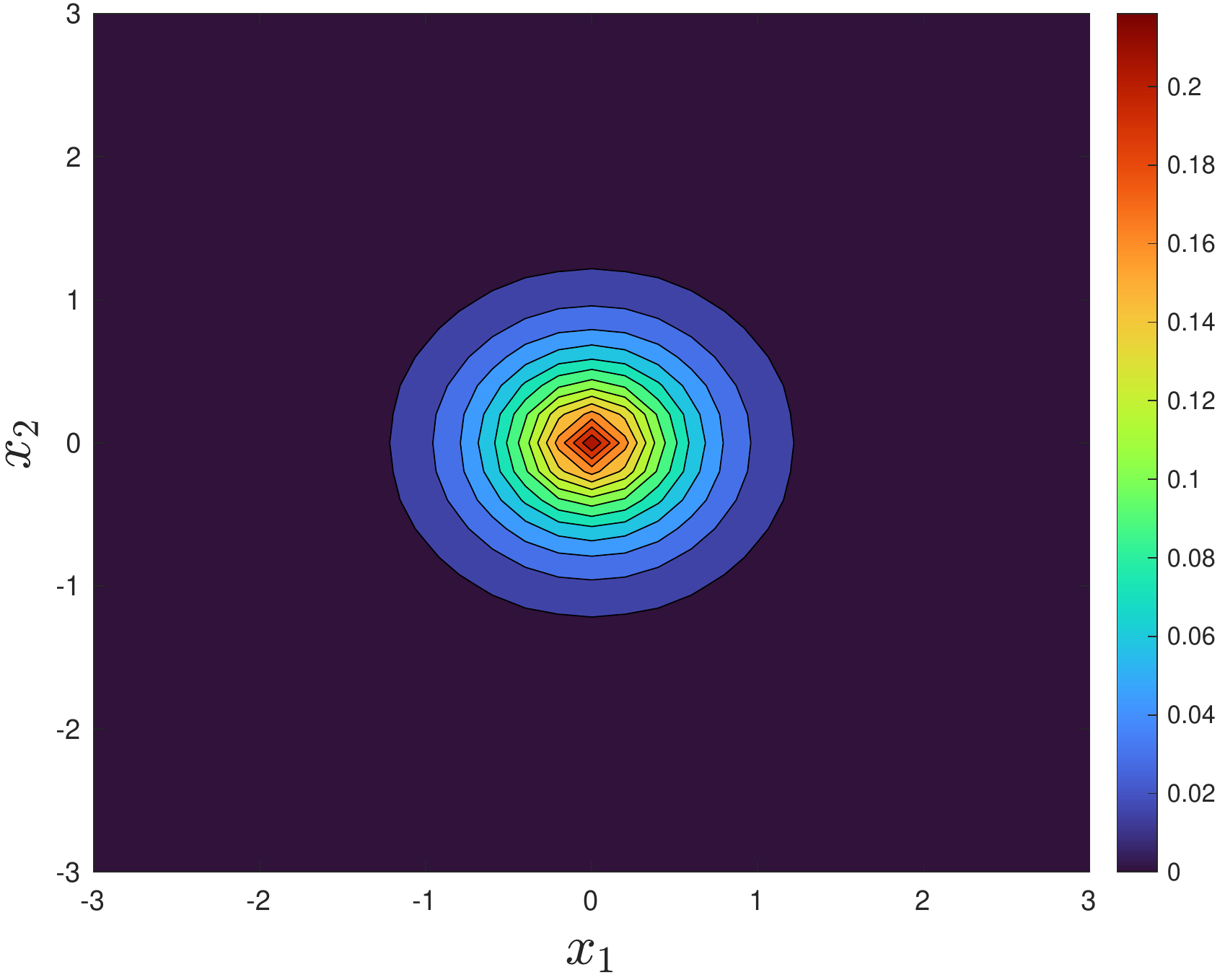}}}
\subfigure[ $W_1(x, k)$ for 1s orbital.\label{1s_wigner}]{
{\includegraphics[width=0.48\textwidth,height=0.27\textwidth]{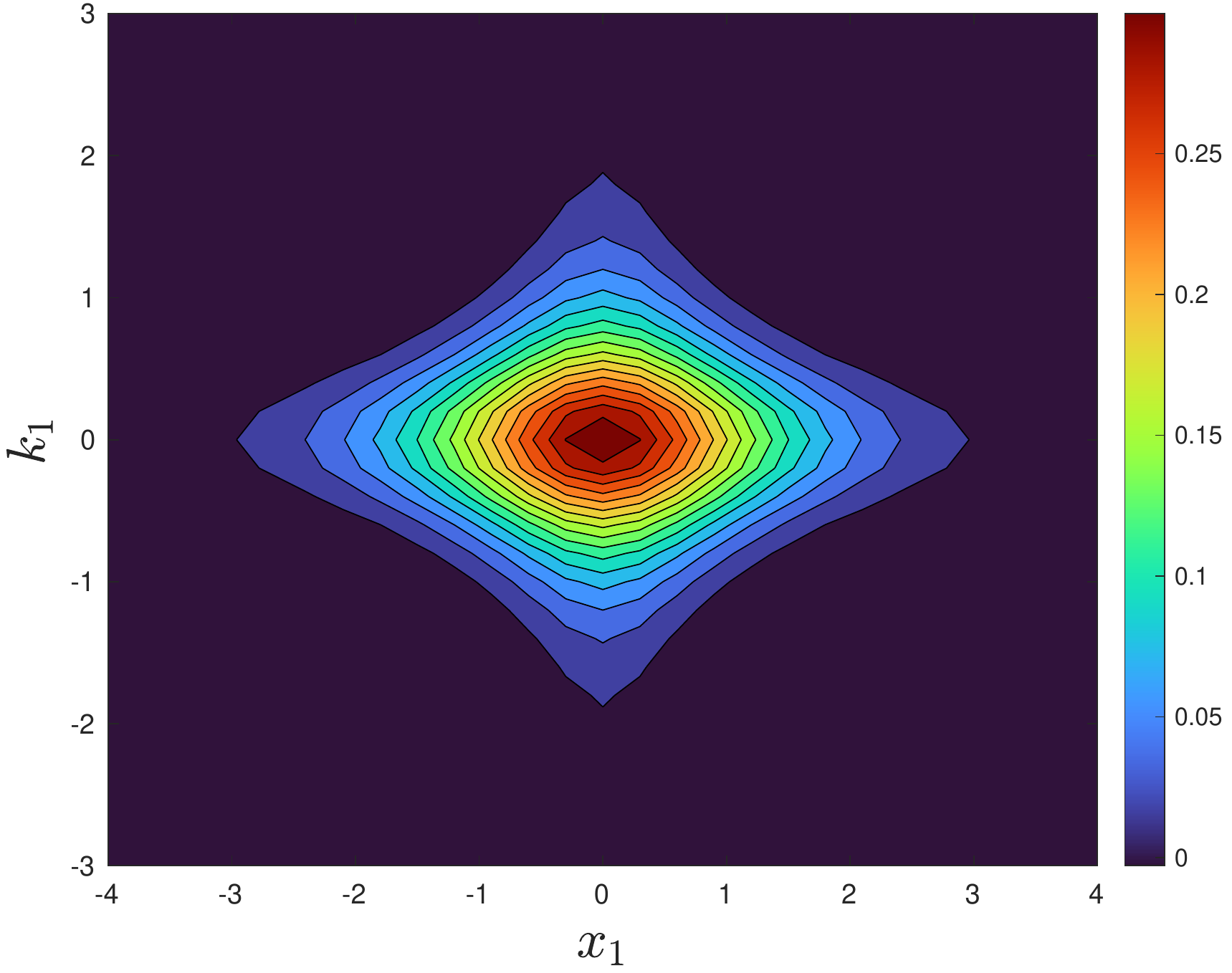}}}
\\
\centering
\subfigure[The heavy tail in momental space.\label{1s_tail}]{
{\includegraphics[width=0.48\textwidth,height=0.27\textwidth]{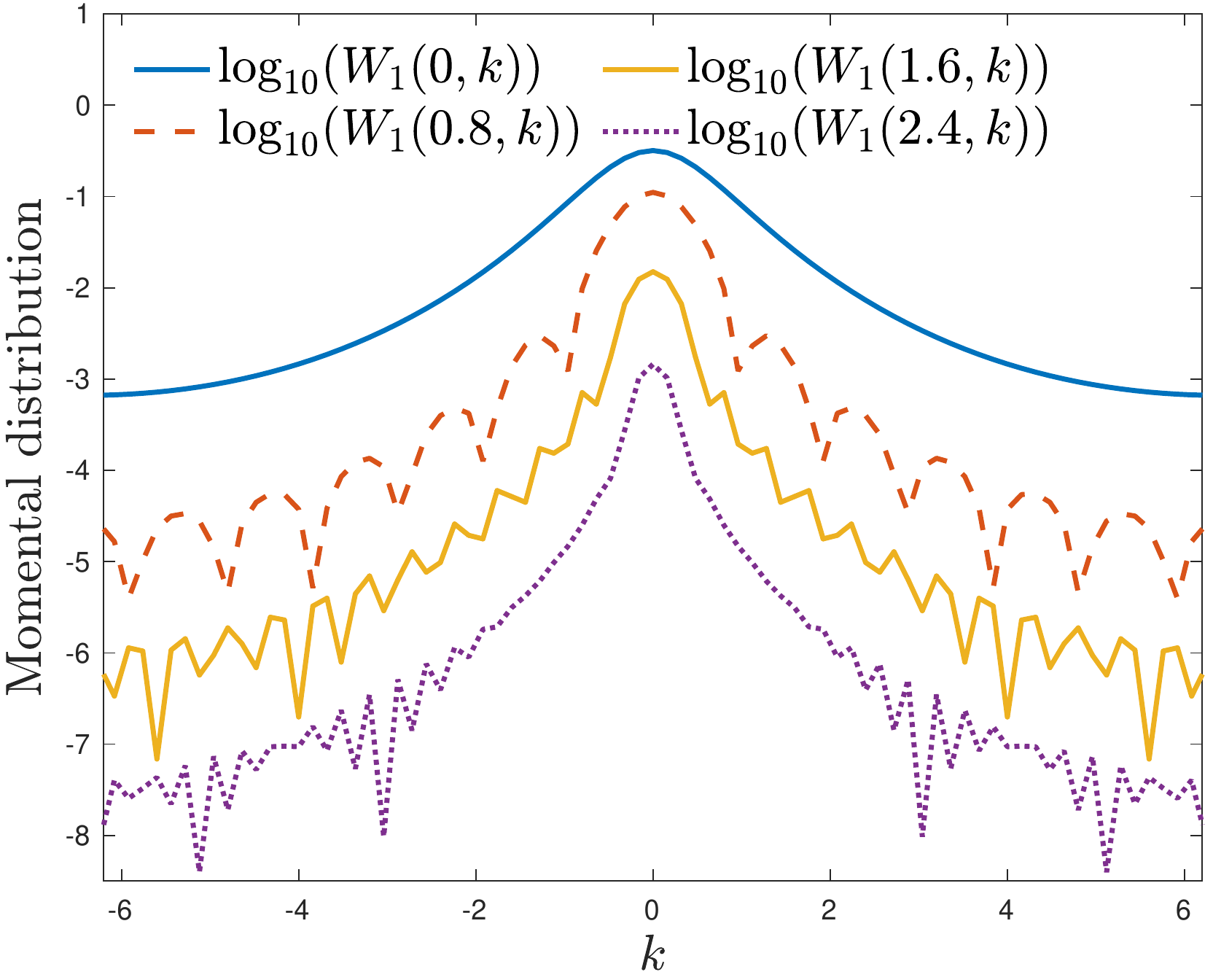}}}
\subfigure[$W_1^{\textup{num}} - W_1^{\textup{ref}}$ at $t = 5$a.u. ($N_k = 64$).\label{1s_error_visualization_Nk64}]
{\includegraphics[width=0.48\textwidth,height=0.27\textwidth]{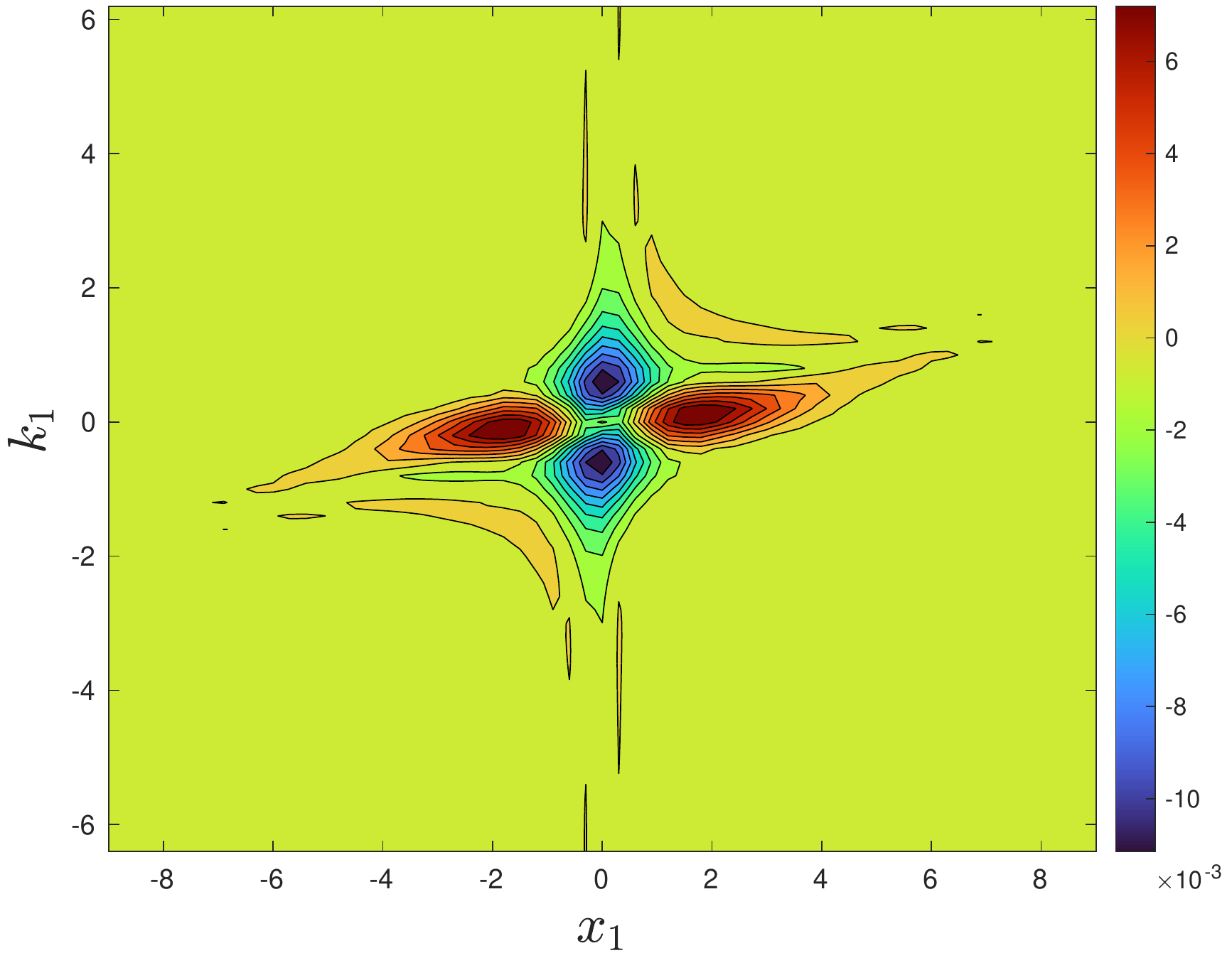}}
\caption{\small  The Hydrogen 1s Wigner function:  A visualization of  the Hydrogen 1s orbital, the reduced Hydrogen 1s Wigner function $W_1(x, k)$ and the numerical errors $W_1^{\textup{num}} - W_1^{\textup{ref}}$ at $t = 5$a.u. Small errors are observed near the $\bk$-boundary as $f_{\textup{1s}}(\bx, \bk)$ has a heavy tail in $\bk$-space, which have influences on the convergence rate of TKM and mass conservation.
\label{1s_visualization}}
\end{figure}

We turn to evaluate the performance of CHASM in 6-D problems. The Hydrogen Wigner function  is very useful for dynamical testing as it is the stationary solution of the Wigner equation.  For the 1s orbital, $\phi_{\textup{1s}}(\bx) = \frac{1}{2\sqrt{2} \pi^2} \exp( - |\bx|)$,  the Wigner function is given by Eq.~\eqref{def.Wigner_function} with $\rho(\bx_1, \bx_2) = \phi_{\textup{1s}}(\bx_1) \phi_{\textup{1s}}^\ast(\bx_2)$.
Although it is too complicated to obtain an explicit formula, the Hydrogen Wigner function of 1s state can be highly accurately approximated by the discrete Fourier transform of Eq.~\eqref{def.Wigner_function}:  For $\bk_{\bzeta} = \bzeta \Delta k$,
\begin{equation*}
 f_{\textup{1s}}(\bx, \bk_{\bzeta}) \approx \sum_{\eta_1 = -\frac{N_{y}}{2}}^{\frac{N_{y}}{2}-1} \sum_{\eta_2 = -\frac{N_{y}}{2}}^{\frac{N_{y}}{2}-1} \sum_{\eta_3 = -\frac{N_{y}}{2}}^{\frac{N_{y}}{2}-1} \phi_{\textup{1s}}(\bx - \frac{\bm{\eta} \Delta y}{2}) \phi_{\textup{1s}}^\ast(\bx + \frac{\bm{\eta} \Delta y}{2}) \me^{-  \mi (\bzeta \cdot \bm{\eta} ) \Delta k \Delta y }  (\Delta y)^3.
\end{equation*}
By taking $\Delta y = \frac{2\pi}{N_k \Delta k}$, it can be realized by FFT (we use $N_y =128$). The spatial density of 1s orbital on $(x_1$-$x_2$) plane and the reduced Wigner function $W_1(x, k)$ projected on $(x_1$-$k_1$) plane are visualized in Figures \ref{1s_xdist} and \ref{1s_wigner}, respectively.

The storage of 6-D grid mesh requires a tremendous amount of computer memory and hinders the benchmarks under very fine grid mesh. To alleviate such problem, we have to adopt SINGLE precision to save halves of memory, which is adequate for cubic spline interpolations, but still adopt DOUBLE precision for TKM. The computational domain is $\mathcal{X} \times \mathcal{K} = [-9, 9]^3 \times [-6.4, 6.4]^3$ with a fixed spatial spacing $\Delta x = 0.3$ ($N_{x} = 61$), where the accuracy of spline interpolation has been already tested in 2-D example. The natural boundary condition is again adopted at two ends.

\begin{figure}[!h]
\centering
\subfigure[Evolution of  $\varepsilon_{\infty}(t)$ under different $N_k$.]
{\includegraphics[width=0.48\textwidth,height=0.27\textwidth]{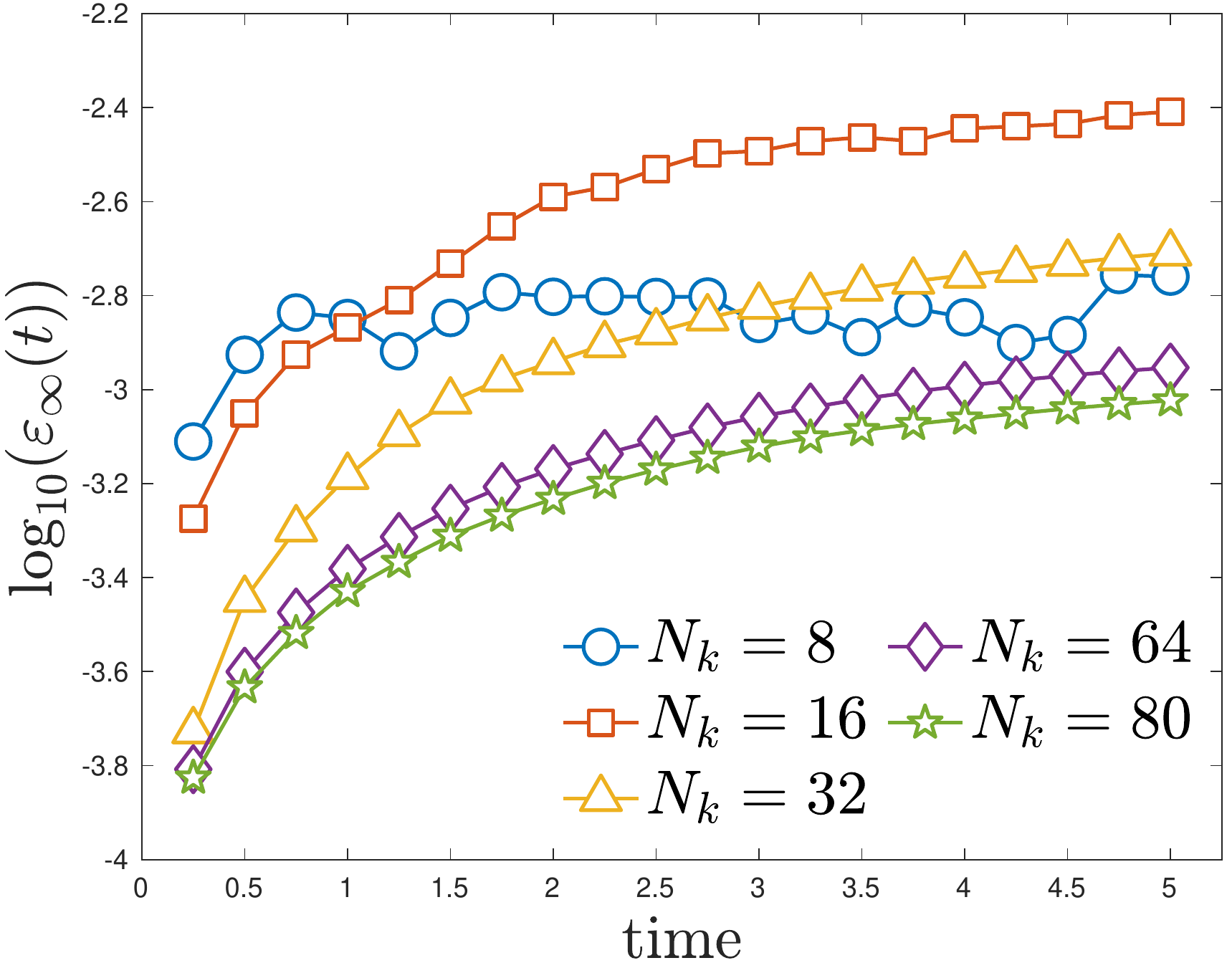}}
\subfigure[Evolution of  $\varepsilon_{2}(t)$ under different $N_k$.]
{\includegraphics[width=0.48\textwidth,height=0.27\textwidth]{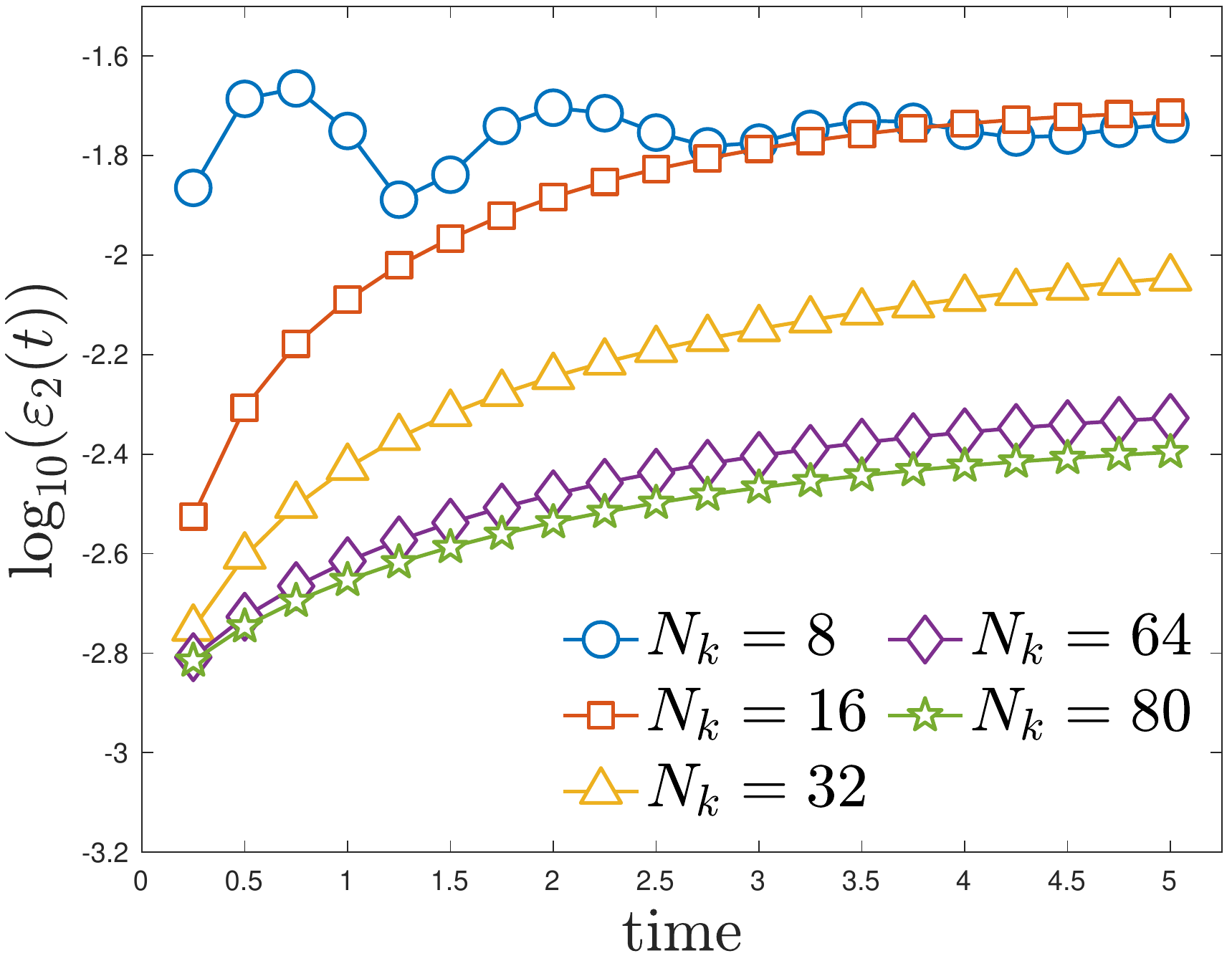}}
\\
\centering
\subfigure[Convergence with respect to $N_k$. \label{1s_convergence}]
{\includegraphics[width=0.48\textwidth,height=0.27\textwidth]{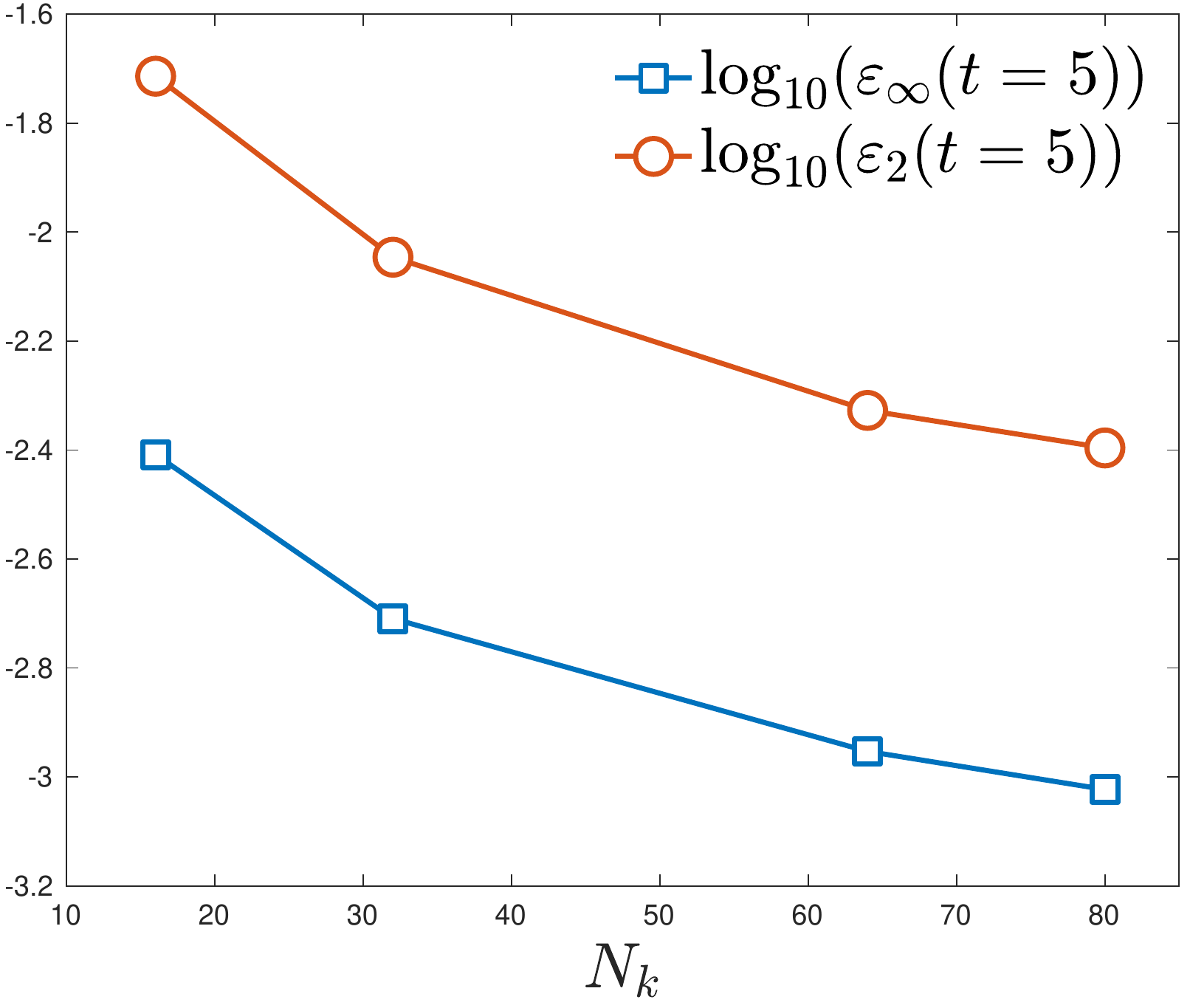}}
\subfigure[Evolution of  $\varepsilon_{\textup{mass}}(t)$.\label{1s_mass}]
{\includegraphics[width=0.48\textwidth,height=0.27\textwidth]{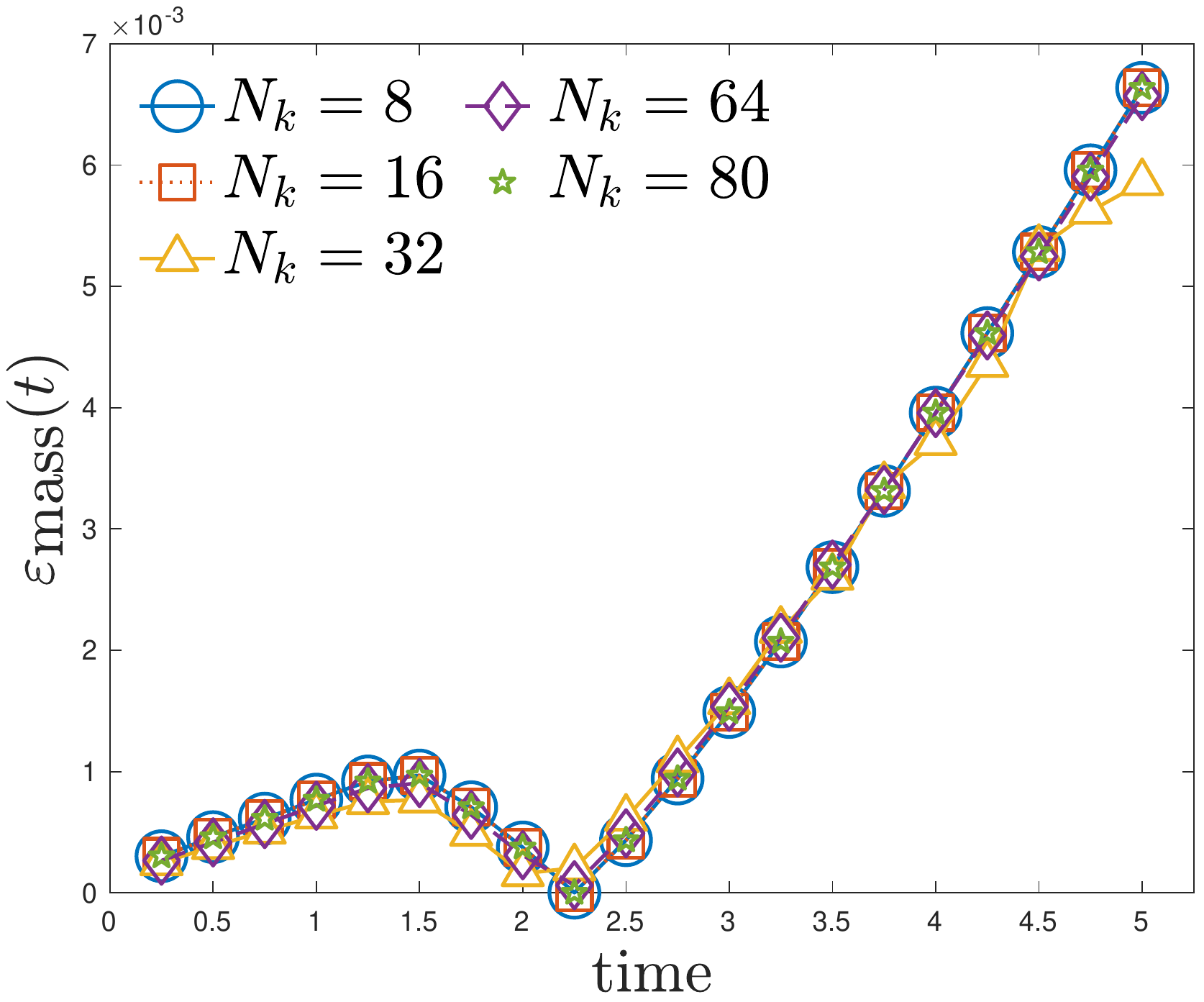}}
\caption{\small The Hydorgen 1s Wigner function:  The performance of TKM under different $\Delta k$, with $\Delta x = 0.3$. The convergence of TKM is verified, albeit with lower convergence rate due to errors caused by the spatial spline interpolation and the heavy tail of $f_{\textup{1s}}(\bx, \bk)$ is $\bk$-space.\label{1s_convergence_mass_LPC}}
\end{figure}

We mainly investigate the convergence of TKM with respect to $N_k$ by five groups: $N_k = 8, 16, 32, 64,80$ ($ \Delta k = 1.6, 0.8, 0.4, 0.2,0.16$). The domain is evenly divided into $4\times 4 \times 4$ patches and distributed by $64$ processors, and each processor provides 4 threads for shared-memory parallelization using the OpenMP library. Other parameters are set as: the stencil length in PMBC is $n_{nb} = 15$ and time step is $\tau = 0.025$. The numerical convergence and the deviation in total mass of LPC1 are presented in Figure \ref{1s_convergence_mass_LPC}, and numerical errors for reduced Wigner function $W_{1}^{\textup{num}} - W_{1}^{\textup{ref}}$ under $N_k = 64$ are visualized in Figure \ref{1s_error_visualization_Nk64}, respectively. From the results, we can make the following observations.

{\bf Convergence with respect to $\Delta k$:} The convergence of TKM is clearly verified in Figure \ref{1s_convergence}, albeit its convergence rate is slower than expectation due to  the mixture of various error terms. Nonetheless, CHASM can still achieve $\varepsilon_{\infty}(5) = 1.11\times10^{-3}$ and $\varepsilon_{2}(5) = 4.706\times 10^{-3}$ under $61^3 \times 64^3$ grid mesh, where $\max(|f_{\textup{1s}}(\bx, \bk)|) =1/\pi^3\approx 3.23\times 10^{-2}$. These metrics further reduce to $\varepsilon_{\infty}(5) = 9.48\times 10^{-4}$ and $\varepsilon_{2}(5) = 4.02\times 10^{-3}$ when $N_k = 80$. We have also tested the Strang splitting scheme for  $N_k = 64$ and obtained $\varepsilon_{\infty}(5) = 2.0\times 10^{-3}$, $\varepsilon_{2}(5) = 7.0\times 10^{-3}$, which are significantly larger than the results of LPC1 (see Section 4.2 of our supplementary material \cite{XiongZhangShao2022}).

{\bf Deviation of total mass}: A slight deviation of the total mass is observed due to the break of Eq.~\eqref{mass_conserve}. From Figure \ref{1s_mass}, one can see that $\varepsilon_{\textup{mass}}(5)$  of LPC1 is $0.66\%$, while that of the Strang splitting is $1.35\%$ (see Section 4.2 of \cite{XiongZhangShao2022}).

Two reasons may explain the above observations. On one hand, $f_{\textup{1s}}(\bx, \bk)$ exhibits a heavy tail in $\bk$-space. In Figure \ref{1s_tail}, the reduced Wigner function $W_1(\bx, \bk)$  is about $10^{-3}$ near $\bk$-boundary, indicating that $f_{\textup{1s}}(\bx, \bk)$ is not truly compactly supported in $[-6.4, 6.4]^3$. Thus the overlap with the periodic image may produce small oscillations near the $\bk$-boundary, which is also visualized in Figure \ref{1s_error_visualization_Nk64}. On the other hand, the solution might also be contaminated by the interpolation errors in the spatial space, which are about $10^{-3}$ for $\Delta x = 0.3$ and $T = 5$a.u. as presented in Figure~\ref{harmonic_comp_dx_nb20}.


\subsection{Electron dynamics interacting with one proton}

\begin{figure}[!h]
\centering
\subfigure[$W_1(x, k, t)$ (left) and $W_2(x, k, t)$ (right) at $t=1$a.u.]{
{\includegraphics[width=0.32\textwidth,height=0.18\textwidth]{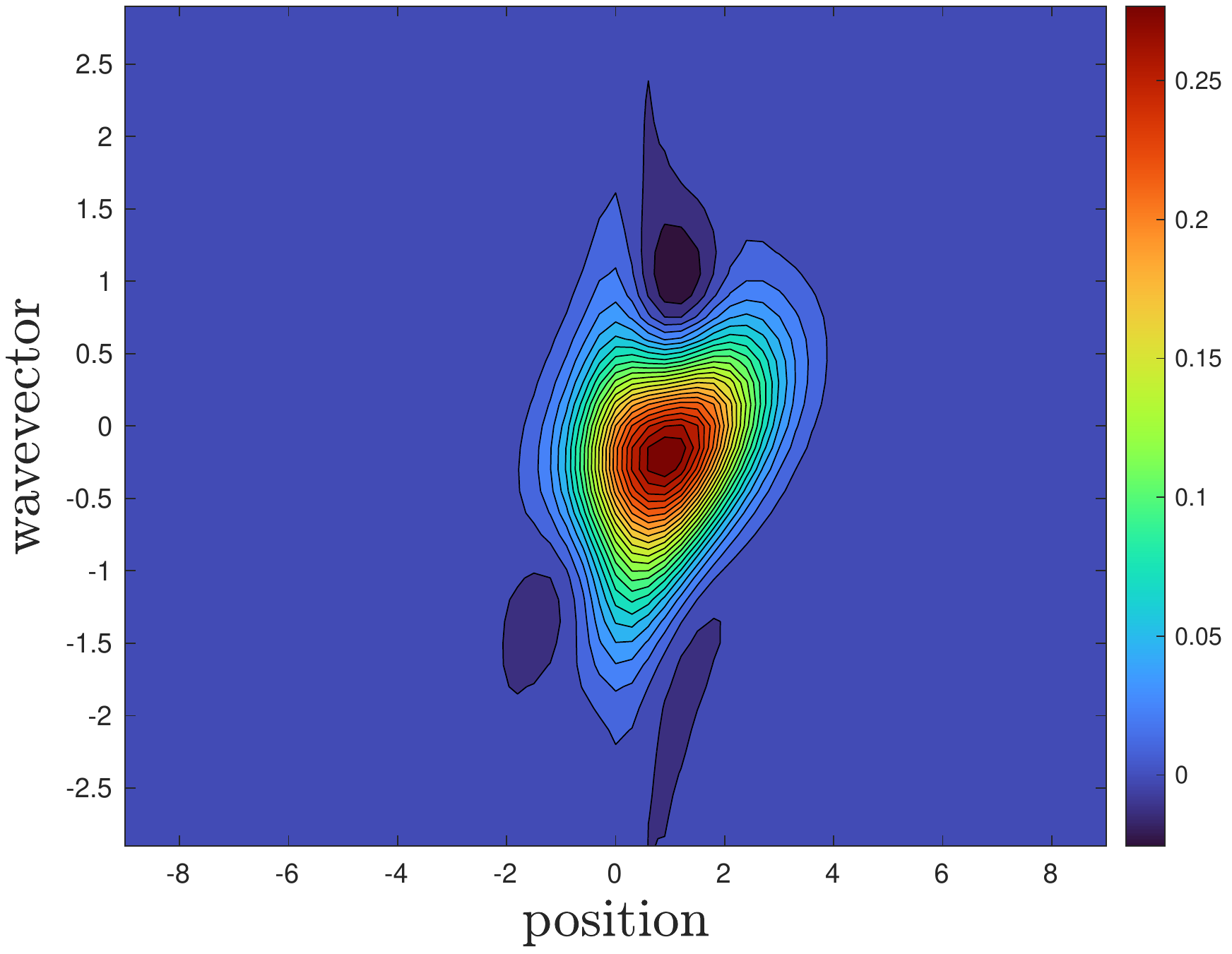}}
{\includegraphics[width=0.32\textwidth,height=0.18\textwidth]{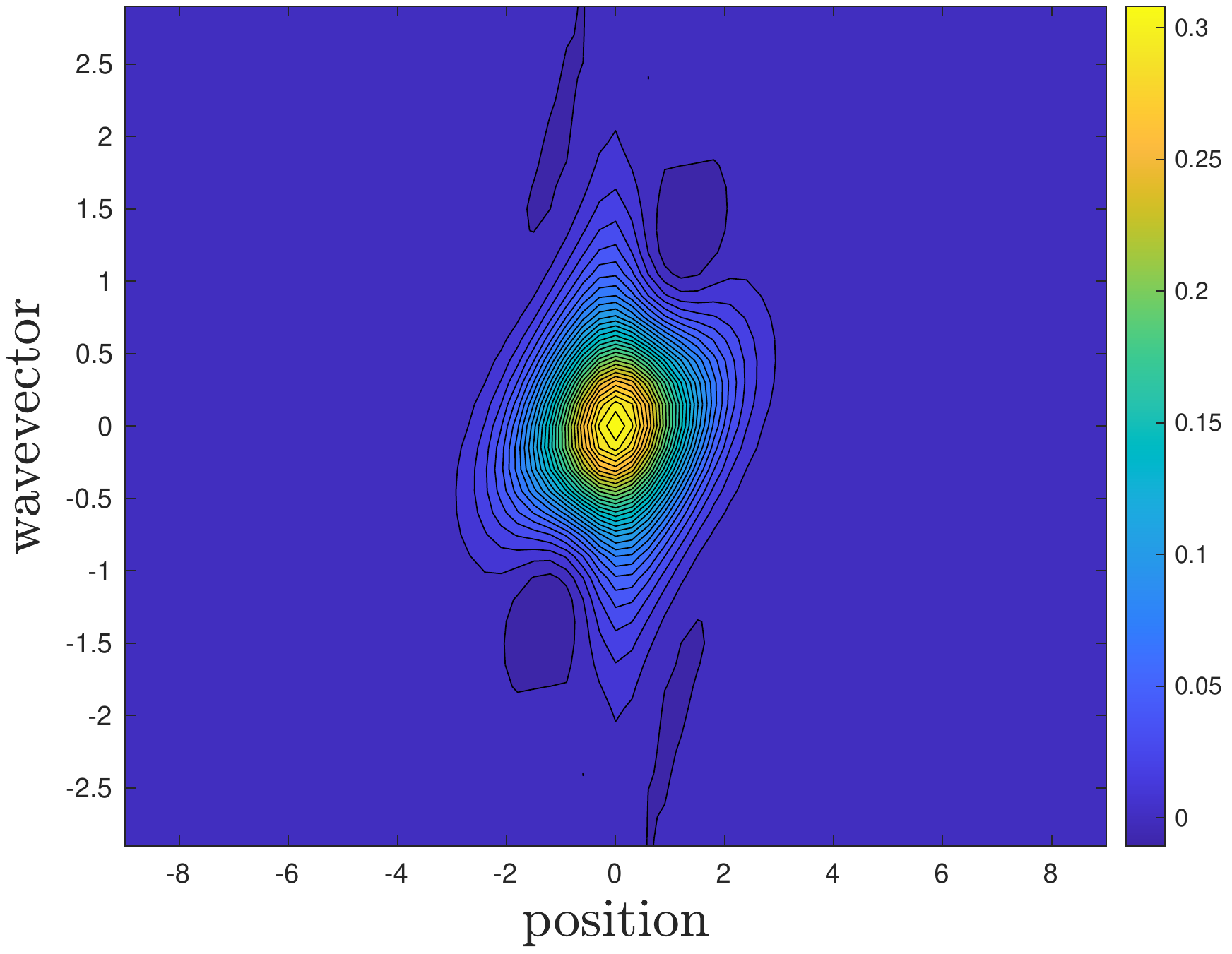}}}
\subfigure[$P(x_1, x_2, t)$ at $t=0.5$a.u.\label{xdist_t005}]{
{\includegraphics[width=0.32\textwidth,height=0.18\textwidth]{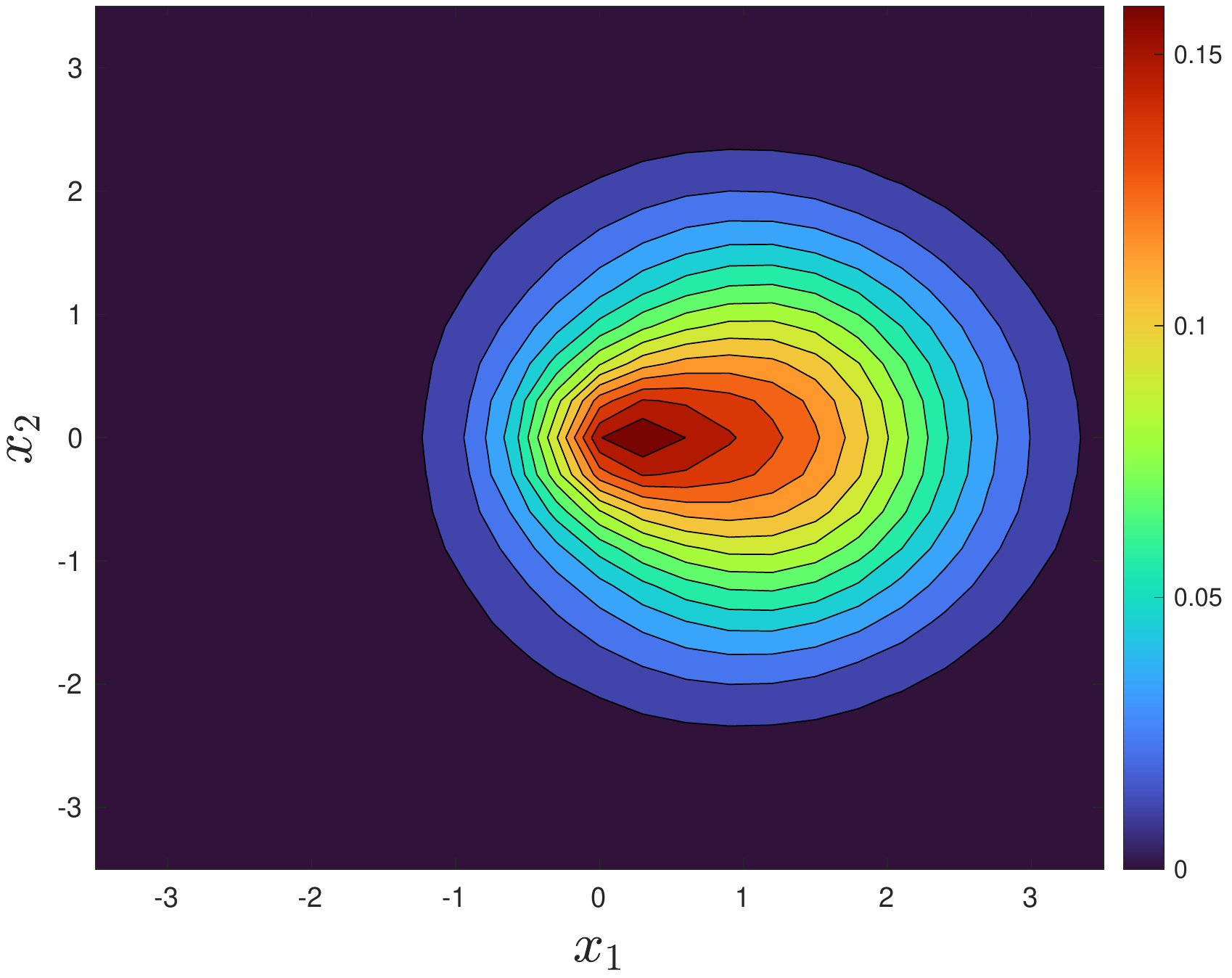}}}
\centering
\subfigure[$W_1(x, k, t)$ (left) and $W_2(x, k, t)$ (right) at $t=2$a.u.]{
{\includegraphics[width=0.32\textwidth,height=0.18\textwidth]{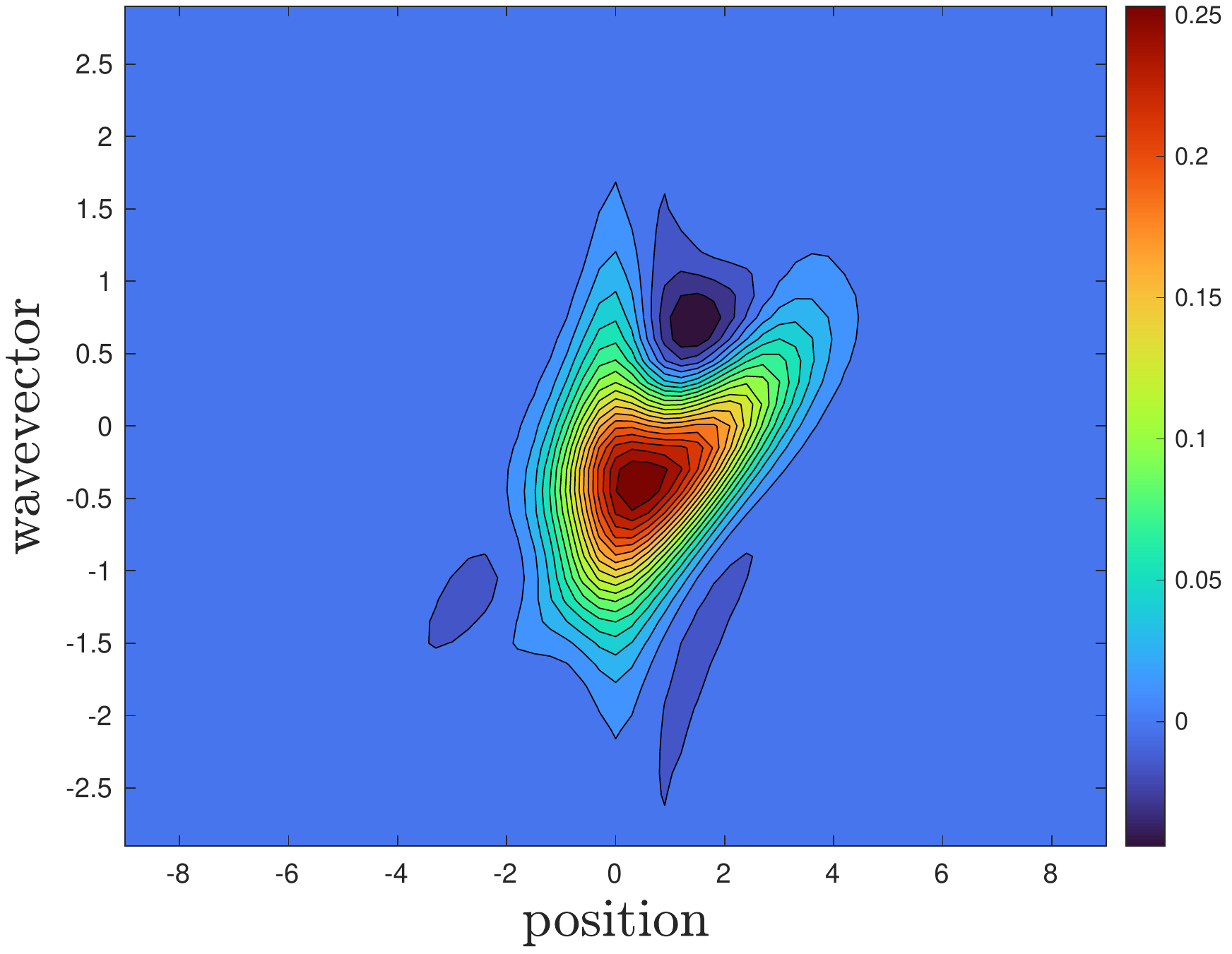}}
{\includegraphics[width=0.32\textwidth,height=0.18\textwidth]{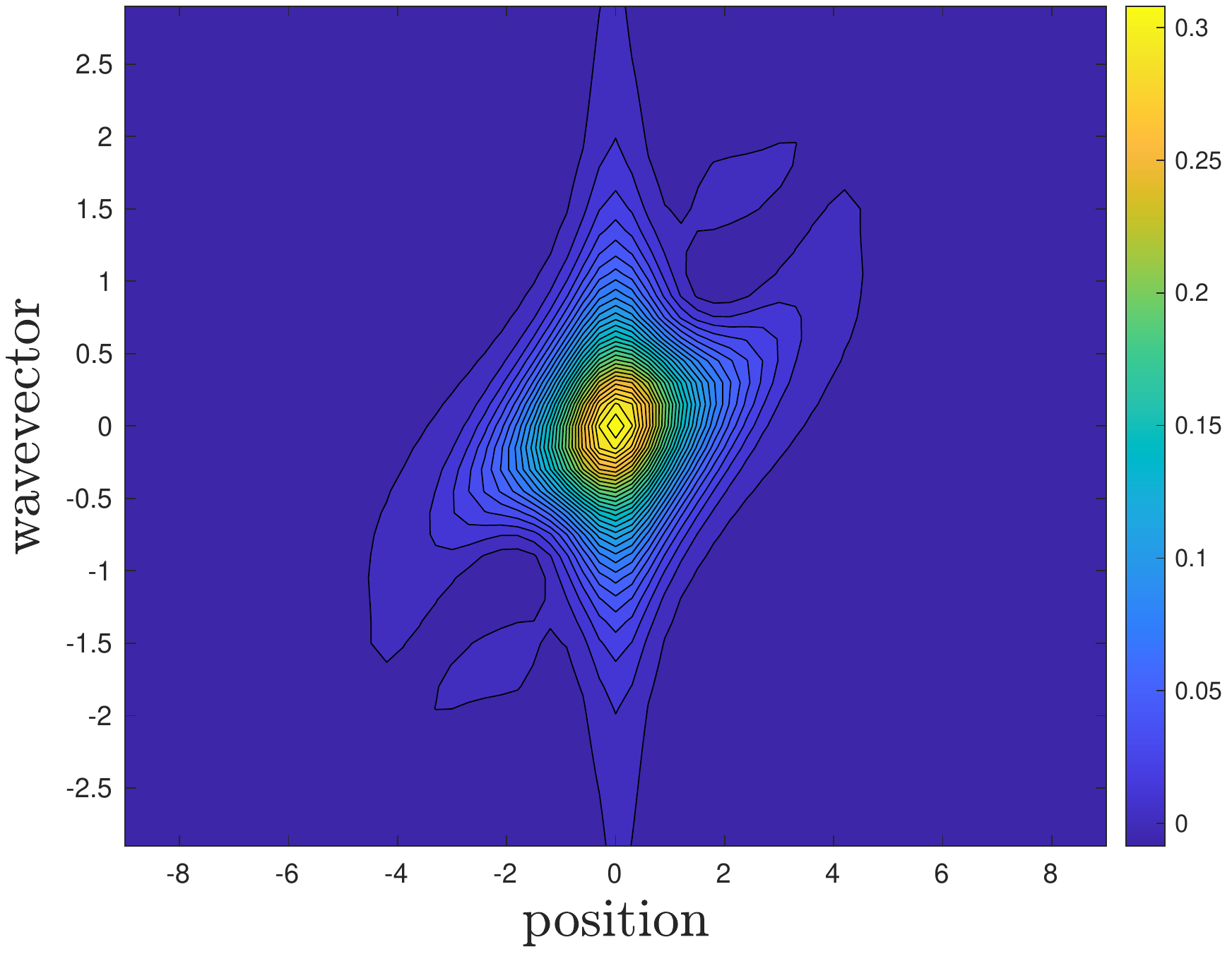}}}
\subfigure[$P(x_1, x_2, t)$ at $t=1$a.u.]{
{\includegraphics[width=0.32\textwidth,height=0.18\textwidth]{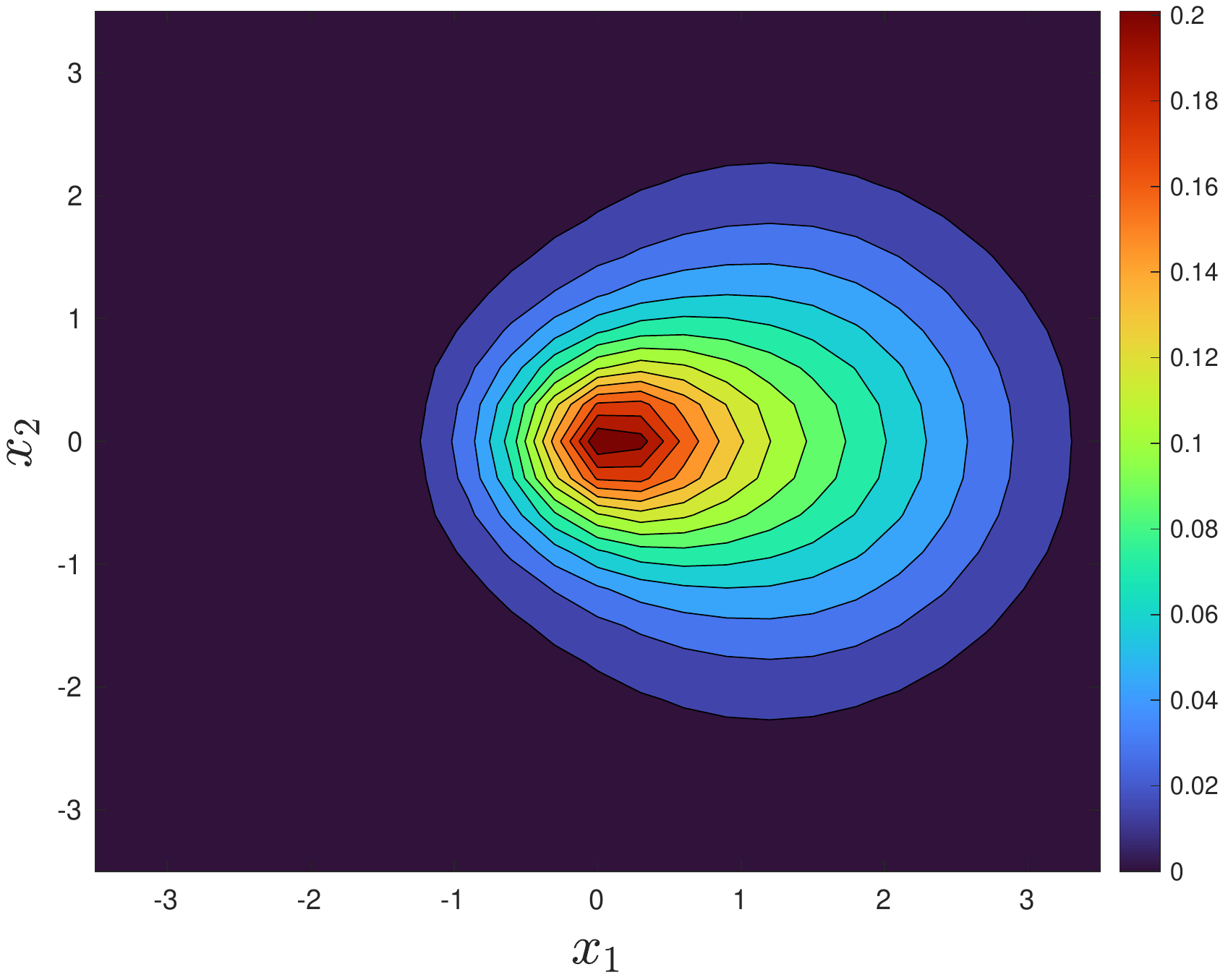}}}
\\
\centering
\subfigure[$W_1(x, k, t)$ (left) and $W_2(x, k, t)$ (right) at $t=4$a.u.]{
{\includegraphics[width=0.32\textwidth,height=0.18\textwidth]{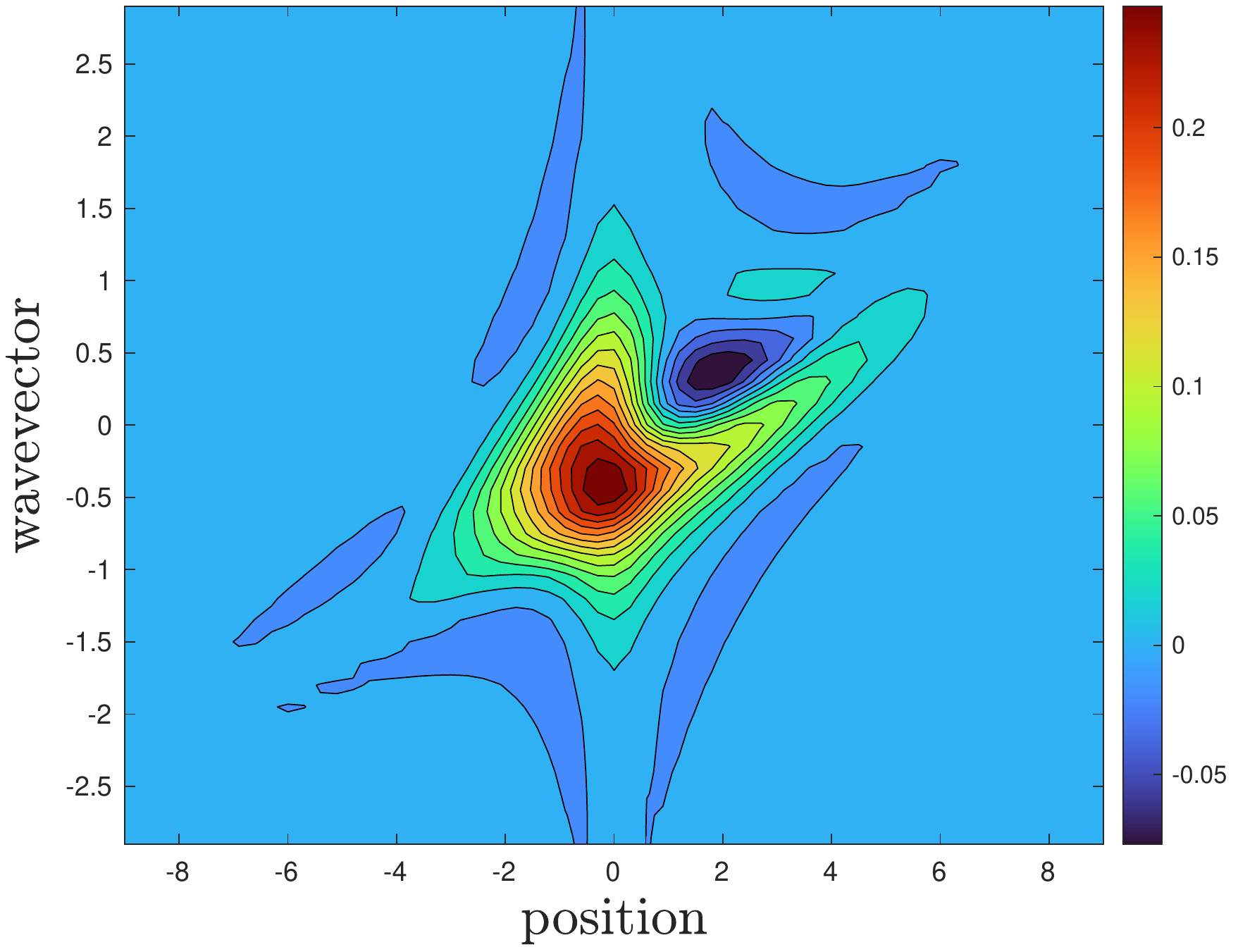}}
{\includegraphics[width=0.32\textwidth,height=0.18\textwidth]{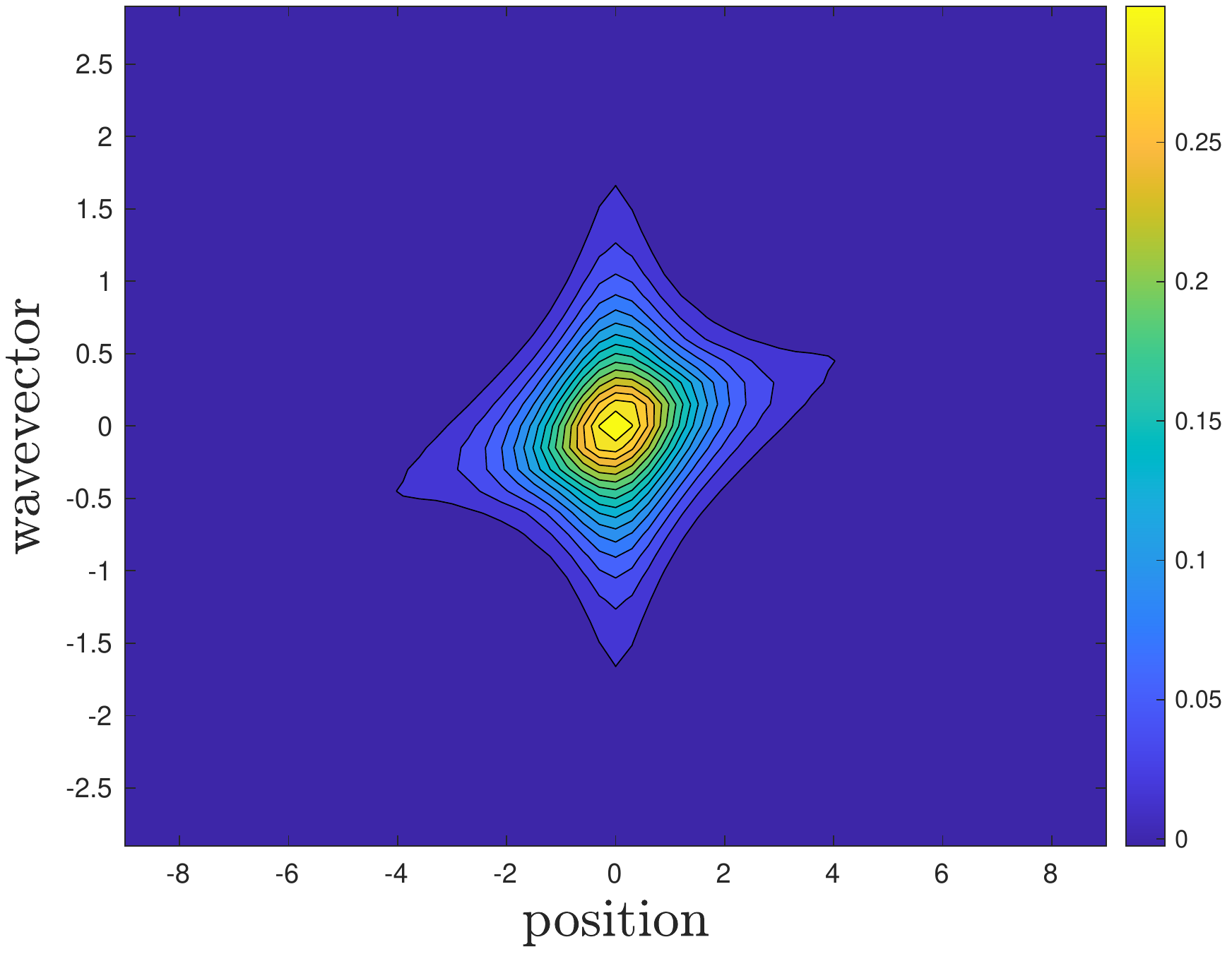}}}
\subfigure[$P(x_1, x_2, t)$ at $t=2$a.u.]{
{\includegraphics[width=0.32\textwidth,height=0.18\textwidth]{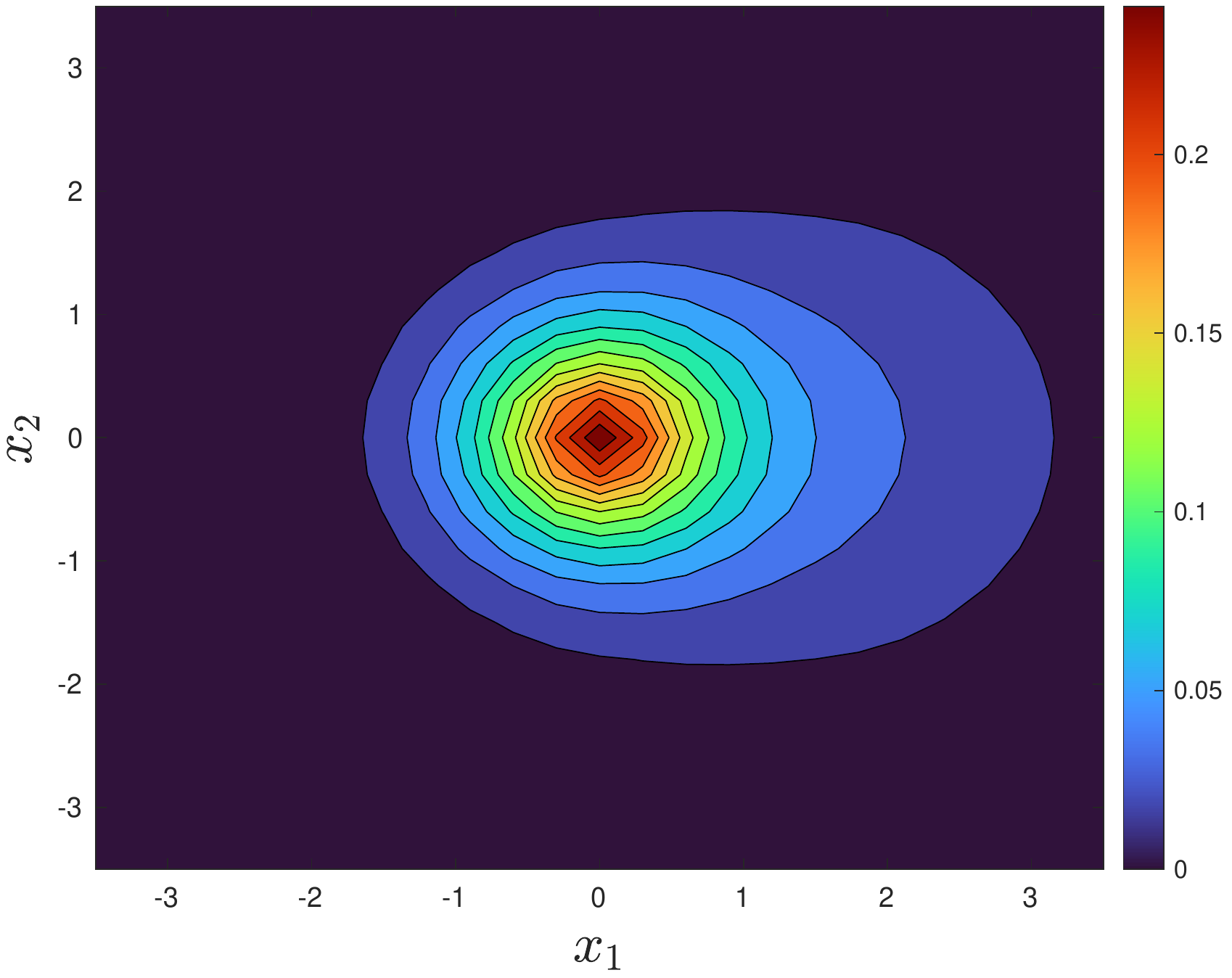}}}
\\
\centering
\subfigure[$W_1(x, k, t)$ (left) and $W_2(x, k, t)$ (right) at $t=8$a.u.\label{xdist_t050}]{
{\includegraphics[width=0.32\textwidth,height=0.18\textwidth]{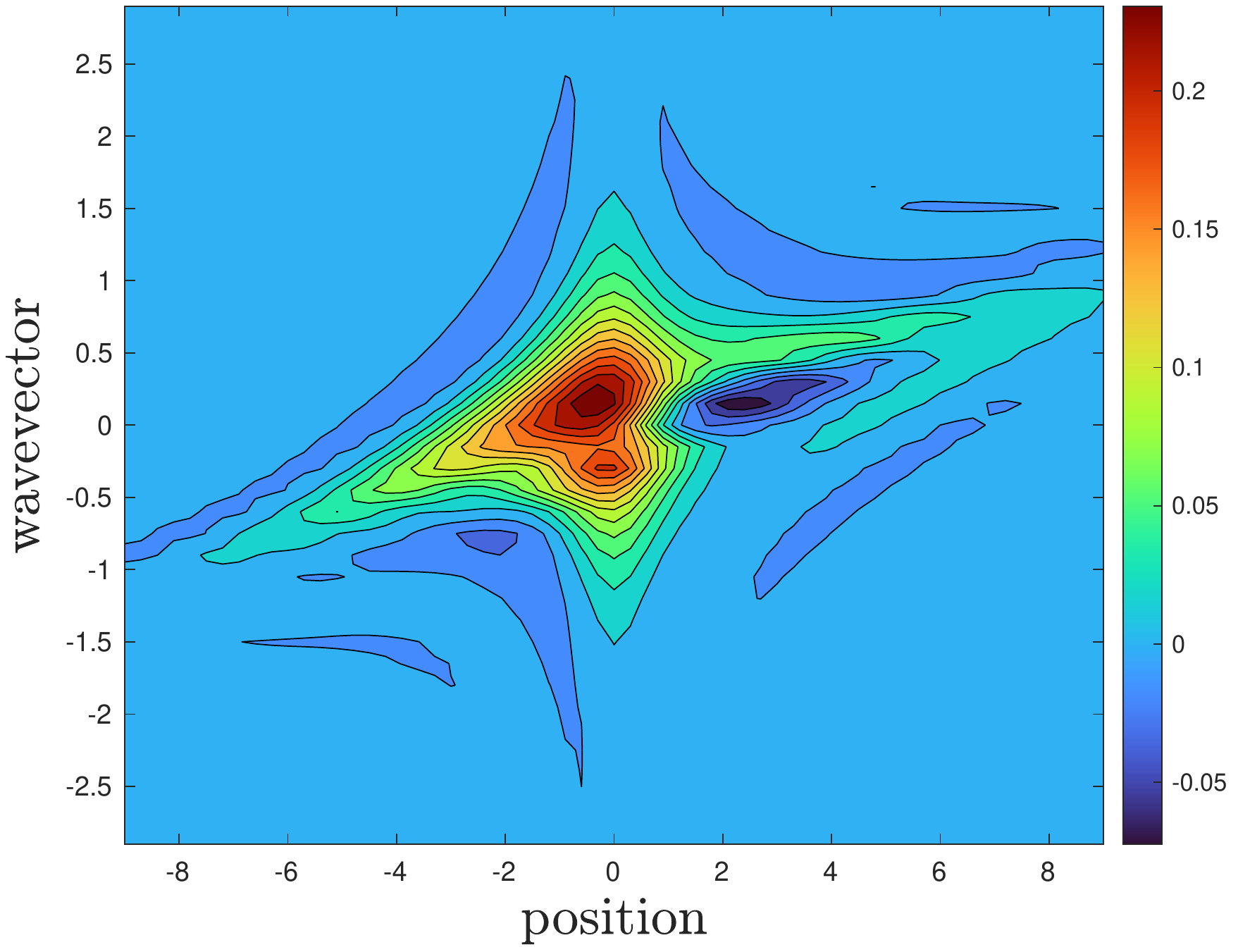}}
{\includegraphics[width=0.32\textwidth,height=0.18\textwidth]{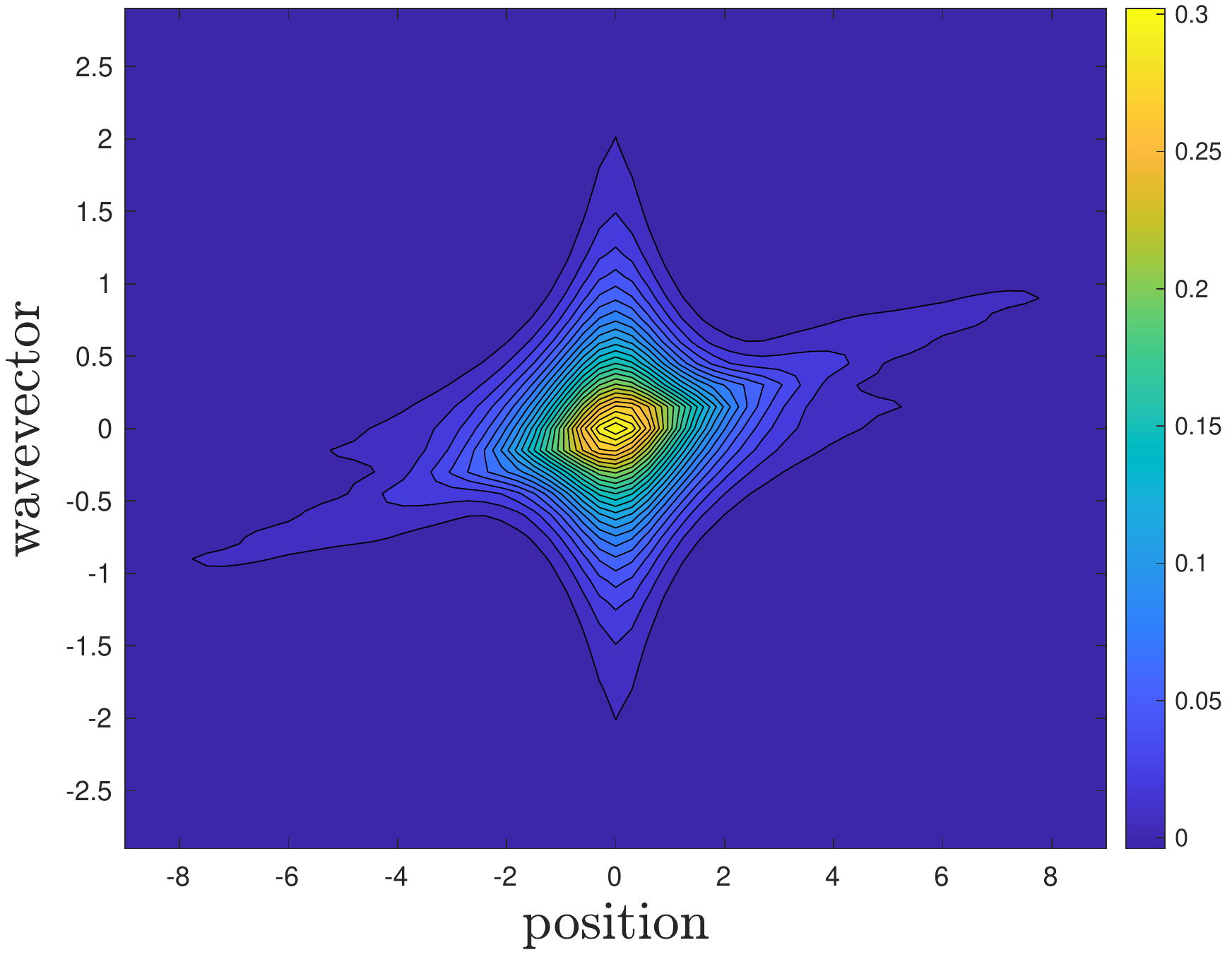}}}
\subfigure[$P(x_1, x_2, t)$ at $t=5$a.u.]{
{\includegraphics[width=0.32\textwidth,height=0.18\textwidth]{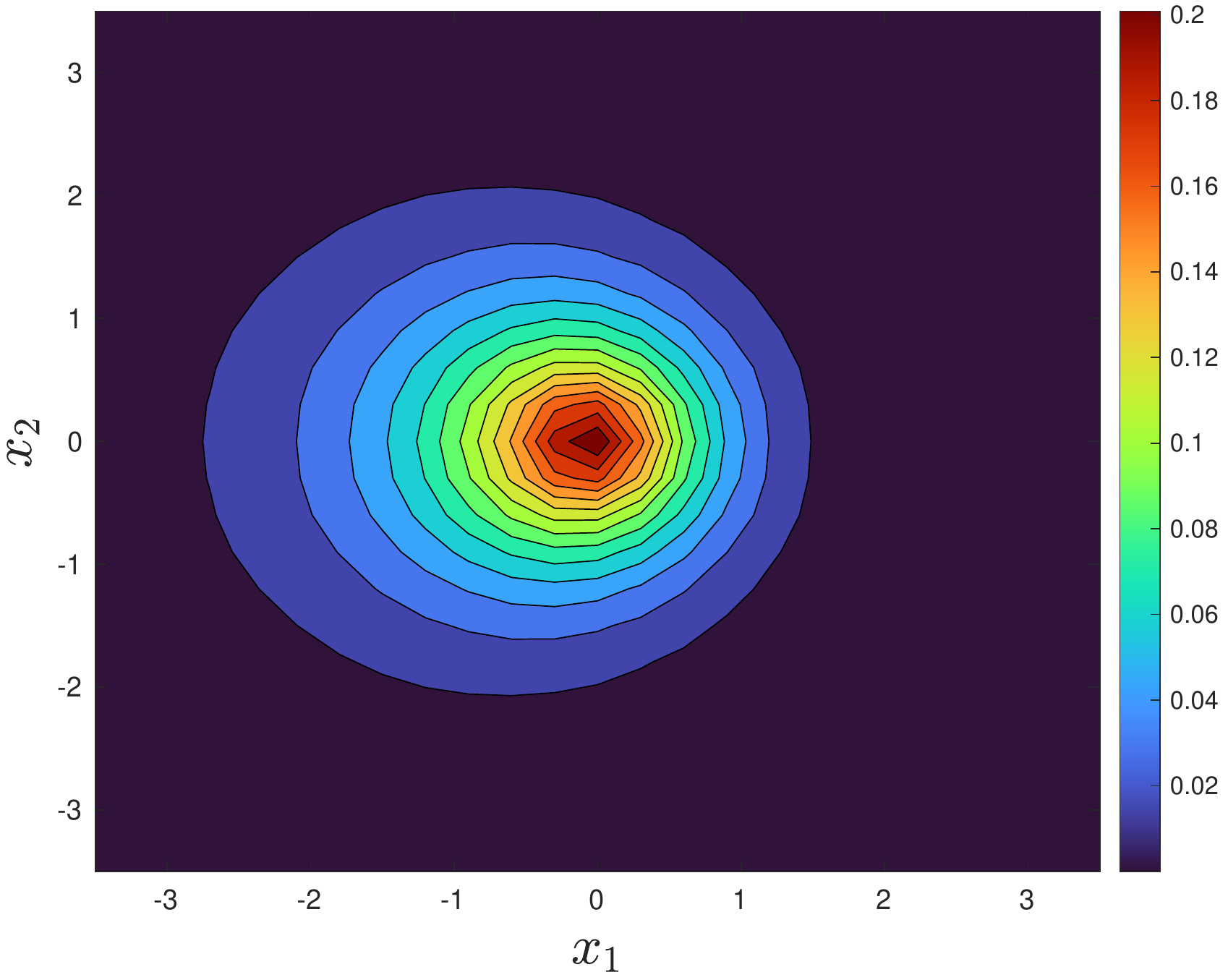}}}
\\
\centering
\subfigure[$W_1(x, k, t)$ (left) and $W_2(x, k, t)$ (right) at $t=12$a.u.]{
{\includegraphics[width=0.32\textwidth,height=0.18\textwidth]{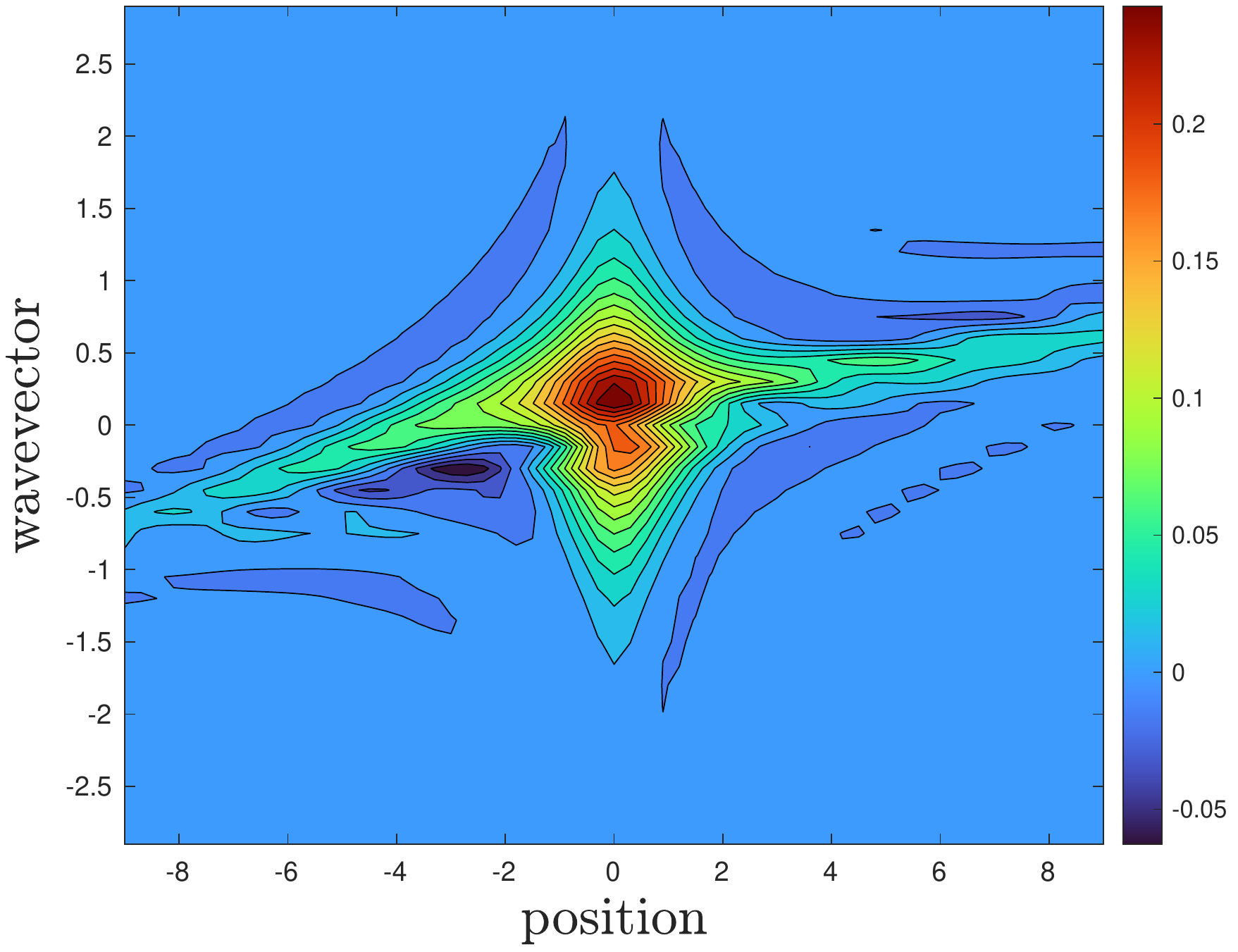}}
{\includegraphics[width=0.32\textwidth,height=0.18\textwidth]{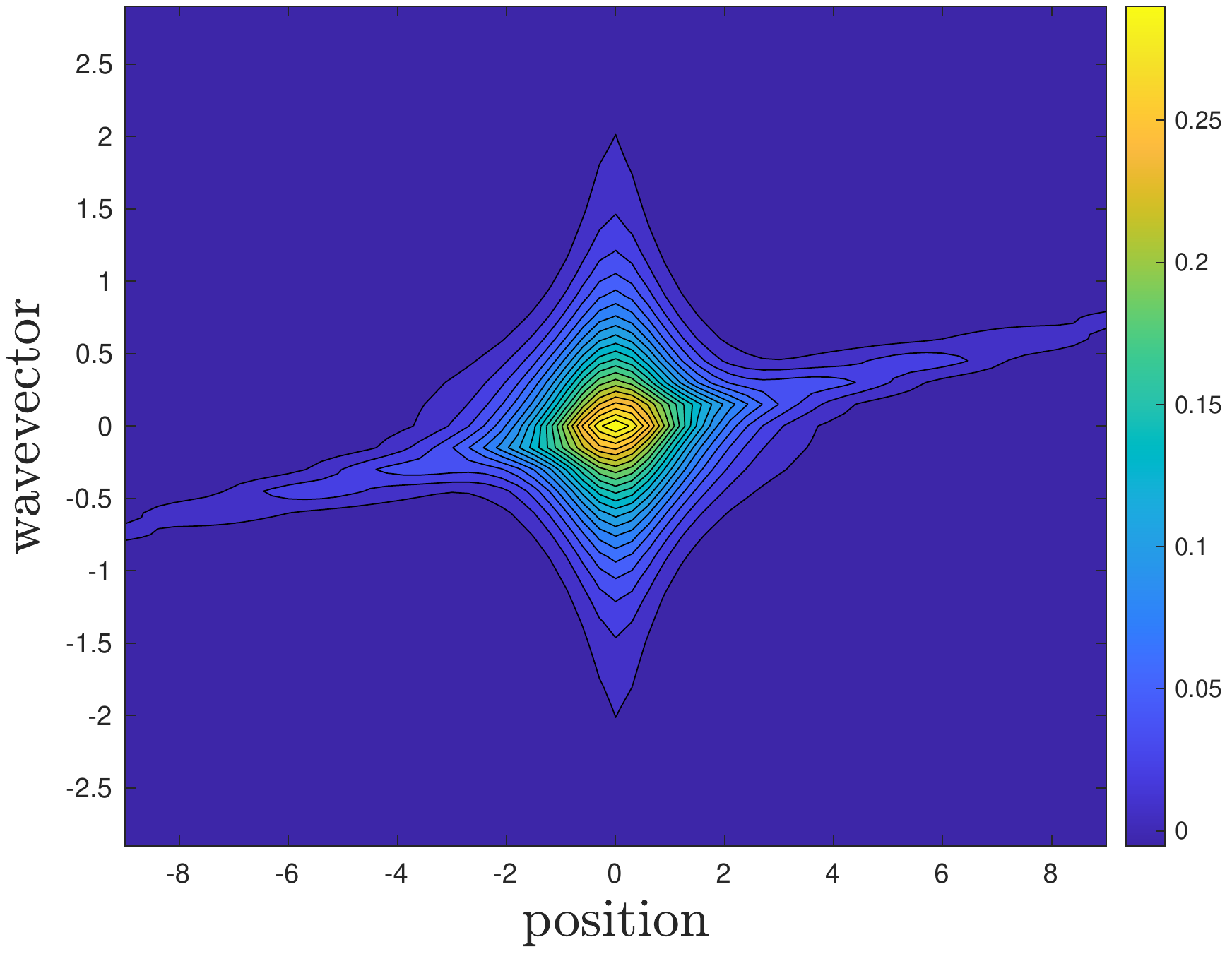}}}
\subfigure[$\langle x_1(t)\rangle$ and $\langle k_1(t)\rangle$.\label{averaged_pos_wvn}]{
{\includegraphics[width=0.32\textwidth,height=0.18\textwidth]{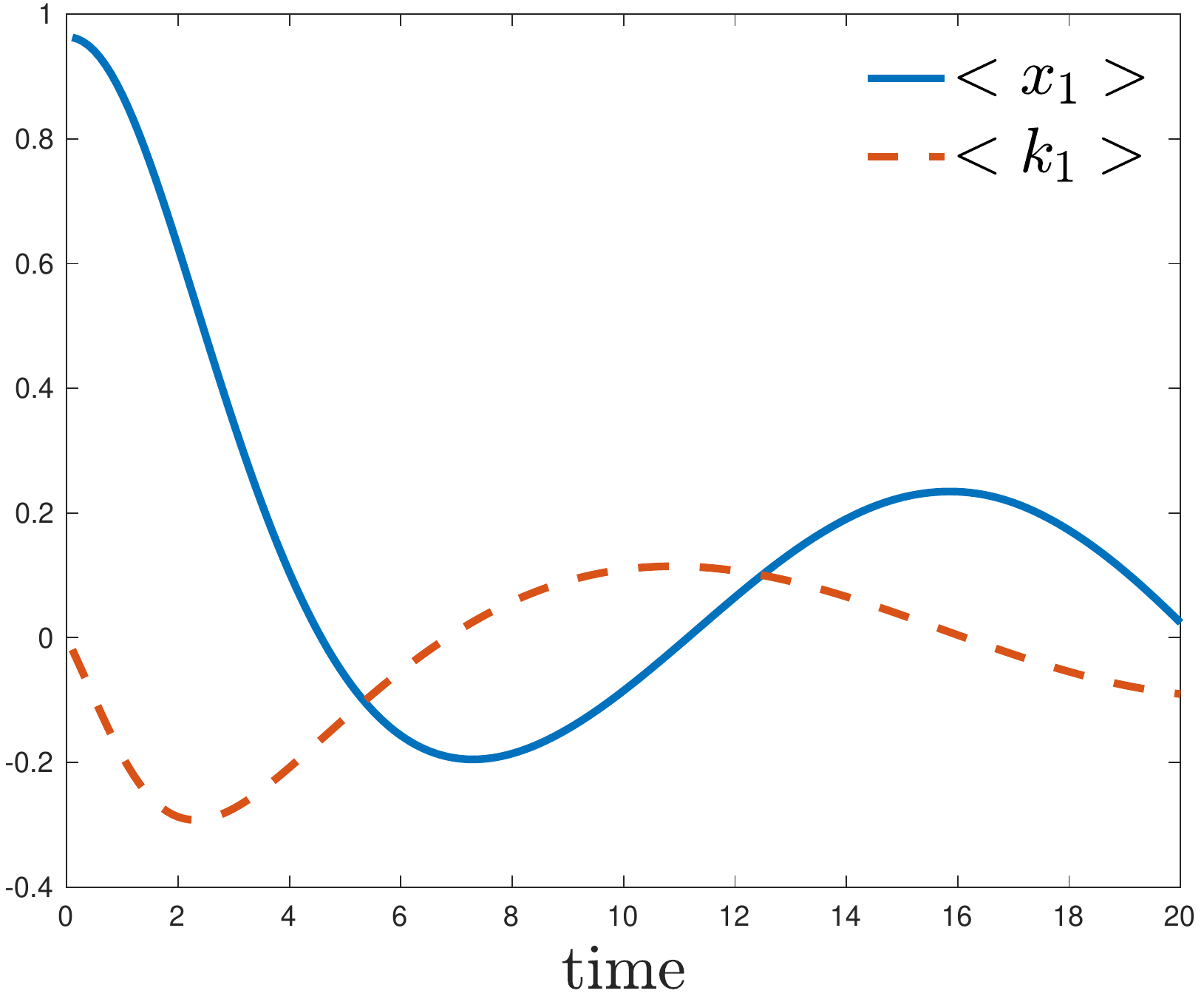}}}
\caption{\small  Electron-proton interaction:  Snapshots of the reduced Wigner functions on $(x_1$-$k_1)$ plane (left) and on $(x_2$-$k_2)$ plane (middle), the spatial marginal distribution (right) and the averaged position and momentum. \label{qcs_time_evolution}}
\end{figure}

With above preparations, we can simulate several typical quantum systems and try to reveal the proton-electron coupling and the uncertainty principle under the Wigner function representation. The following example is motivated from the strong-field ionization process studied in \cite{TianWangEberly2017,HanGeFangYuGuoMaDengGongLiu2019}. The computational domain $[-9, 9]^3 \times [-4.8, 4.8]^3$ under a $81^3 \times 64^3$ uniform grid is decomposed into $4^3$ patches with $n_{nb} = 15$. The time step is $\tau = 0.025$.

\begin{example}\label{example_one_proton}
\textup{
Consider a electron interacting with a proton fixed at $(0, 0, 0)$. The initial condition is $f_0(\bx, \bk) = \pi^{-3} \me^{-\frac{1}{2} ((x_1-1)^2 + x_2^2 + x_3^2) - 2(k_1^2 + k_2^2 + k_3^2)}$, where the Gaussian wavepacket describes the coherent state.
}
\end{example}

{\bf Spatial unharmonic oscillation:}  As presented in the third column of Figure~\ref{qcs_time_evolution},  the electron wavepacket is soon attracted by the proton and then oscillates near the origin, and it presents an evident unharmonic oscillation pattern  in the spatial space under the Coulomb interaction.  We record the average position $\langle x_1(t) \rangle$ and momentum $\langle k_1(t) \rangle$ in Figure \ref{averaged_pos_wvn} and indeed observe that the amplitude of oscillations damp away in time, which is distinct from the harmonic  trajectories.

{\bf Uncertainty principle:}  The time evolutions of $W_1(x, k, t)$ and $W_2(x, k, t)$ are plotted in the first two columns of Figure \ref{qcs_time_evolution}. Since the electron initially deviates from the origin in $x_1$-direction, $W_1(x, k, t)$ exhibits a highly asymmetric pattern and becomes more and more oscillating. The uncertainty principle is visualized by the negative parts of the Wigner function, which seem to be concentrated on the region opposite to the moving direction. By contrast, $W_2(x, k, t)$ is always symmetric, and only small negative components are observed.

\subsection{$H^+_2$ system: Electron dynamics interacting with two protons} A more challenging problem is to put an electron in the delocalized potential produced by two protons, motivated from the Hydrogen tunneling phenomenon \cite{PakHammesSchiffer2004}. The computational domain is $[-9, 9]^3 \times [-4.8, 4.8]^3$ with a $61^3 \times 64^3$ uniform grid mesh, which is decomposed into $4\times 4\times 4$ patches with $n_{nb} = 15$.

\begin{example}
\textup{
Suppose there are two protons with fixed position $\bx_A^- = (-R, 0, 0)$ and $\bx_A^+ = (R, 0, 0)$, $R = 0.614161$a.u. (0.325 Angstrom), so that the potential is $V(\bx) = -\frac{1}{|\bx - \bx_A^-|} - \frac{1}{|\bx - \bx_A^+|}$.
The initial Gaussian wavepacket is set as $f_0(\bx, \bk) = \pi^{-3} \me^{-\frac{1}{2} \left(x_1^2 + x_2^2 + x_3^2\right) - 2 \left(k_1^2 + k_2^2 + k_3^2\right)}$.
}
\end{example}

\begin{figure}[!h]
\centering
\subfigure[$W_1(x, k, t)$ (left), $W_2(x, k, t)$ (middle) and $P(x_1, x_2, t)$ (right) at $t=1$a.u.]{
{\includegraphics[width=0.32\textwidth,height=0.18\textwidth]{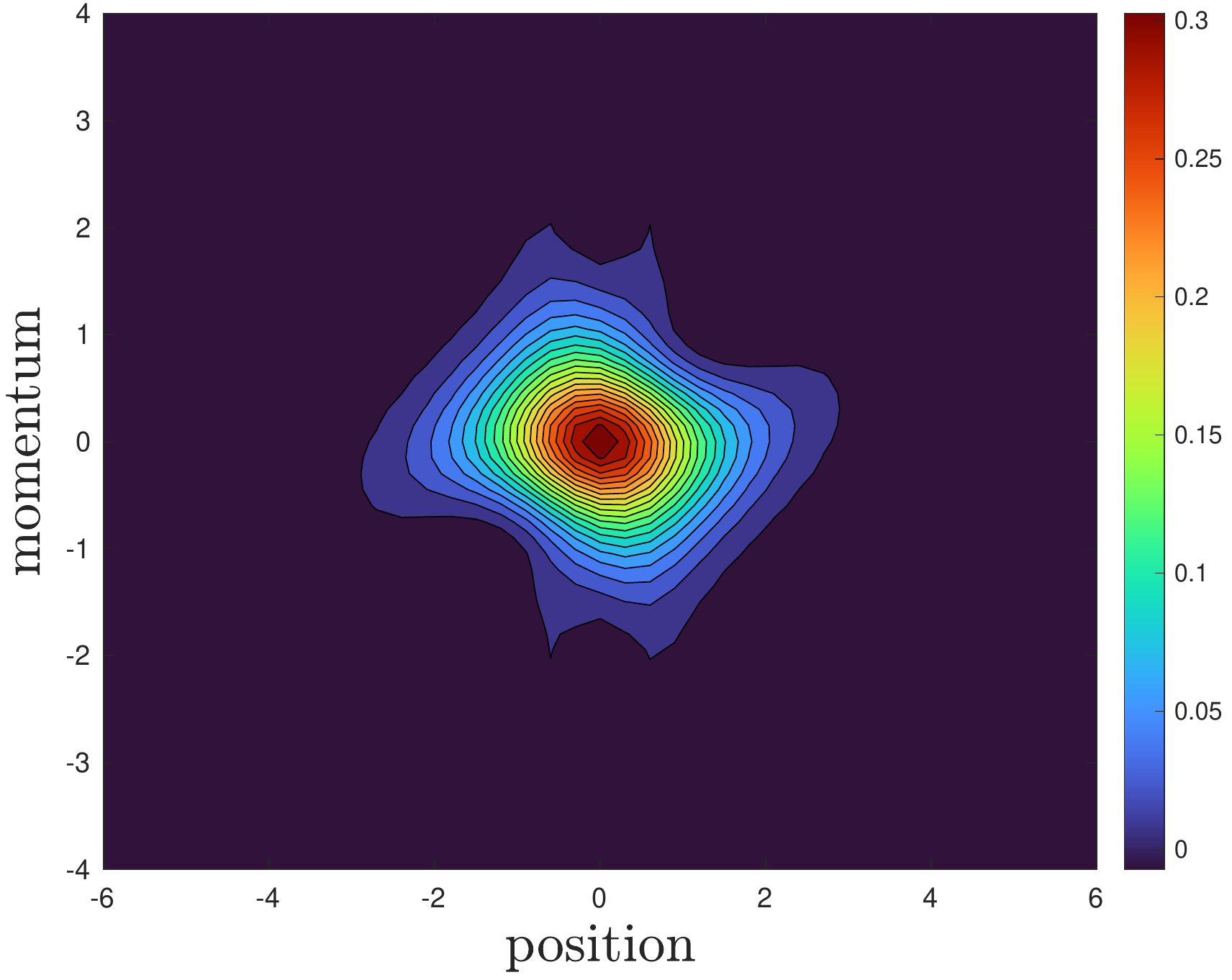}}
{\includegraphics[width=0.32\textwidth,height=0.18\textwidth]{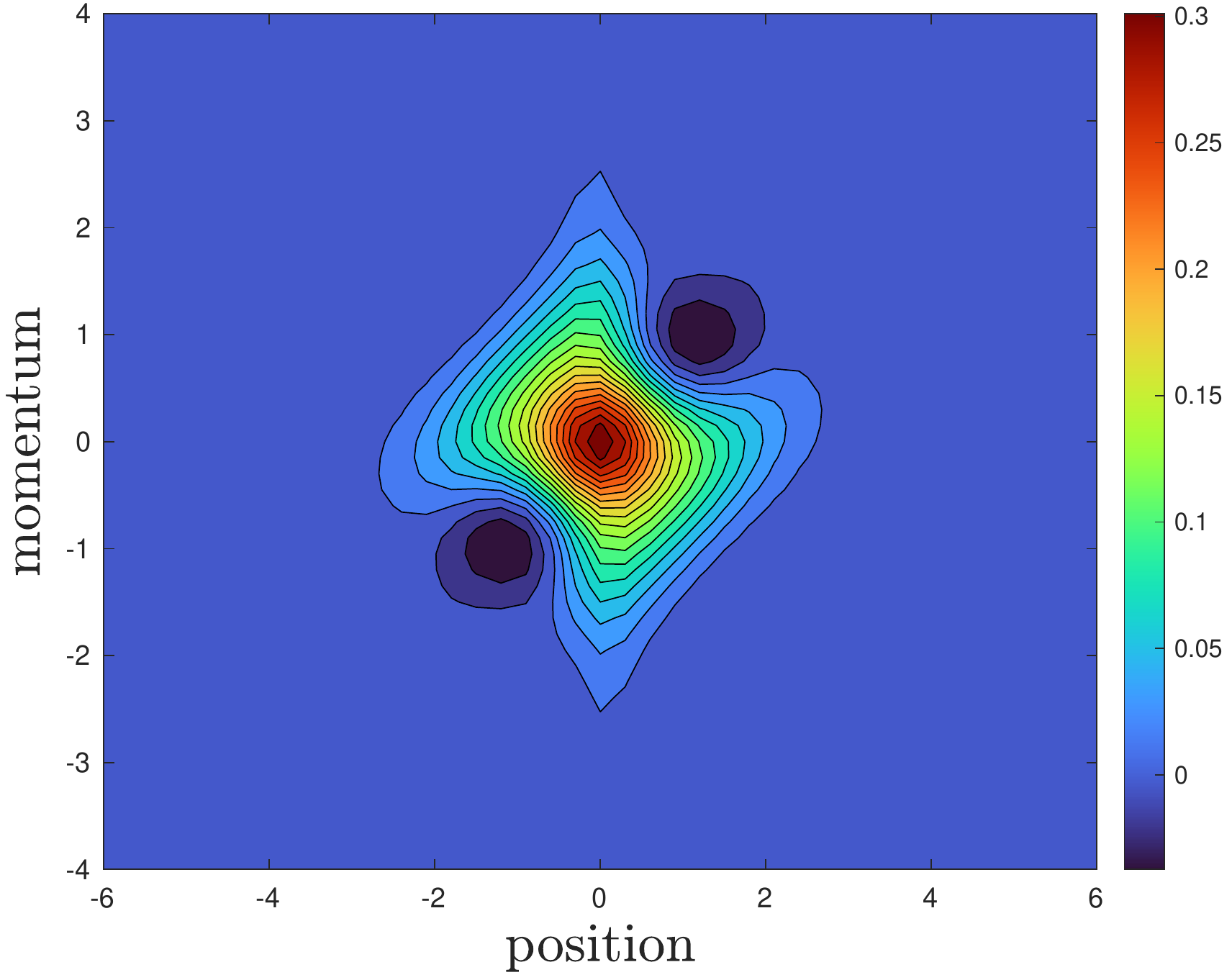}}
{\includegraphics[width=0.32\textwidth,height=0.18\textwidth]{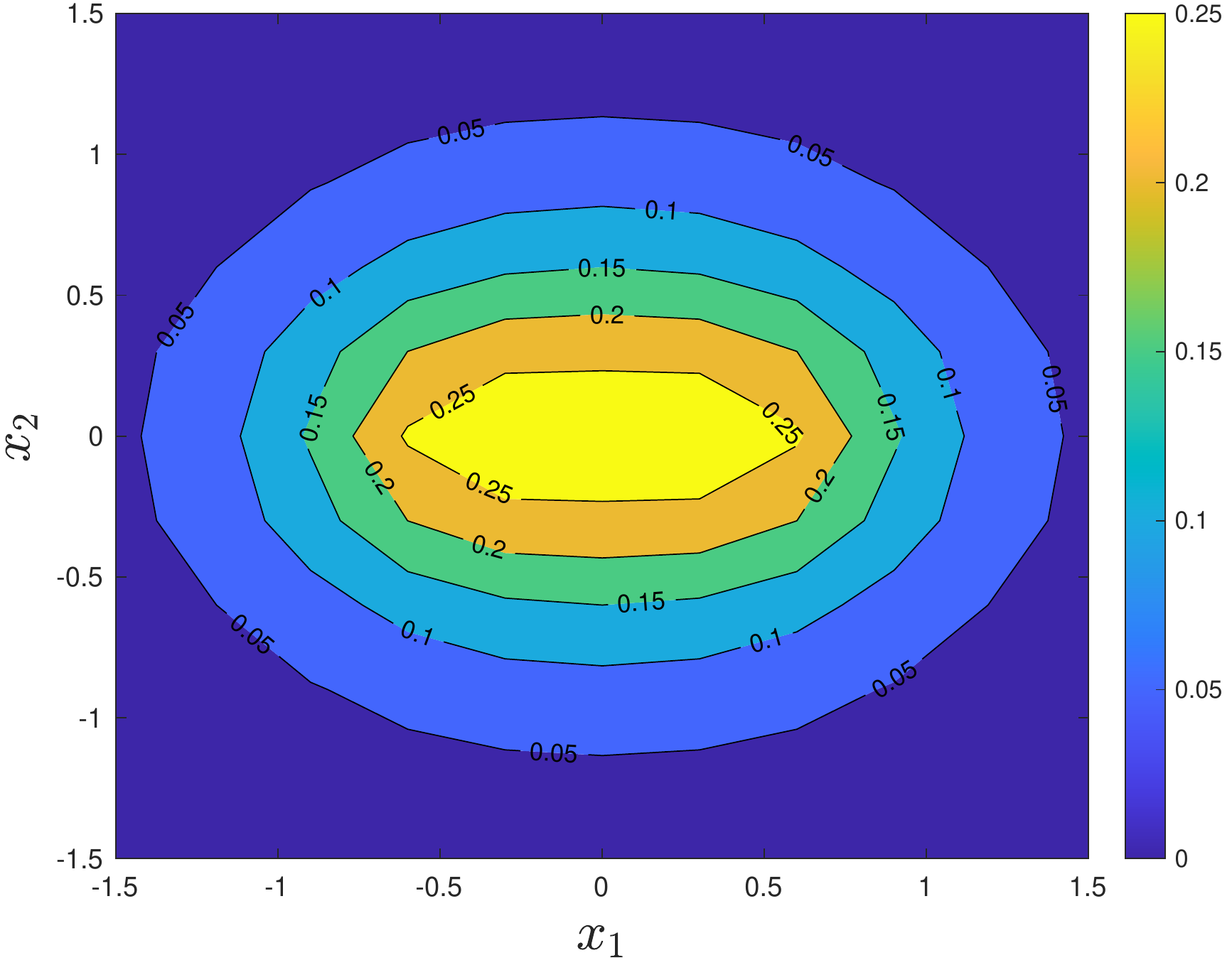}}}
\centering
\subfigure[$W_1(x, k, t)$ (left), $W_2(x, k, t)$ (middle) and $P(x_1, x_2, t)$ (right) at $t=2$a.u.]{
{\includegraphics[width=0.32\textwidth,height=0.18\textwidth]{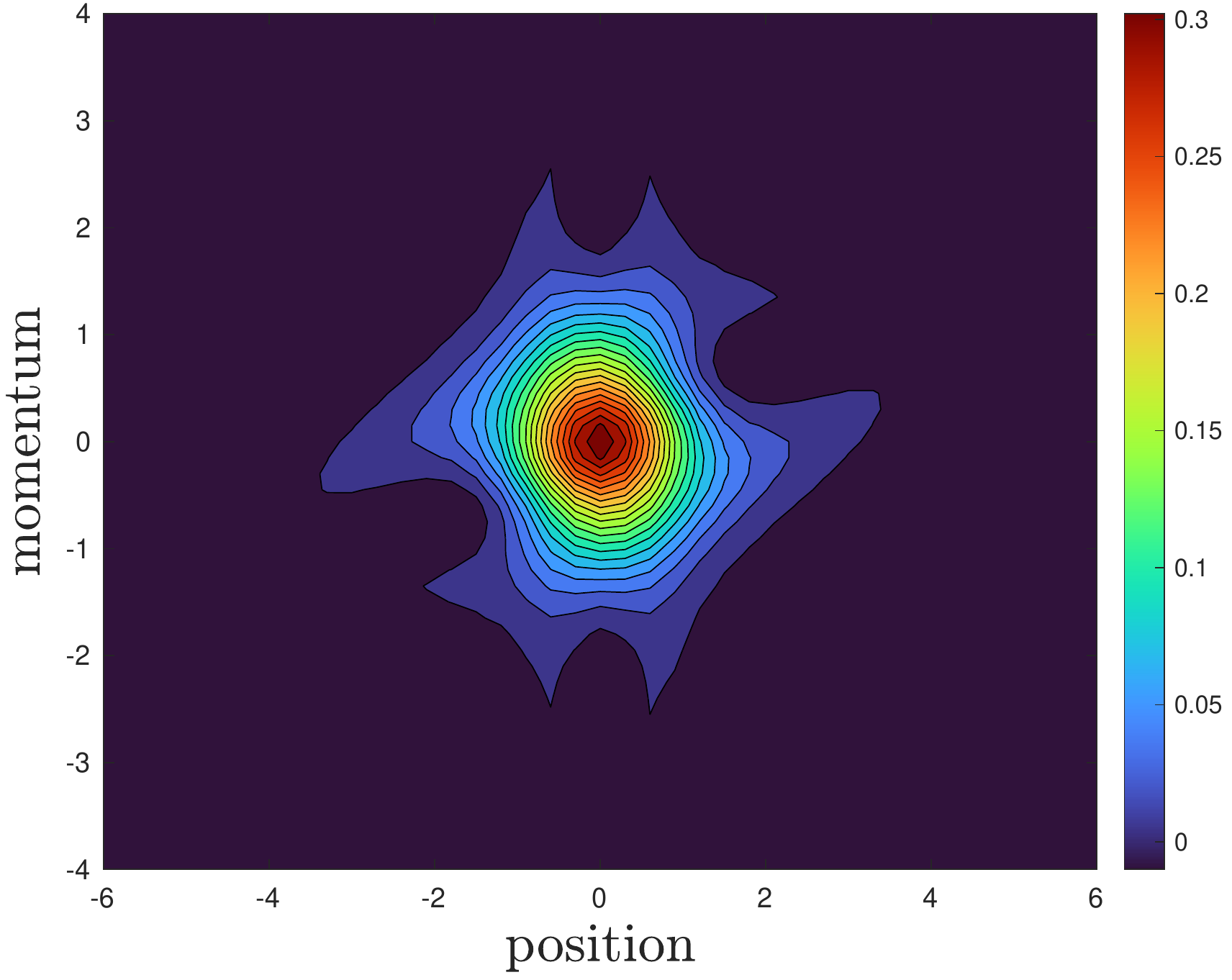}}
{\includegraphics[width=0.32\textwidth,height=0.18\textwidth]{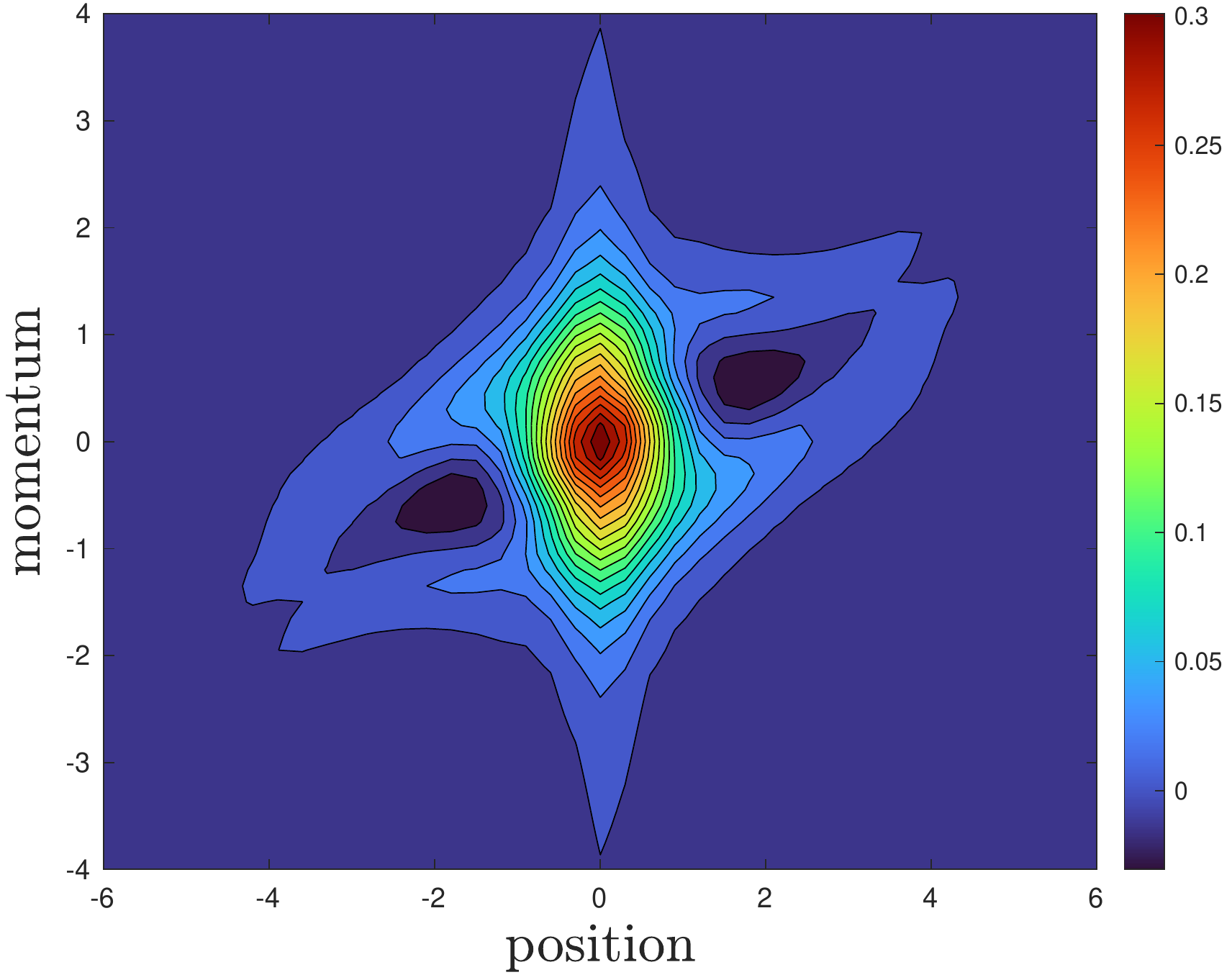}}
{\includegraphics[width=0.32\textwidth,height=0.18\textwidth]{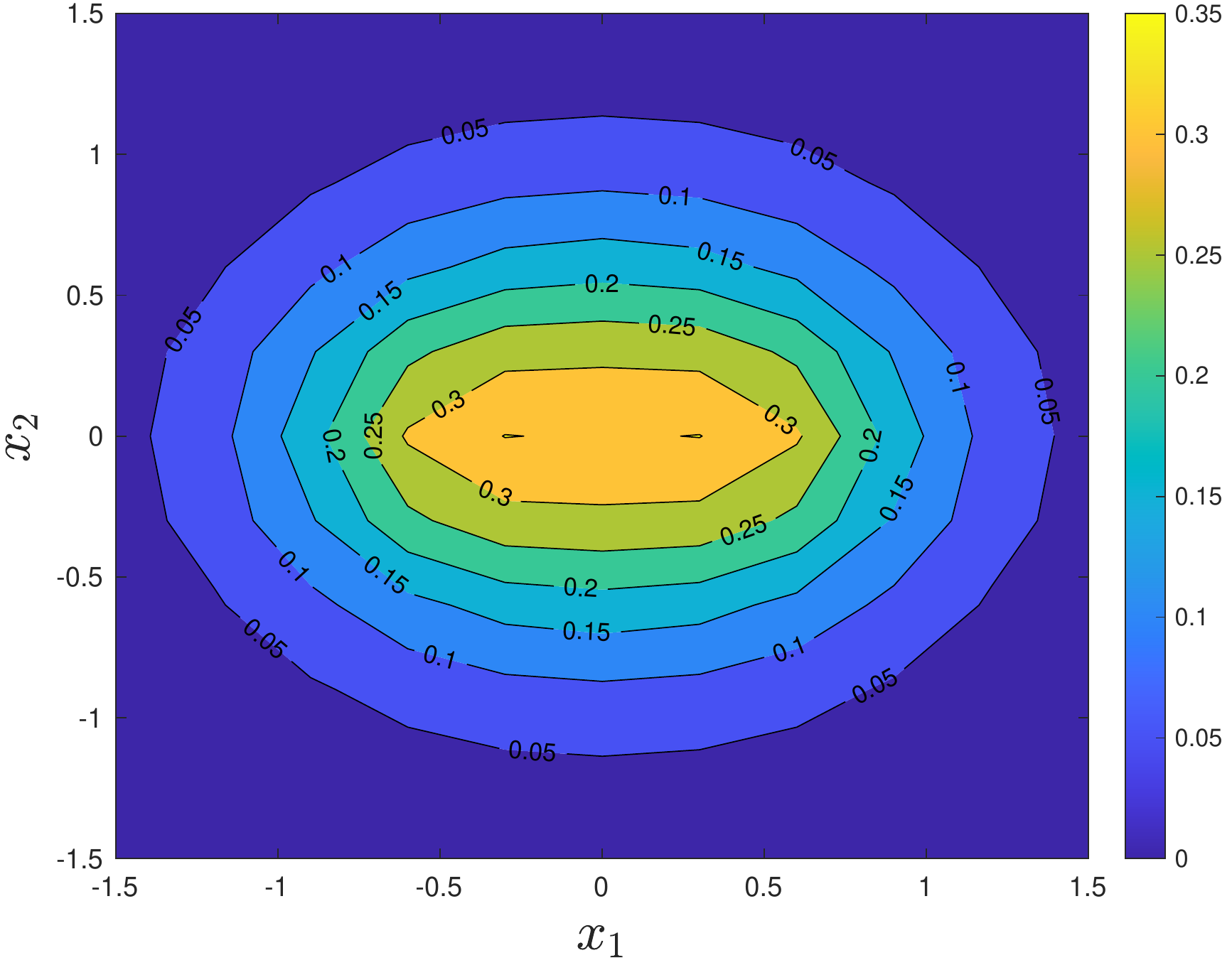}}}
\\
\centering
\subfigure[$W_1(x, k, t)$ (left), $W_2(x, k, t)$ (middle) and $P(x_1, x_2, t)$ (right) at $t=4$a.u.]{
{\includegraphics[width=0.32\textwidth,height=0.18\textwidth]{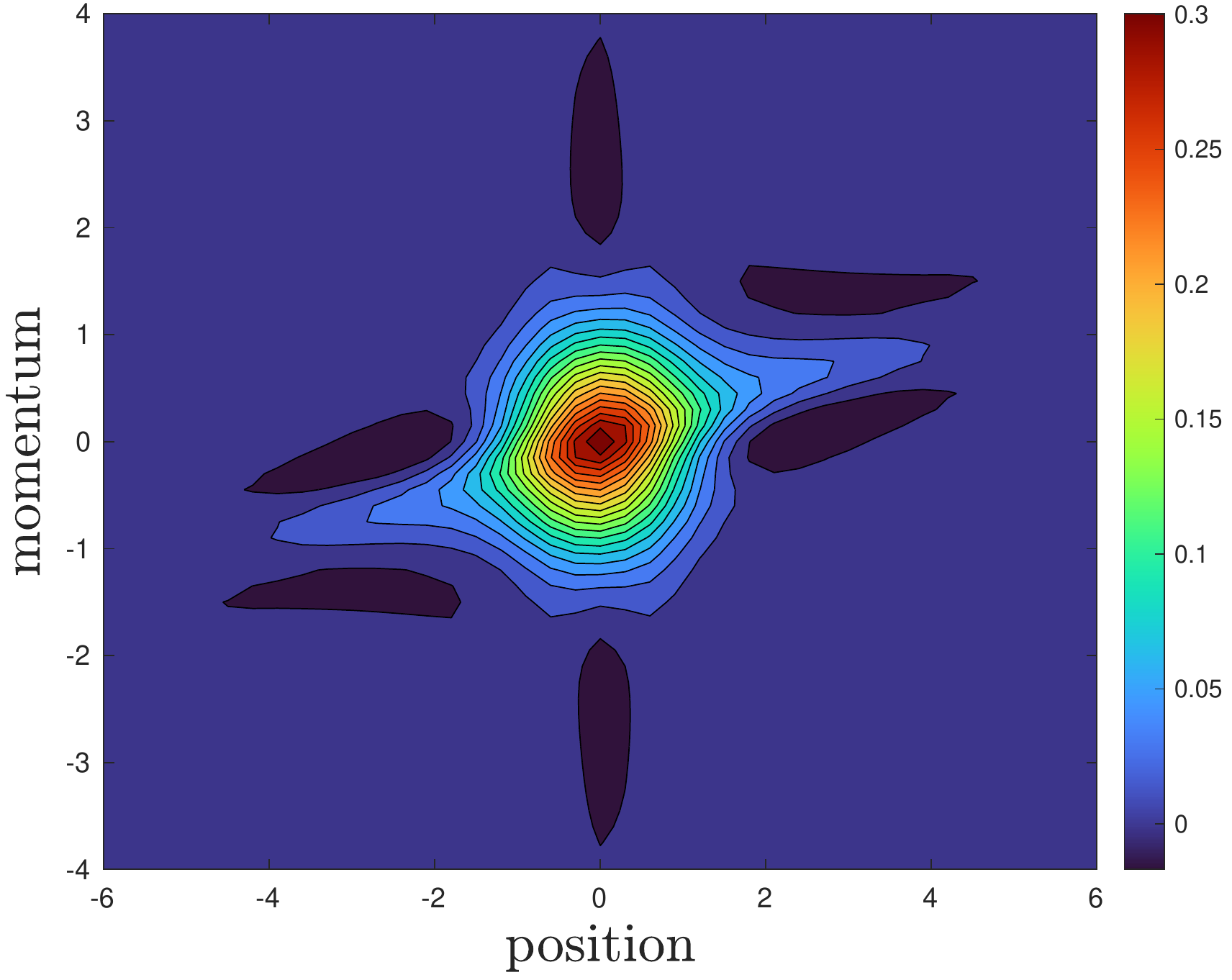}}
{\includegraphics[width=0.32\textwidth,height=0.18\textwidth]{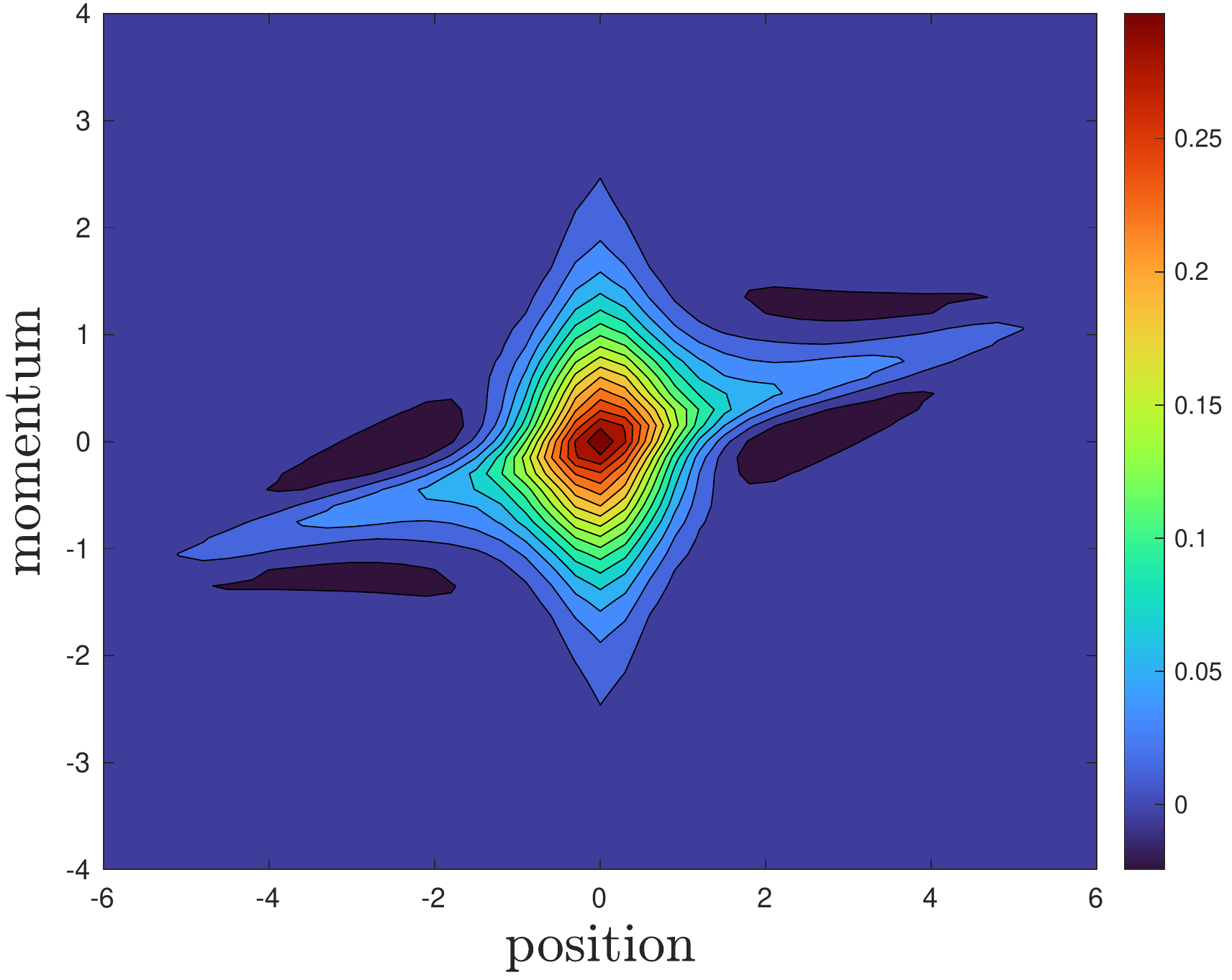}}
{\includegraphics[width=0.32\textwidth,height=0.18\textwidth]{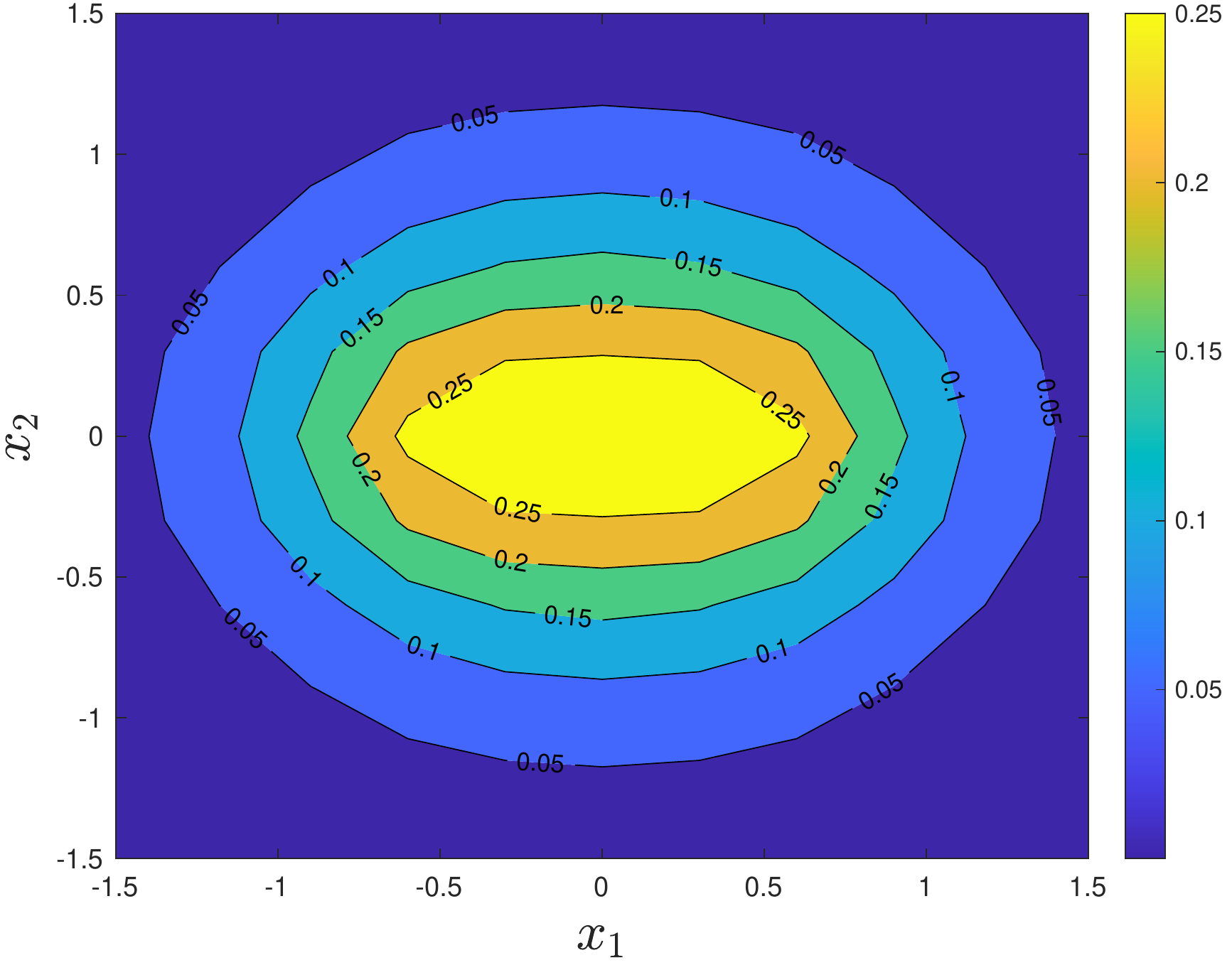}}}
\\
\centering
\subfigure[$W_1(x, k, t)$ (left), $W_2(x, k, t)$ (middle) and $P(x_1, x_2, t)$ (right) at $t=8$a.u.]{
{\includegraphics[width=0.32\textwidth,height=0.18\textwidth]{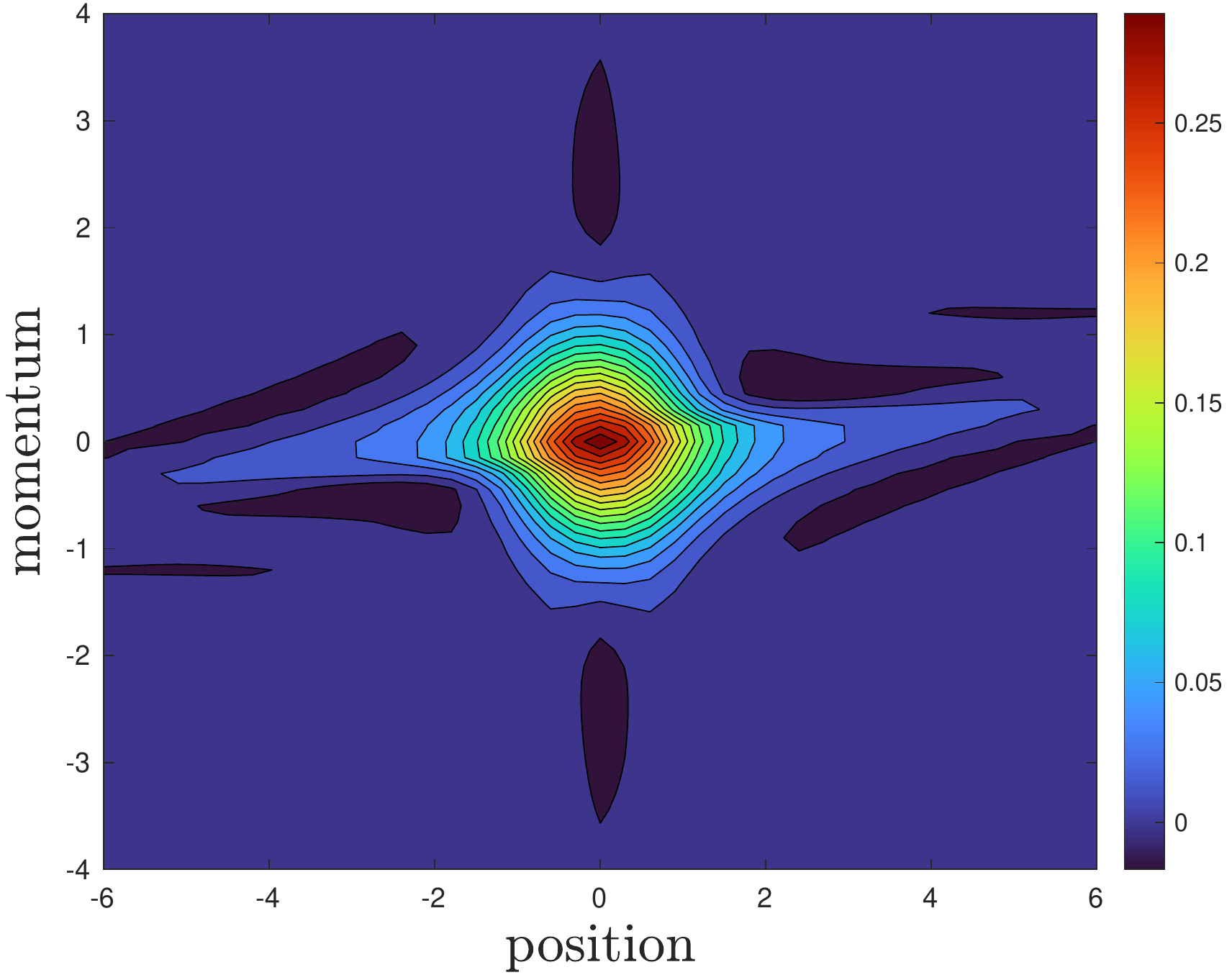}}
{\includegraphics[width=0.32\textwidth,height=0.18\textwidth]{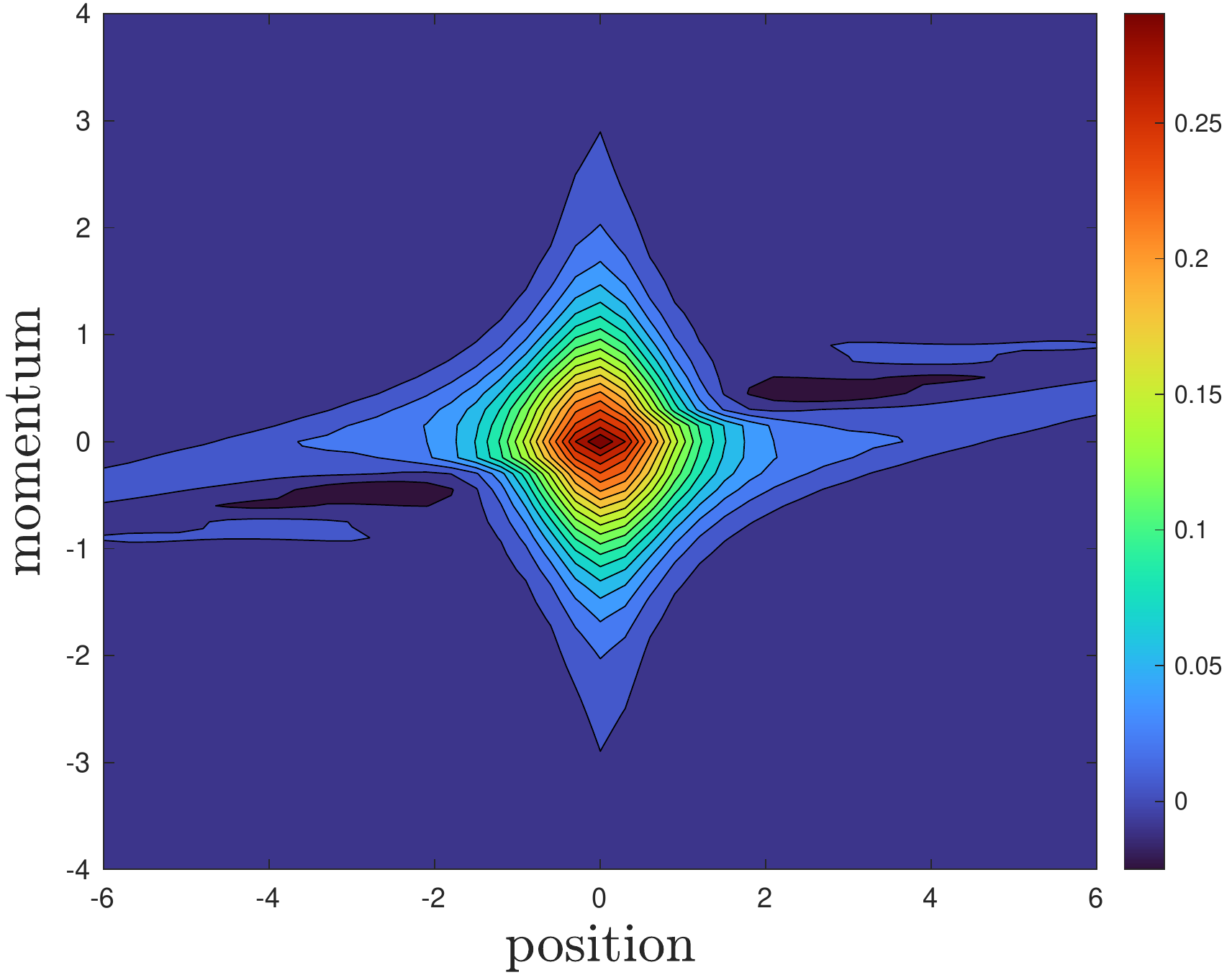}}
{\includegraphics[width=0.32\textwidth,height=0.18\textwidth]{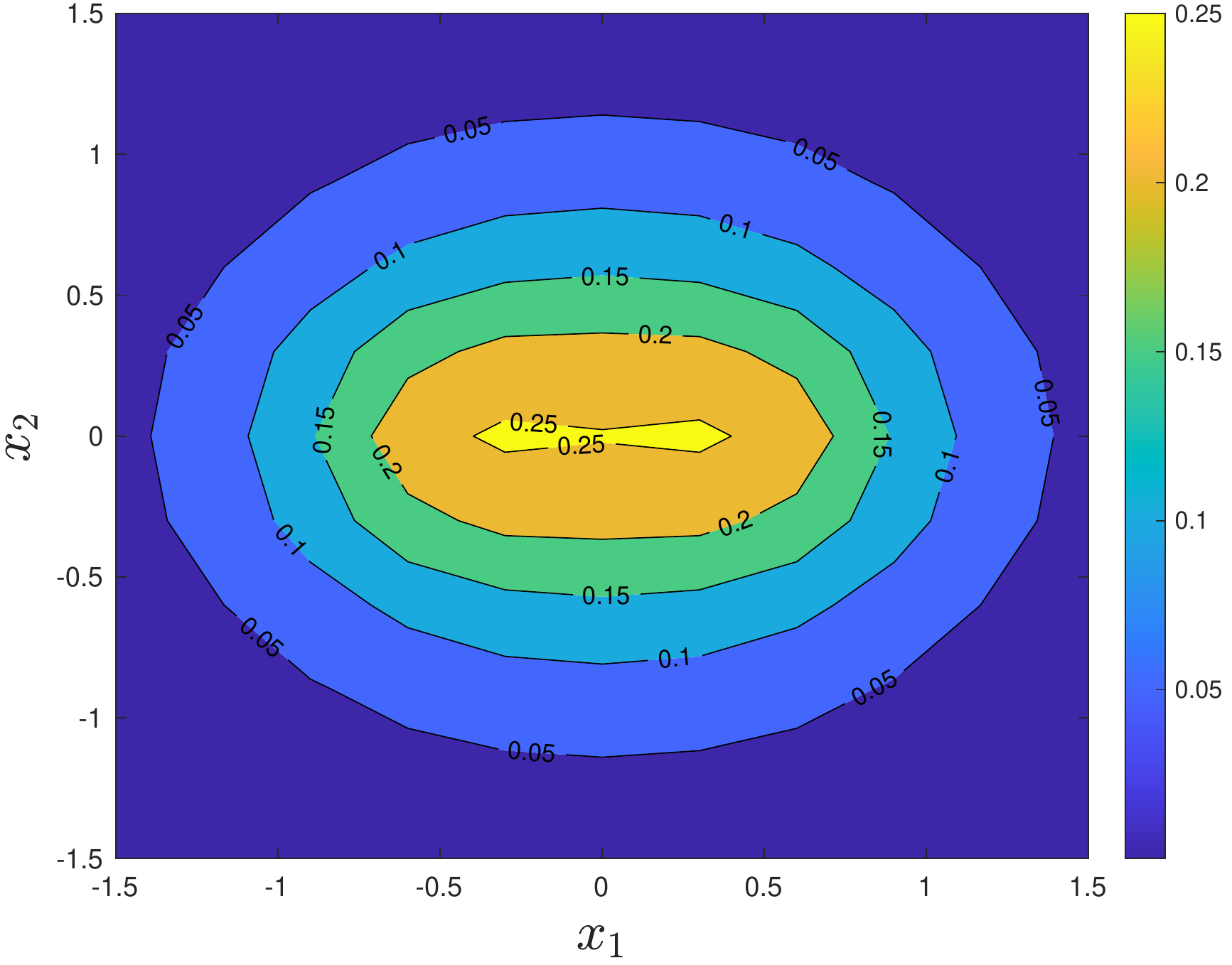}}}
\\
\centering
\subfigure[$W_1(x, k)$ (left) and $W_2(x, k)$ (right) at $t=12$a.u.]{
{\includegraphics[width=0.32\textwidth,height=0.18\textwidth]{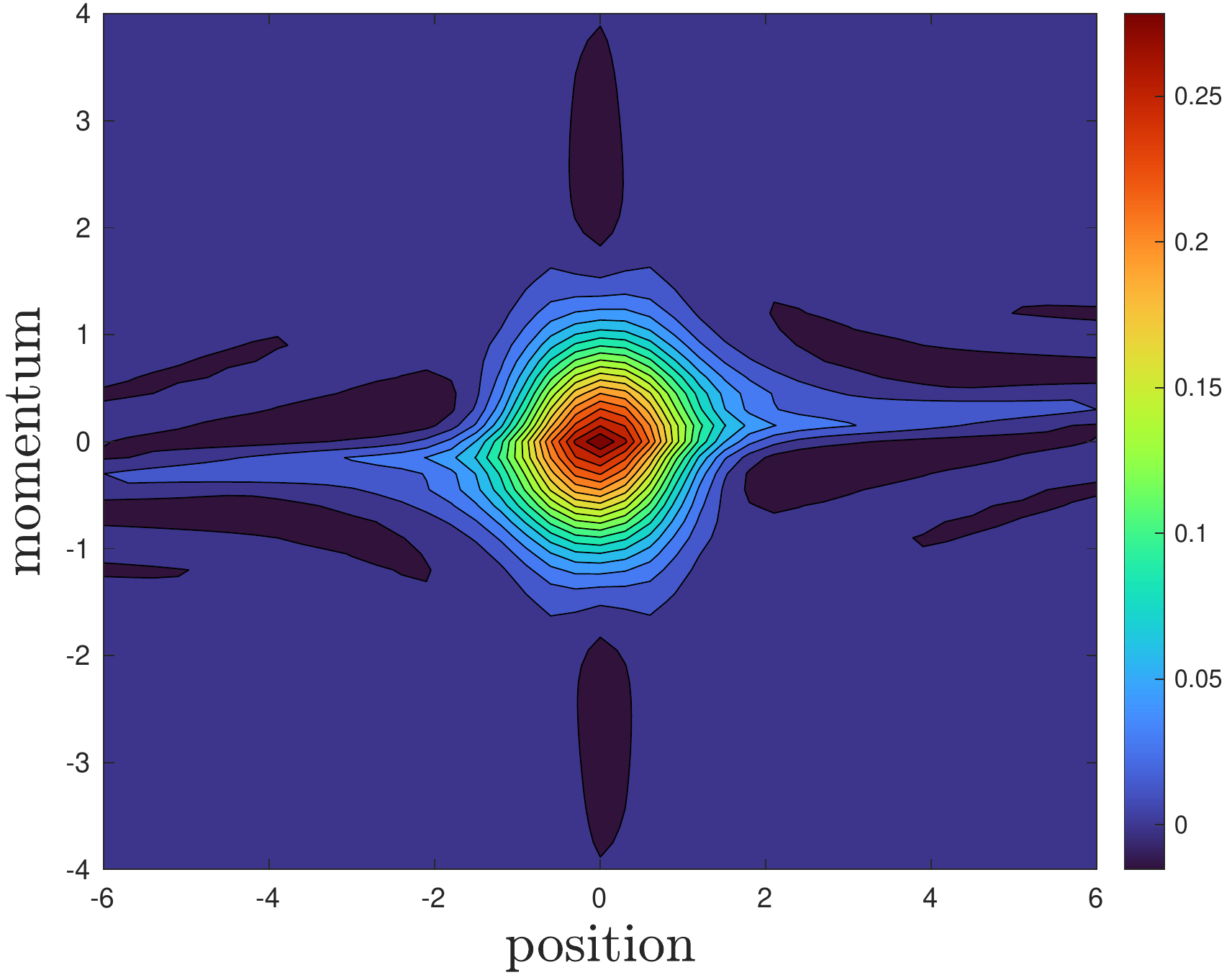}}
{\includegraphics[width=0.32\textwidth,height=0.18\textwidth]{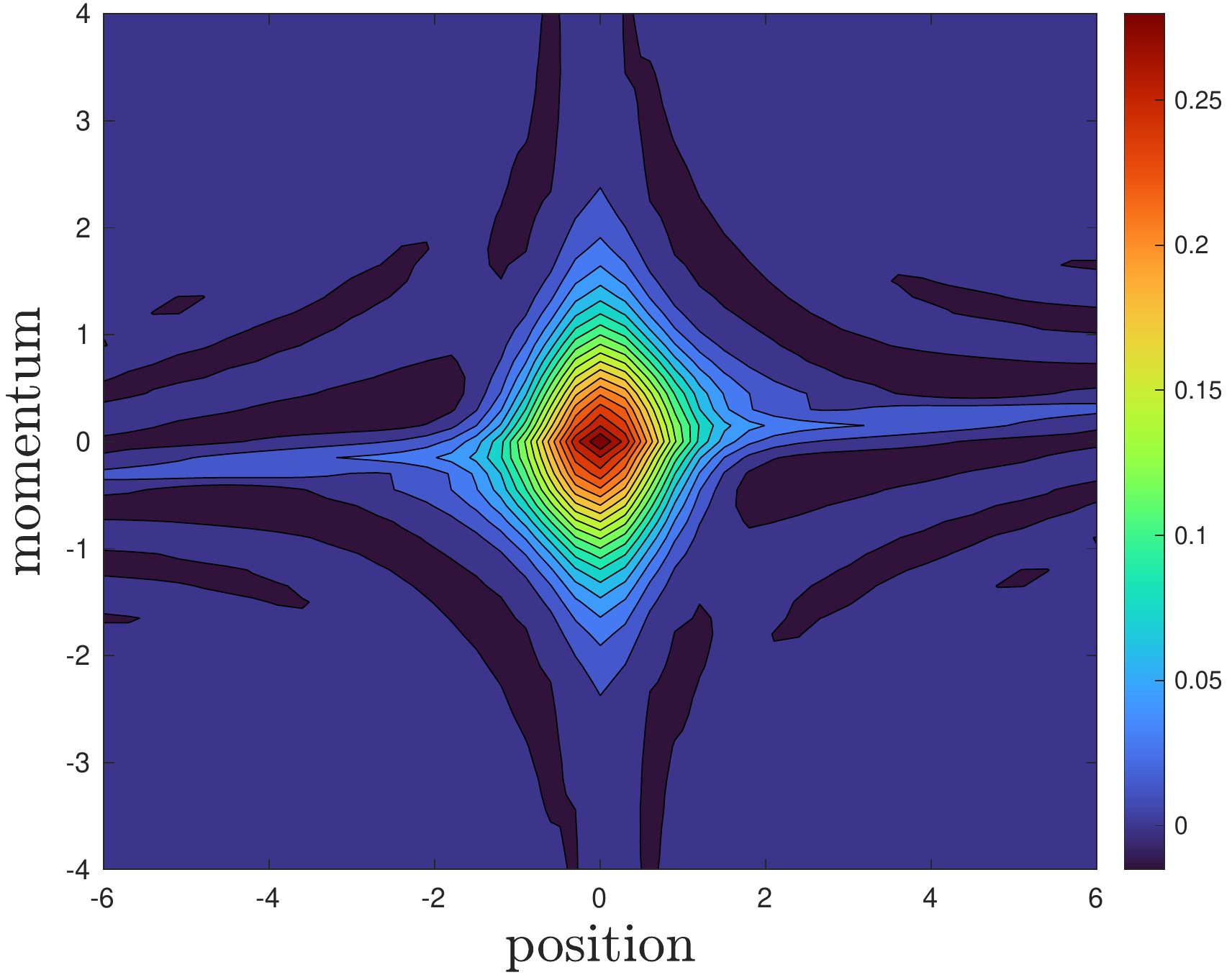}}}
\subfigure[Projection on $x_1$ direction.\label{xdist_marginal}]{
{\includegraphics[width=0.32\textwidth,height=0.18\textwidth]{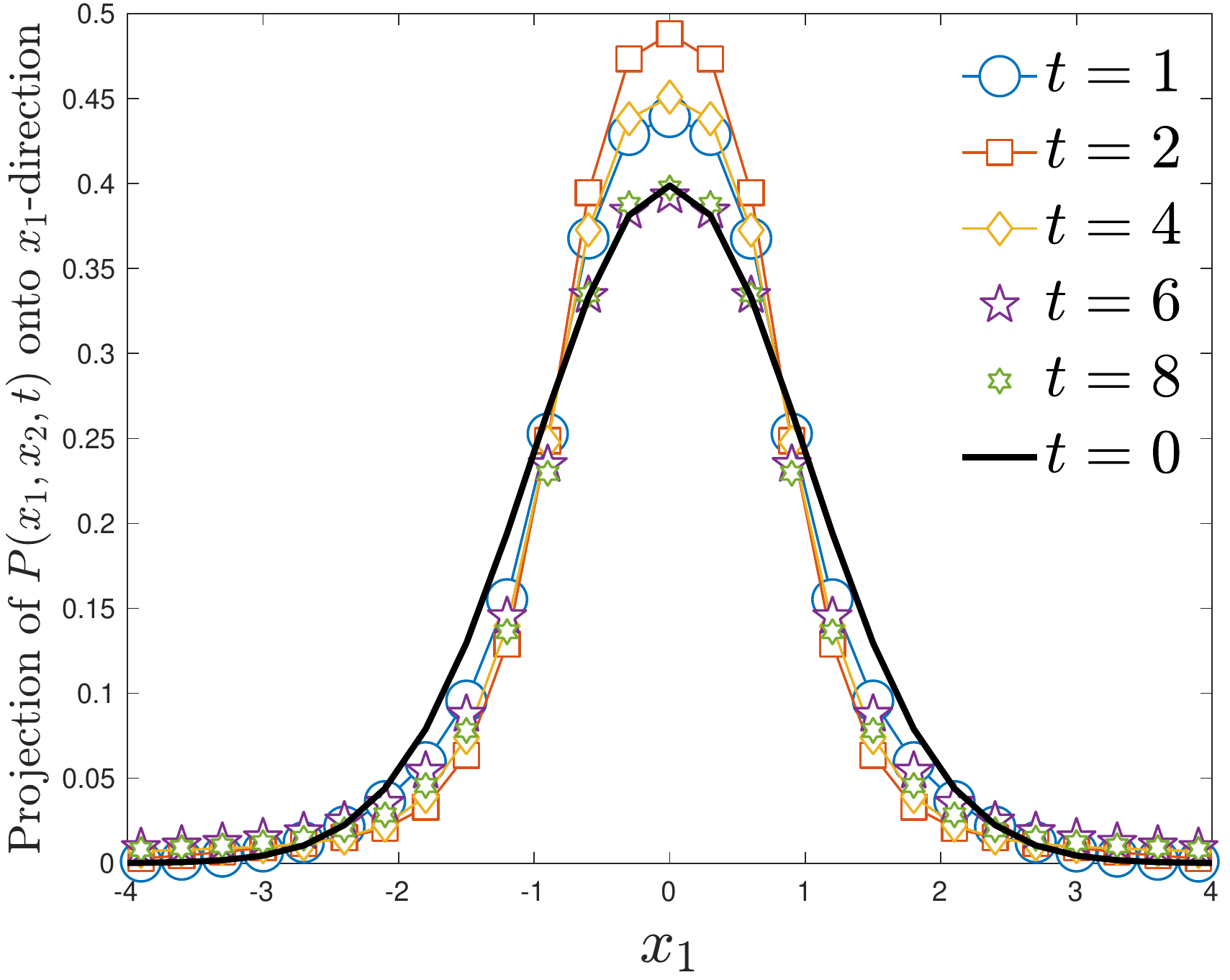}}}
\caption{\small   $H^+_2$ system:  Snapshots of the reduced Wigner functions on $(x_1$-$k_1)$ plane (left) and on $(x_2$-$k_2)$ plane (middle), and the spatial marginal distribution (right). \label{H2_time_evolution} }
\end{figure}

{\bf Spatial concentration}: The time evolutions of $P(x_1, x_2, t)$ are plotted in Figure \ref{H2_time_evolution}. In particular, Figure \ref{xdist_marginal} gives the projection of $P(x_1, x_2, t)$ onto $x_1$-direction, i.e., $\int_{\mathbb{R}} P(x_1, x_2, t) \D x_2$. It is seen that the electron is almost trapped in the field produced by two delocalized protons, and the wavepacket at $t = 1$a.u. is evidently more concentrated near the origin than the initial Gaussian. The peak of spatial marginal distribution reaches the maximum at $t = 2$a.u. Afterward, it gradually descends until $8$a.u., and begins to oscillate around a stable level. Clearly, the spatial marginal distribution has a fatter tail compared with the initial Gaussian profile.

{\bf Quantum tunneling}: In fact, the spatial concentration seems to be an outcome of the quantum uncertainty and tunneling. From the reduced Wigner functions in Figure \ref{H2_time_evolution}, one can see (1) the electron has certain probability to escape from the attractive potentials by two protons; (2) The quantum Coulomb interactions produce some negative regions, indicating that the electron with certain momentum is forbidden to escape; (3) The concentration of $P(x_1, x_2, t)$ seems to be related to the negative parts of the Wigner function as they ``squeeze'' the Gaussian wavepacket inside and force the electron to occupy the centre region with larger probability, while the heavy tail corresponds to the wavepacket that escapes from the attractive potentials.

\subsection{Implementation and parallelization}
\label{sec:parallel}

Finally, we provide details of parallel implementations in Table \ref{cpu_time}, including the memory requirement for storing a 6-D tensor in single precision, the computational time and corresponding platform.  

All the simulations are performed via our own Fortran implementation, with a mixture of MPI and OpenMP library to realize the distributed and shared-memory parallelization, respectively, and the domain is decomposed to $4^3$ patches ($2^3$ patches for the group with mesh size $41^3 \times 32^3$). It notes that the simulations under the mesh size $41^3 \times 32^3$ or $61^3 \times 32^3$ can be performed by a single computer without any difficulty in data storage, while other groups have to be performed on multiple computers due to the severe limitation of memory.

\begin{table}[!h]
  \centering
  \caption{\small The memory requirement of storing a 6-D tensor of size  $N_x^3 \times N_k^3$ in single precision, the computational time of LPC1 scheme up to $T = 5$a.u. ($\tau = 0.025$a.u., 200 steps) and the corresponding running  platform.  \label{cpu_time}}
\label{notation}
 \begin{lrbox}{\tablebox}
  \begin{tabular}{c|c|c|c|c}
\hline\hline
$N_x^3 \times N_k^3$	& Memory & High-performance Computing Platform 	&	Cores	&Time(h)\\
\hline
$41^3 \times 32^3$	&	$8.41$GB	& AMD 5950X (3.40GHz, 16C32T), 128GB Memory			&32		&13.27\\
$61^3 \times 32^3$	&	$27.71$GB	& AMD 2990WX (3.00GHz, 32C64T), 256GB Memory			&64		&66.16\\
 $61^3 \times 64^3$	&	$274.88$GB	& E5-2697A v4 (2.60GHz,16C32T), 256GB Memory $\times 8$		&256		&66.79\\
  $61^3 \times 80^3$	&	$432.93$GB	& E5-2697A v4 (2.60GHz,16C32T), 256GB Memory $\times 8$		&256		&88.67\\
 $81^3 \times 64^3$	&	$557.26$GB	& E5-2680  v4 (2.40GHz,14C28T), 256GB Memory $\times 16$	&448		&66.13\\
\hline\hline
 \end{tabular}
\end{lrbox}
\scalebox{0.82}{\usebox{\tablebox}}
\end{table}

\begin{figure}[h]
\centering
\includegraphics[width=0.48\textwidth,height=0.27\textwidth]{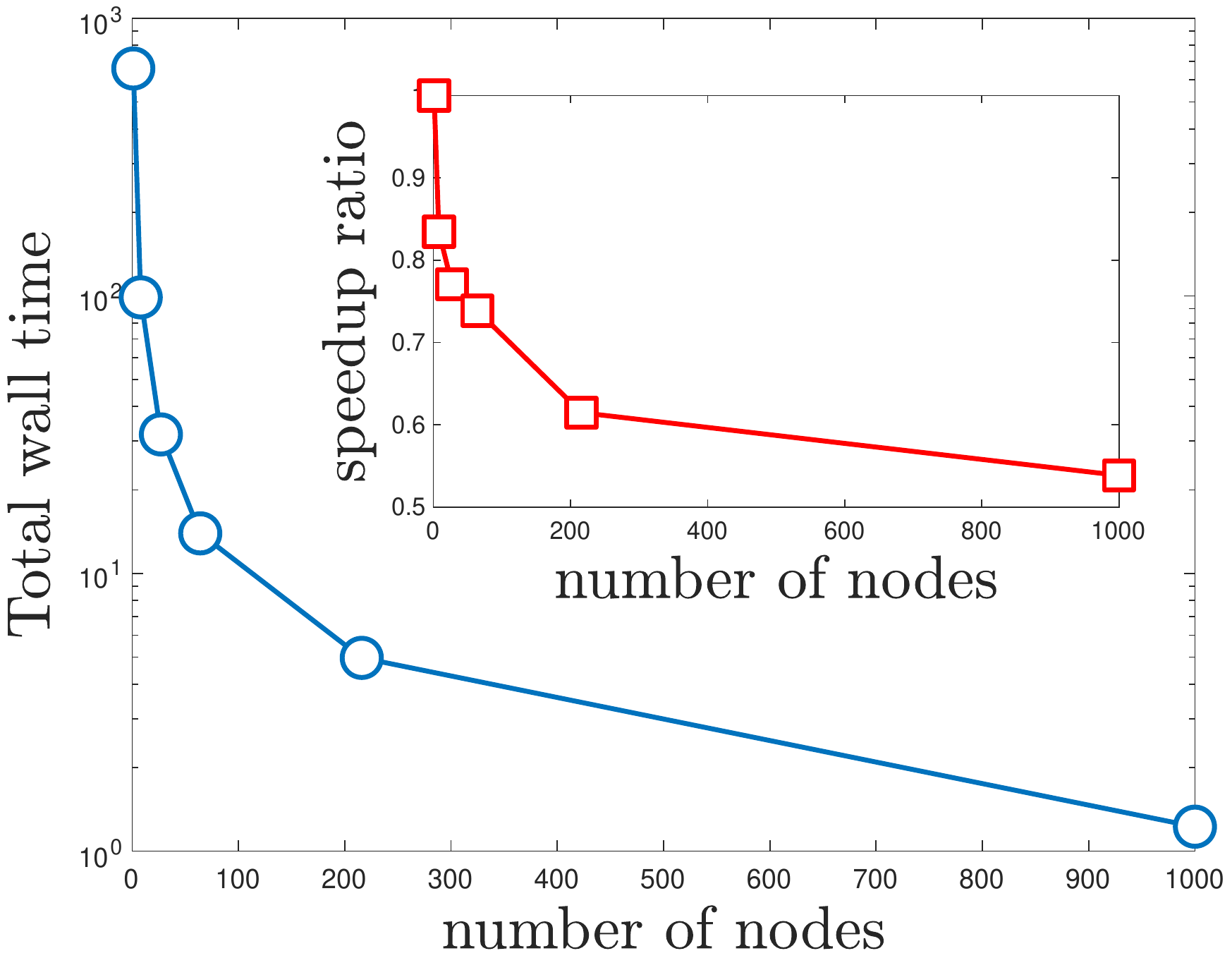}
\includegraphics[width=0.48\textwidth,height=0.27\textwidth]{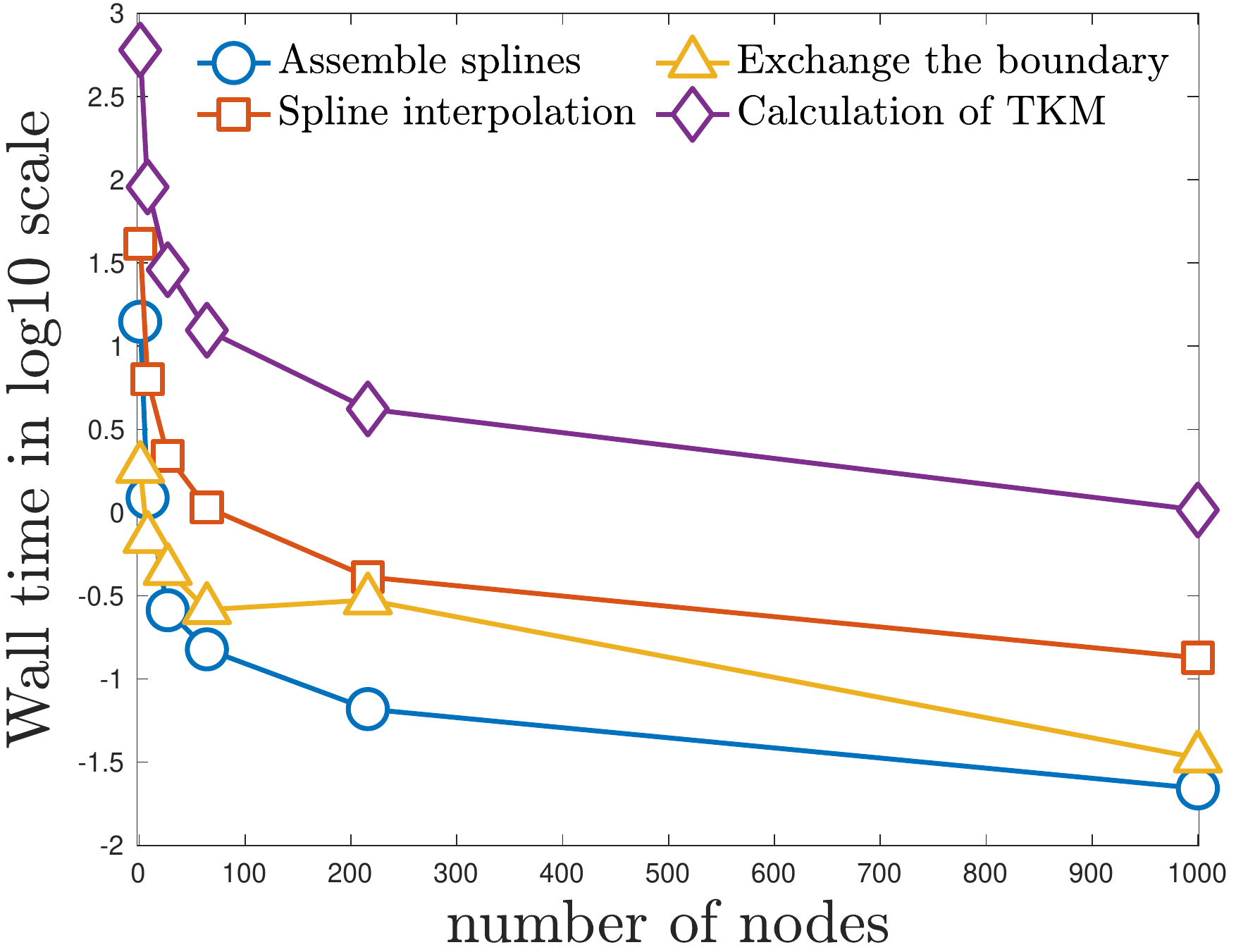}
\caption{\small Parallelization: CHASM achieves speedup ratio at least $53.84\%$ with the grid mesh $61^3\times 16^3$ distributed in $1000$ nodes, which is  further boosted when larger $N_k$ is used.
\label{fig_speed_ratio}}
\end{figure}

We have also tested the scalability of CHASM up to 1000 nodes and 16 threads per task (16000 cores in total) by simulating one-step Euler integration  under the grid mesh $61^3 \times 16^3$. The speedup ratio is presented in Figure \ref{fig_speed_ratio}. CHASM achieves the speedup ratio at least $53.84\%$ under $10\times 10 \times 10$ decomposition, where the calculation of $\pdo$ occupies most of computational time. Since the nonlocal calculation turns out to be the bottleneck in complexity, which scales as $\mathcal{O}(N_k^3\log N_k)$ according to Table \ref{TKM_convergence_data}, it is expected that CHASM can  achieve higher speedup ratio as $N_k$ increases.

\section{Conclusion and discussion}
\label{sec.discussion}

Numerical algorithms for high-dimensional Wigner equation have drawn a growing attention, but the lack of reliable reference solutions poses a major bottleneck to their design and evaluations. For 6-D Wigner-Coulomb dynamics, we propose a massively parallel scheme, termed CHAracteristic-Spectral-Mixed (CHASM). It exploits the local spline interpolation and the truncated kernel method to tackle the local spatial advection and nonlocal pseudodifferential operator with weakly singular symbol, respectively. CHASM may provide accurate references for a relatively new branch of particle-based stochastic Wigner simulations \cite{KosinaNedjalkovSelberherr2003,MuscatoWagner2016,ShaoXiong2019}, which may be potentially extended to even realistic many-body quantum systems (D $=$ 12) and further overcome the curse of dimensionality.

 It deserves to mention that the proposed scheme can be straightforwardly applied to other  6-D problems, including the Vlasov equation \cite{CrouseillesLatuSonnendrucker2009,Kormann2015,KormannReuterRampp2019} and the Boltzmann equation \cite{DimarcoLoubereNarskiRey2018} due to their strong similarities. In addition, several issues, including the generalization of CHASM to the fully nonlinear Wigner-Poission-Boltzmann equation and the GPU implementation, will be discussed in our future work.
 
 \section*{Acknowledgement}
This research was supported by the National Natural Science Foundation of China (No.~1210010642), the Projects funded by China Postdoctoral Science Foundation (No.~2020TQ0011, 2021M690227) and the High-performance Computing Platform of Peking University. 
SS is partially supported by Beijing Academy of Artificial Intelligence (BAAI). The authors are sincerely grateful to  Haoyang Liu and Shuyi Zhang at Peking University for their technical supports on computing environment, which have greatly facilitated our numerical simulations.

%
%



\end{document}